\documentclass[useAMS,usenatbib]{mn2e}

\usepackage{graphicx} 
\usepackage[usenames,dvipsnames]{color}
\usepackage{latexsym}
\usepackage{amssymb}
\usepackage{amsmath}
\usepackage{soul}

\usepackage{hyperref}	
\hypersetup{colorlinks=true,linkcolor=blue,citecolor=blue,filecolor=blue,urlcolor=blue}

\newcommand{\mo}{{\rm M}_\odot}

\title[Cosmological vertical waves]{A fully cosmological model of a Monoceros-like ring}
\author[F. A. Gomez et al.]{Facundo A. G\'omez$^{1}$\thanks{E-mail: fgomez@mpa-garching.mpg.de}, Simon D. M. White$^{1}$, Federico Marinacci$^{2}$, Colin T. Slater$^{3}$, \newauthor 
Robert J. J. Grand$^{4,5}$, Volker Springel$^{4,5}$, and R{\"u}diger Pakmor$^{4}$\\
$^{1}$Max-Planck-Institut f\"ur Astrophysik, Karl-Schwarzschild-Str. 1, D-85748, Garching, Germany\\
$^{2}$Department of Physics, Kavli Institute for Astrophysics and Space Research, MIT, Cambridge, MA 02139, USA\\
$^{3}$Department of Astronomy, University of Washington, Box 351580, Seattle, WA 98195, USA\\
$^{4}$Heidelberger Institut f\"ur Theoretische Studien, Schloss-Wolfsbrunnenweg 35, 69118 Heidelberg, Germany\\
$^{5}$Zentrum f\"ur Astronomie der Universitat Heidelberg, Astronomisches Recheninstitut, Monchhofstr. 12-14, 69120 Heidelberg, Germany\\
}

\begin{document}

\date{}

\pagerange{\pageref{firstpage}--\pageref{lastpage}} \pubyear{2015}

\maketitle

\label{firstpage}

\begin{abstract}
We study the vertical structure of a stellar disk obtained from a fully cosmological high-resolution hydrodynamical simulation of the 
formation of a Milky Way-like galaxy. At the present day, the disk's mean vertical height shows a well-defined and strong pattern, 
with amplitudes as large as 3 kpc in its outer regions. This pattern is the result of a satellite -- host halo -- disk interaction 
and reproduces, qualitatively, many of the observable properties of the Monoceros Ring. 
In particular we find disk material at the distance of Monoceros ($R \sim$ 12--16 kpc, galactocentric) 
extending far above the mid plane ( 30$^{\circ}$, $\langle Z \rangle \sim$ 1--2 kpc) in both hemispheres, as well as well-defined arcs 
of disk material at heliocentric distances $\gtrsim 5$ kpc. The pattern was first excited $\approx 3$ Gyr ago as an $m=1$
mode that later winds up into a leading spiral pattern. Interestingly, the main driver behind this perturbation is a low-mass low-velocity
fly-by encounter. The satellite has total mass, pericentre distance and pericentric velocity of $\sim 5\%$ of the host, $\sim 80$ kpc, and 
215 km/s, respectively. The satellite is not massive enough to directly perturb the galactic disk but we show that the density
field of the host dark matter halo responds to this interaction resulting in a strong amplification of the perturbative effects. 
This subsequently causes the onset and development of the Monoceros-like feature.

\end{abstract}


\section{Introduction}
\label{sec:introduction}

Several recent studies based on different astrometric catalogues have revealed a complex vertical structure in the Galactic disk.
These results have accumulated evidence in favor of an asymmetry about the Galactic mid-plane. The first of this series of studies 
was presented by \citet{2012ApJ...750L..41W}. Using a sample of 11,000 main sequence stars from the Sloan Extension for Galactic Understanding
and Exploration (SEGUE) survey, they showed that the Solar Neighborhood has a north-south asymmetry in both the number count and the mean 
vertical velocity distribution of stars. The study shows a 10 per cent number count deficit (excess), i.e. 
(north $-$ south)/(north $+$ south), 
at  $\pm 400$ pc ($\pm 800$ pc) from the Galactic mid-plane and 
a vertical velocity gradient of $\sim$ 3--5 km s$^{-1}$ kpc$^{-1}$. While the asymmetry 
in the vertical velocity suggests a breathing mode, i.e., rarefaction and compression of the
Galactic disk, the stellar number counts clearly suggest a bending mode, i.e., a local displacements of the disk from the midplane. 
These results were later supported by \cite{2013ApJ...777...91Y}, who carefully characterized the effects that uncertainties and potential biases 
in the data set could have on the discovery of vertical perturbations. In the following year, the presence of bulk vertical motions in the disk 
was confirmed by studies based on the Radial Velocity Experiment (RAVE) and the LAMOST/LEGUE survey 
\citep{2013ApJ...777L...5C,2013MNRAS.436..101W}. \citet{2013MNRAS.436..101W} mapped out bulk motions as a function of Galactocentric
radius and found evidence for compressional motion outside the solar circle and rarefaction inside. The observed peak vertical bulk velocities
are as large as 15 km s$^{-1}$.

The presence of a bending mode has also been supported by several studies. Using a sample of  main sequence turn-off stars provided by  
Pan-STARRS1 \citep[][]{2010SPIE.7733E..0EK}, \citet[][hereafter S14]{2014ApJ...791....9S} mapped the stellar distribution at galactic
latitudes $|b| \lesssim 30^{\circ}$, up to heliocentric distances of $\approx 17$ kpc. Their study revealed a much more complex 
morphology of the Monoceros (Mon) Ring than previously known. As discussed by S14, the Mon ring was originally identified as an overdensity of stars 
at $\sim 10$ kpc from the Sun, spanning Galactic latitudes from $b \sim +35^{\circ}$ to the edge of the Sloan Digital Sky Survey (SDSS)
footprint at $b \sim +20^{\circ}$ and in Galactic longitude extending between $l = 230^{\circ}$ and $l = 160^{\circ}$ 
\citep{new02,2003ApJ...594L.115R, yanny03,belu06}. S14 showed that this structure, which can be observed in both the North and South 
Galactic hemispheres, exhibits stream-like features and sharp edges. 
More importantly, they saw a North-South asymmetry, with the southern and northern parts dominating the regions closer
and further from the Sun, respectively. 

More recently, \citet{2015ApJ...801..105X}  showed that the SDSS number counts of main-sequence stars at Galactic latitudes 
$110^{\circ} < l < 229^{\circ}$ exhibit a clear and radially extended oscillatory behaviour. 
This corresponds to variations of the stellar density asymmetry whose amplitude
increases as a function of distance from the Sun, in the direction of the Galactic anticenter.  In particular, they identified four 
``substructures'' that represent the locations of peaks in the oscillations of the disk midplane. These four peaks are located at an 
approximate distance of 10.5 kpc (north), 13 kpc (south), 16.5 kpc (north) and 21 kpc (south) from the Galactic center. The last two peaks
correspond to the Mon ring and Triangulum-Andromeda (TriAnd) cloud  \citep{2004ApJ...615..732R,2004ApJ...615..738M}, respectively. 
This interpretation of the data was supported by \citet{2015arXiv150308780P}, who studied the number ratio of RR Lyrae to M 
giant stars, $f_{\rm RR:MG}$, associated with the TriAnd and TriAnd2 clouds \citep{2007ApJ...668L.123M}. They find that these two ring-like
over-densities, the latter located at $(R,Z) \approx (30,-10)$ kpc, have  stellar populations quite unlike any of the known satellites of 
the Milky Way and more similar to stars born in the much deeper potential of the Galactic disk.

Given the wealth of observations suggesting an asymmetric disk vertical structure it is interesting to explore possible
formation scenarios and to identify the main physical mechanisms driving these oscillations.  Naively one would expect vertical
modes to be excited by external perturbations. However, a series of recent studies have shown that breathing modes can arise
as a result of the interaction between disk stars and the bar \citep{2015arXiv150507456M} or spiral arms 
\citep{2014MNRAS.440.2564F, 2014MNRAS.443L...1D}. Even though these perturbations are likely to be taking place in the Galactic disk, they cannot 
however account for the observed bending modes. In contrast, \citet[][]{2014MNRAS.440.1971W} \citep[see also][]{2015MNRAS.450..266W} 
showed that a satellite galaxy plunging through the 
mid-plane of the disk can simultaneously excite both types of mode. The relative strength of these modes depends strongly, among 
other factors, on the vertical velocity of the passing satellite in comparison to that of the disk's stars. 

Several numerical models have been studied in the past to understand the effects that a satellite-disk interaction can have on the vertical 
structure of the disk \citep[e.g.,][]{1993ApJ...403...74Q,1999MNRAS.304..254V,2008MNRAS.391.1806V,2009ApJ...700.1896K,2014ApJ...789...90K,
2015MNRAS.446.1000F}.
\citet{2011Natur.477..301P} presented a set of simulations of the interaction between the Milky Way disk and the Sagittarius 
dwarf spheroidal (Sgr). Their study showed that such interaction can qualitatively reproduce many of the morphological features observed in
the Galactic disk, including a Mon ring-like feature. \citet[][hereafter G13]{2013MNRAS.429..159G} re-analyzed these simulations and showed 
that they can also reproduce the  observed local North-South asymmetry, as well as perturbations  in the in-plane velocity 
field observed on a sample of SEGUE F/G dwarf stars \citep{2012MNRAS.423.3727G}. In these simulations, the local north-south asymmetry
is a manifestation of a global oscillating mode perturbing the disk located at galactocentric distances $R 
\gtrsim 6$ kpc. \citet{2015arXiv150308780P} re-ran these simulations with higher numerical resolution to better sample the outer galactic 
regions. They find that rings of offset material with respect to the mid-plane of the disk can be found at distances as large as 30 kpc, 
reminiscent of the TriAnd clouds. Despite this success, these simulations fail to deposit enough disk material at the observed 
galactocentric heights, especially in the outer regions. 

Vertical perturbations in a galactic disk can also arise from the disk-host dark halo interaction \citep[see][for a recent 
review]{2013pss5.book..923S}. Either a misaligned DM halo (with respect to the 
disk) or the late time accretion of material can provide enough torque to form strong warps 
\citep[e.g.][]{1999ApJ...513L.107D,1999MNRAS.303L...7J,2006MNRAS.370....2S,2012MNRAS.426..983D,2013MNRAS.428.1055A,2014arXiv1411.3729Y}.
In addition, as shown by \citet{2000ApJ...534..598V}, even an encounter with a low-mass, low-velocity fly-by that penetrates 
the outer regions of a galaxy can generate asymmetric features in the host DM halo density field. These density perturbations can 
be efficiently transmitted to the inner parts of the primary system and subsequently perturb an embedded galactic disk. Such 
perturbations can induce the formation of vertical patterns \citep[e.g.][]{1998MNRAS.299..499W}.

In this work we study the vertical structure of a Milky Way-like galaxy formed in fully cosmological hydrodynamical simulation. The simulated disk 
analyzed here is obtained from the highest numerical resolution simulation of the suite presented by \citet[][hereafter M14]{2014MNRAS.437.1750M}.
As discussed by M14, this system is one of the best Milky Way-look alikes among the suite. It is  strongly disk--dominated with a 
realistic rotation curve, a surface density profile close to exponential, and both an age distribution and a size
that are consistent with expectations from large galaxy surveys in the local Universe. In Section~\ref{sec:simu} we briefly review the main
properties of the simulation. We characterize the time evolution of the disk's vertical structure in Section~\ref{sec:relevance} and show that,
at the present day, it shows a strong and well defined vertical pattern that matches many of the observational properties of 
the Mon ring. In Section~\ref{sec:pert} we identify and characterize the main driving source of this vertical perturbation. We summarize and discuss
our results in Section~\ref{sec:discussion}.

\section{The Simulations}
\label{sec:simu}

In this work we analyze a cosmological hydrodynamical simulation of the 
formation of a Milky Way-like galaxy. In what follows we briefly summarize 
the methodology and the main characteristics of this simulation. 

Our work focuses on one of the Milky Way-like galaxies first introduced in M14. Following  previously adopted nomenclature,
we will refer to this galaxy as Aq-C-4. 
The simulations were run adopting a $\Lambda$CDM cosmology with parameters 
$\Omega_{m}  = \Omega_{dm} + \Omega_{b} = 0.25$, $\Omega_{b} = 0.04$, 
$\Omega_{\Lambda}= 0.75$,  $\sigma_{8}  = 0.9$, $n_{s}  =  1$,  and  Hubble  
constant  $H_{0}  =  100~h$  km  s$^{-1}$ Mpc$^{-1}$ = 73  km s$^{-1}$ Mpc$^{-1}$. 
The halo was first  identified  in   a  lower   resolution  version   of  the
Millennium-II Simulation  \citep{bk09} which was carried  out within a
periodic box of  side 125 $h^{-1}$ Mpc. Applying the ``zoom-in'' technique, the galaxy was 
simulated multiple times at different resolutions, increasing in each step the 
mass-resolution by a factor of 8. After setting up the initial dark matter distribution, gas was added by 
splitting each dark matter particle into a dark matter particle and gas cell pair. The masses assigned to each 
are determined from the cosmological baryon mass fraction. The dark matter particle and gas cell in each pair 
are separated by a distance equal to half the mean inter-particle spacing. Their phase-space coordinates are chosen 
such that the centre of mass and velocity of the pair is the same as the original dark matter particle.

The simulations were carried out using the moving-mesh code \textrm{AREPO} 
\citep{2010MNRAS.401..791S}. AREPO solves the gravitational
and collisionless dynamics by using a TreePM approach \citep{springel2005a}. 
To follow the evolution of the gas component this code solves the Euler equations
on an unstructured Voronoi mesh by adopting a finite-volume
discretization. The main advantage of AREPO is that the set of points generating 
its unstructured Voronoi mesh are allowed to move freely, thus inducing a
transformation of the mesh that adapts itself to the characteristics of
the flow. For more details about this method, see \citet[][]{2010MNRAS.401..791S} 
and \citet[][]{2012MNRAS.425.3024V}. 

The simulations include treatments for the critical 
physical processes that govern galaxy formation, such as
gravity, gas cooling/heating, star formation, mass return and metal
enrichment from stellar evolution, and feedback from stars and supermassive
black holes. The parameters that regulate the efficiency of each process have been 
selected by contrasting the results obtained from cosmological simulations to the 
corresponding observational quantities \citep[][M14]{2013MNRAS.436.3031V}. The resulting 
fiducial settings produce a good match to the stellar mass to halo mass function, the 
galaxy luminosity functions, the history of the cosmic star formation rate density and 
several other important observables. 

As discussed by M14, Aq-C-4 is one the most similar to the Milky Way among the models analyzed in their
work. It has a flat rotation curve that peaks at $v_{\rm c}^{\rm max} \approx 250$ km s$^{-1}$ at $\sim 4$ kpc. 
A decomposition of the stellar surface density profile into an exponential disk and a S{\'e}rsic profile yields a disk scale length and 
a bulge effective radius of approximately 3.1 and 0.8 kpc, respectively. The decomposition results in a disk-to-total mass ratio, 
D/T $= 0.93$. A more robust determination of D/T based on a kinematic decomposition using all star particles within 
$0.1~R_{\rm vir}$ yields a value of 0.67. Both results indicate the presence of a dominant disk. The resulting galaxy falls well 
within the observed size-luminosity relation in the {\it B} band and matches the Tully-Fisher relation. Also it exhibits a present 
day gas fraction in reasonable agreement with the Milky Way. Behind this well characterized disk lies a quiet formation history with 
no significant perturbations to its steady slow growth since $z\approx 1$. Interestingly, the latest external perturbation experienced 
by this stellar disk takes place close to $z=0$, approximately 2.7 Gyr ago. As we will show later in Section~\ref{sec:pert}, a fly-by 
encounter with a $\sim 4 \times 10^{10}\mo$ satellite causes the vertical pattern observed in the disk at the present day. The main 
properties of this model at the present day are summarized in Table~\ref{tab:properties}

Throughout this paper, unless otherwise stated, we define a disk particle to be any star particle that {\it i)} shares the same 
sense of rotation as the overall disk, as determined by the sign of the $Z$-component of its angular momentum, {\it ii)} has a vertical
velocity $v_{\rm z} < v_{\rm c}^{\rm max}$ and {\it iii)} is located at $|Z| \leq 15$ kpc. These very simple selection criteria serve 
to significantly suppress contamination from the accreted component of the stellar halo. Tests performed with stronger selection 
criteria showed no significant differences in our results. Furthermore, as previously mentioned, this galaxy is strongly disk dominated. Thus, contamination from the spheroidal component is minimal.  

\begin{figure}
\centering
\includegraphics[width=82mm,clip]{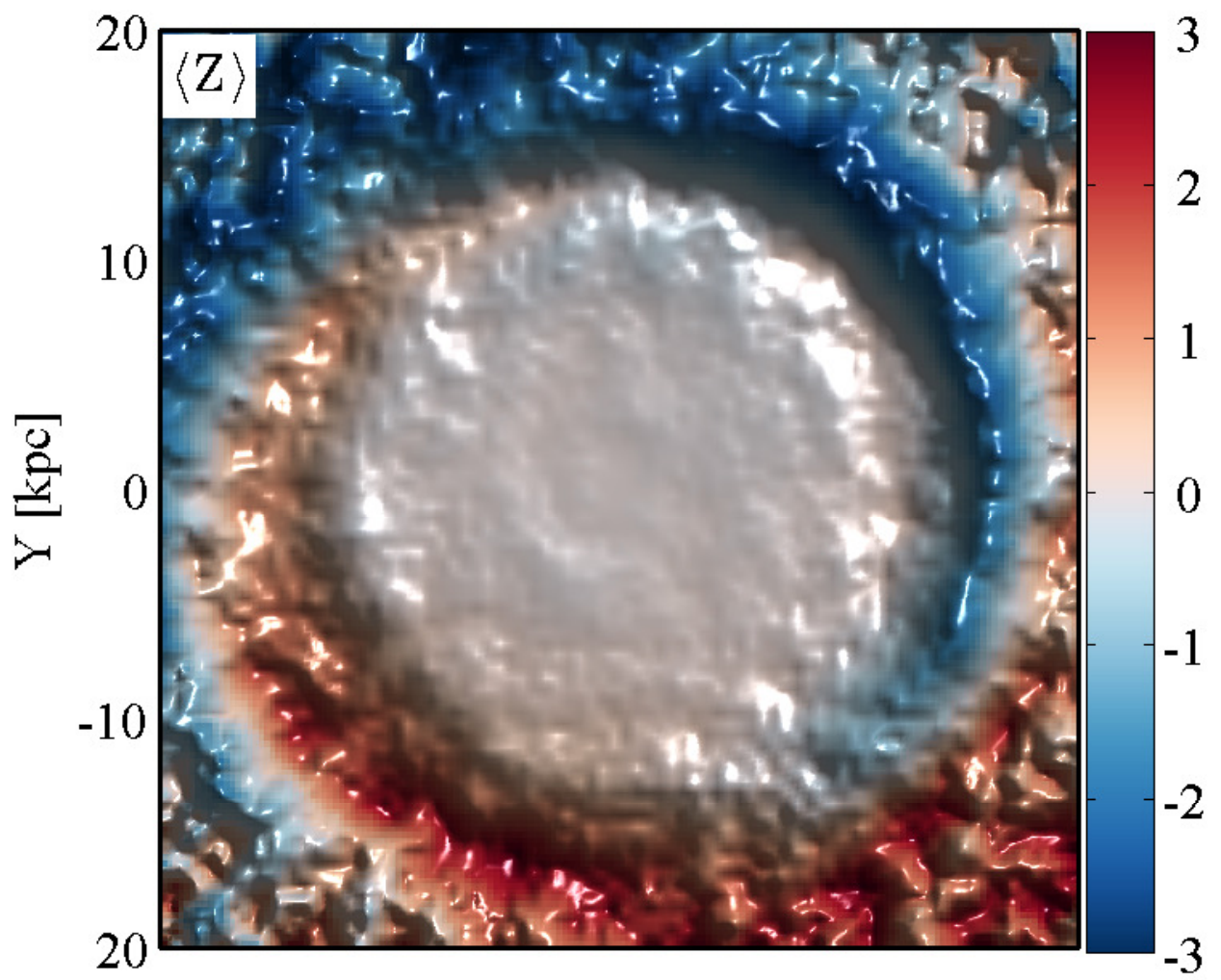}\\
\includegraphics[width=85mm,clip]{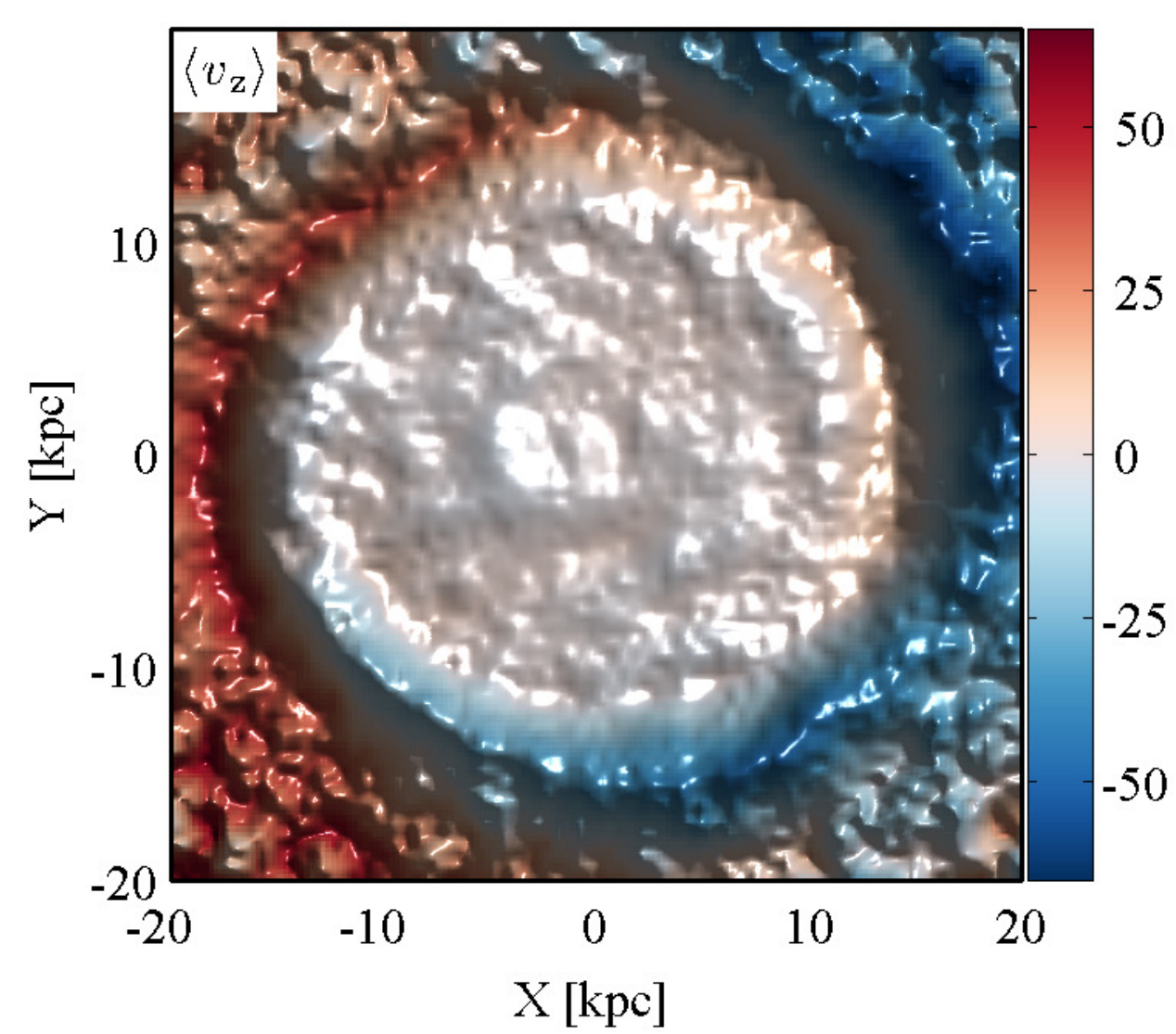}
\caption{{\it Top panel:} Map of the simulated galactic disk's mass-weighted $\langle {\rm Z} \rangle$  at the present day. 
The different colours and the relief indicate different values of $\langle {\rm Z} \rangle$  in kpc. {\it Bottom panel:} As in the top panel but 
for $\langle v_{\rm z} \rangle$. The colour bar is in units of km/s. In both maps a very clear vertical pattern can be observed. Note 
the anticorrelation between $\langle {\rm Z} \rangle$ and $\langle v_{\rm z} \rangle$.}
\label{fig:maps_t0}
\end{figure}

\begin{table*}
\centering
\begin{tabular}{lrrrrrrrrrrr}
\hline
Run & 
$R_{\rm vir}$  &
$M_{\rm gas}$  &
$M_\star$      &
$M_{\rm dm}$   &
$N_{\rm cells}$ &
$N_\star$      &
$N_{\rm dm}$  &
$m_{\rm gas}$ &
$m_{\rm dm}$  &
$\epsilon$ \\
 &
$({\rm kpc})$   & 
$(10^{10}\mo)$ & 
$(10^{10}\mo)$ & 
$(10^{10}\mo)$ & 
$(10^6)$ &
$(10^6)$ &
$(10^6)$ &
$(10^5\mo)$ & 
$(10^5\mo)$ & 
$({\rm pc})$ 
\\
\hline
Aq-C-4 & 234.4 & 8.39 &  5.31 & 145.71 & 1.53 & 1.63 & 5.4 &  0.51 &   2.70  & 340 \\
\hline
\end{tabular}
\caption{Properties of the simulated galaxy at the present day. Following previous convention, we will 
refer to it as Aq-C-4. From left to right, the columns 
give the virial radius defined by a sphere enclosing an overdensity of 200 with
respect to the critical density; the total gas, stellar and dark matter 
masses inside the virial radius; the  numbers of gaseous cells,
star and dark matter particles; the gas and dark matter mass resolutions 
in the  high-resolution region and the softening length in physical units.}
\label{tab:properties}
\end{table*}

\section{Galactic disk's vertical structure}
\label{sec:relevance}

In order to characterize the present day disk's vertical structure, we have carefully aligned the disk with the X–Y plane. This is done by iteratively 
computing, and aligning with the Z-direction, the total angular momentum of the star particles located within 5 kpc radius cylinders 
of decreasing height. The star particles considered for this procedure have been chosen to have ages $\leq 3$ Gyr. In this way we {\it i)} 
 take into account only the coldest component of the stellar disk and {\it ii)} we minimize contamination from the spheroidal (older) 
stellar component. 

\begin{figure}
\centering
\includegraphics[width=85mm,clip]{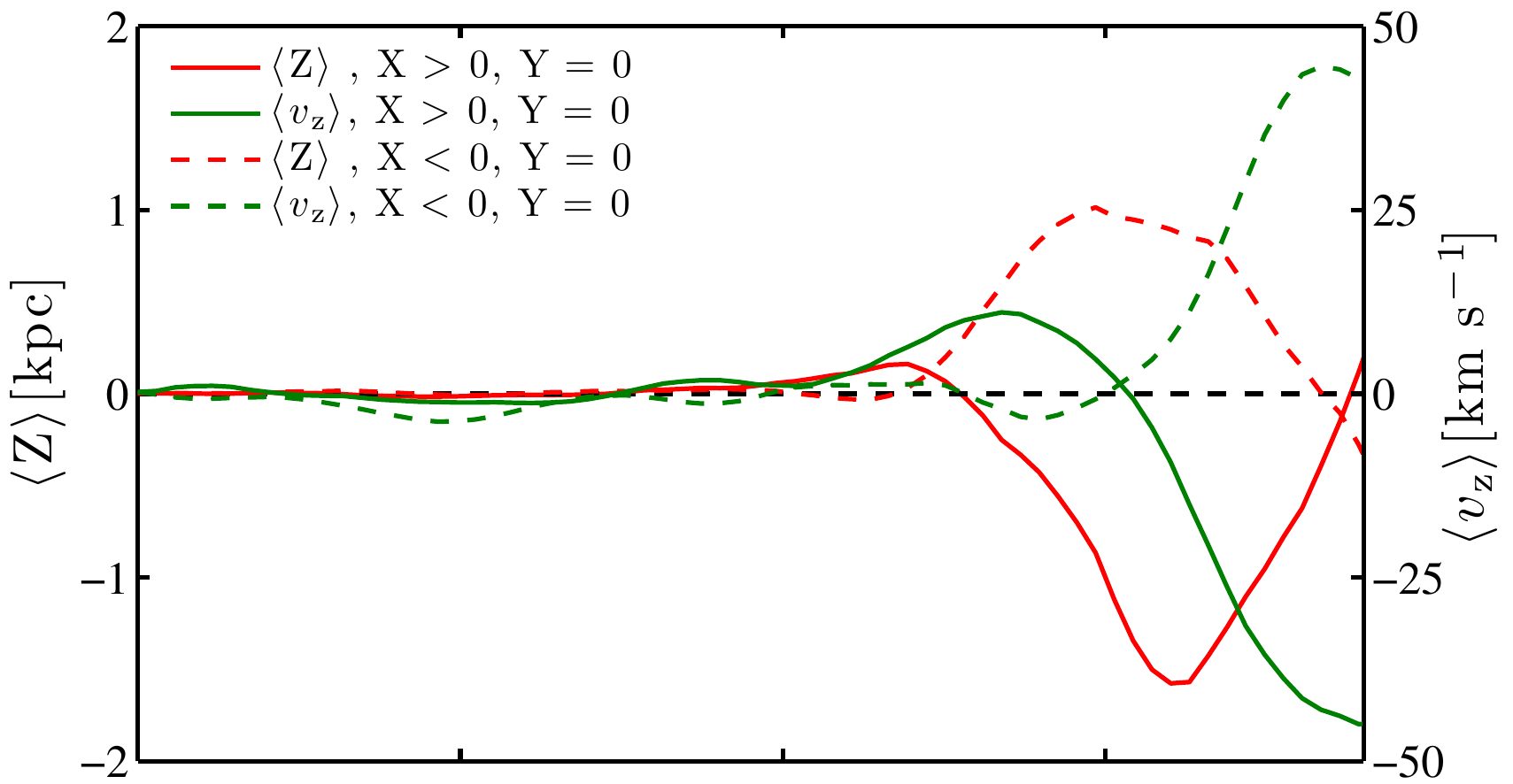}\\
\includegraphics[width=85mm,clip]{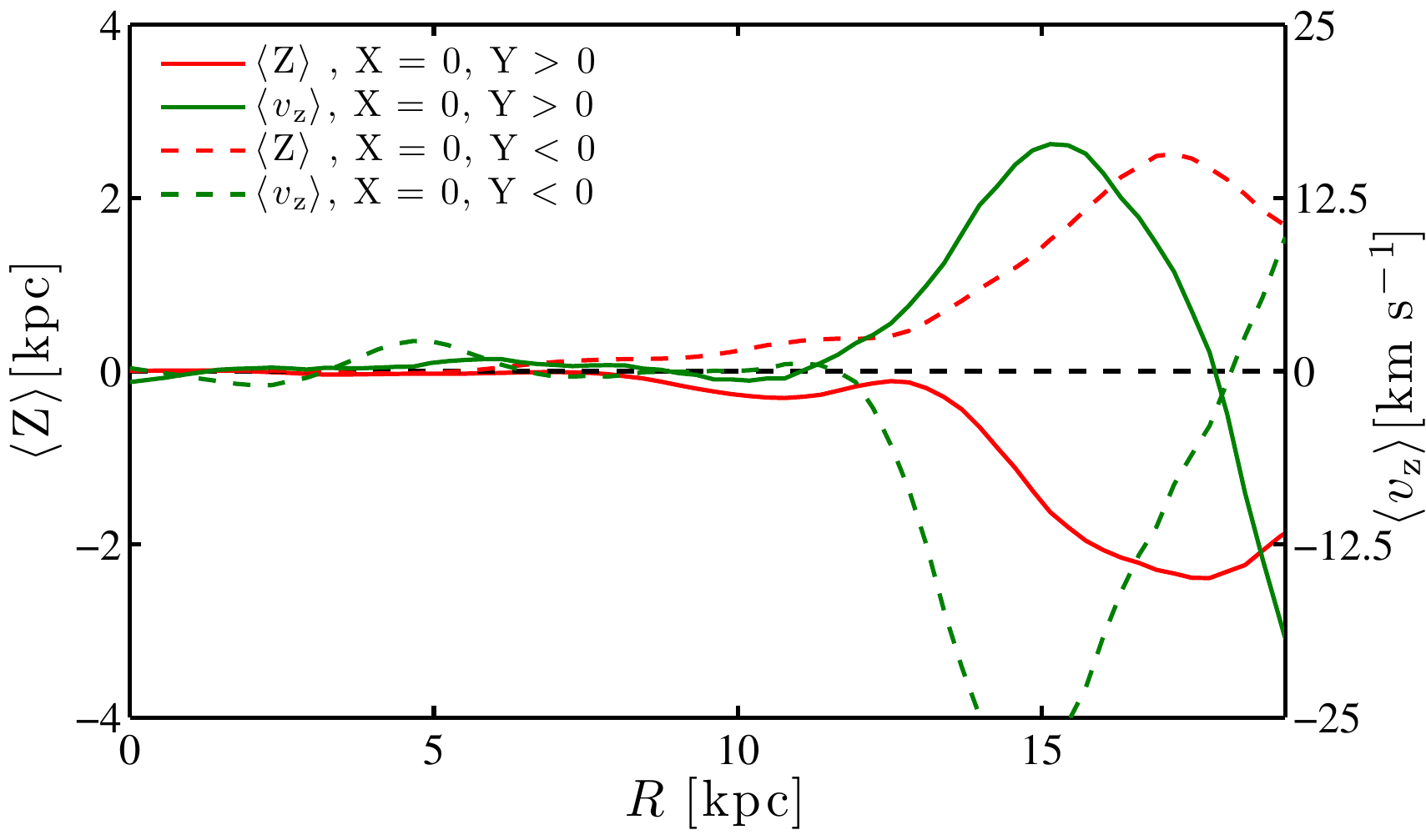}
\caption{Red lines show the variation of the mass-weighted mean height of the disk, $\langle {\rm Z} \rangle$, as a function of
galactocentric radius along two pairs of diametrically opposed galactic longitudes. The solid line indicates the value of 
$\langle {\rm Z} \rangle$  along the positive X, Y$=0$ kpc (top panel) and positive Y, X=$0$ kpc (bottom panel) directions. 
The dashed line shows the results obtained in the opposite directions, $180^{\circ}$ away. The green lines show the same results 
for  mass-weighted mean vertical velocity, $\langle v_{\rm z} \rangle$. In the bottom panel, the axis range for $\langle {\rm Z} \rangle$ 
($\langle v_{\rm z} \rangle$) has been scaled up (down) by a factor of 2 to account for the change in the phase of the 
vertical pattern as a function of galactocentric longitude. Note the specular behaviour presented by both $\langle {\rm Z} \rangle$ and 
$\langle v_{\rm z} \rangle$ along these two pairs of galactocentric lines of sight. Note as well the very clear oscillatory 
behaviour presented by this wave.}
\label{fig:los}
\end{figure}

Following G13, to obtain a map of the mass-weighted mean height, $\langle {\rm Z} \rangle$, we grid the disk with a regular Cartesian mesh of bin
size = 0.5 kpc aligned with the X-Y plane. On each grid node we centre a 1 kpc radius cylinder of 10 kpc height and compute the mass-weighted 
$\langle {\rm Z} \rangle$. 
The top panel of Figure~\ref{fig:maps_t0} shows such a map out to $R = 20$ kpc. The outer galactic disk shows a strong and well-defined 
vertical pattern that becomes noticeable at $R \gtrsim 12$ kpc. The inner galactic regions appear unperturbed. Rather than a set of 
perfect concentric rings, the vertical pattern shows a spiral morphology that winds into the inner regions with its amplitude gradually 
decreasing. 
Note as well, the antisymmetric shape of the vertical pattern about the galactic centre. The bottom panel of Figure ~\ref{fig:maps_t0} shows
a map of the mass-weighted mean vertical velocity of the galactic disk, $\langle v_{\rm z} \rangle$. The vertical pattern is also very clear in
the $\langle v_{\rm z} \rangle$ map. The amplitude of this perturbation is very large, especially in the outer region of the disk, where
$\langle v_{\rm z} \rangle \approx 65$ km/s. There is a significant anti-correlation with respect to the $\langle {\rm Z} \rangle$ map. 
This can be more clearly seen in Figure~\ref{fig:los}, where we show $\langle {\rm Z} \rangle$ (red lines) and $\langle v_{\rm z} \rangle$ 
(green lines) as a function of galactocentric radius along two pairs of diametrically opposed galactic longitudes. 
In this figure we observe a near-antisymmetric behaviour along these diametrically opposed lines in
both $\langle {\rm Z} \rangle$ and $\langle v_{\rm z} \rangle$. This figure
also shows a very clear oscillatory behaviour. At galactocentric radii where $\langle {\rm Z} \rangle$ is extremal, 
$\langle v_{\rm z} \rangle \approx 0$ km/s, whereas when $\langle v_{\rm z} \rangle$  is extremal, $\langle {\rm Z} \rangle \approx 0$ kpc.

\begin{figure*}
\includegraphics[width=36.5mm,clip]{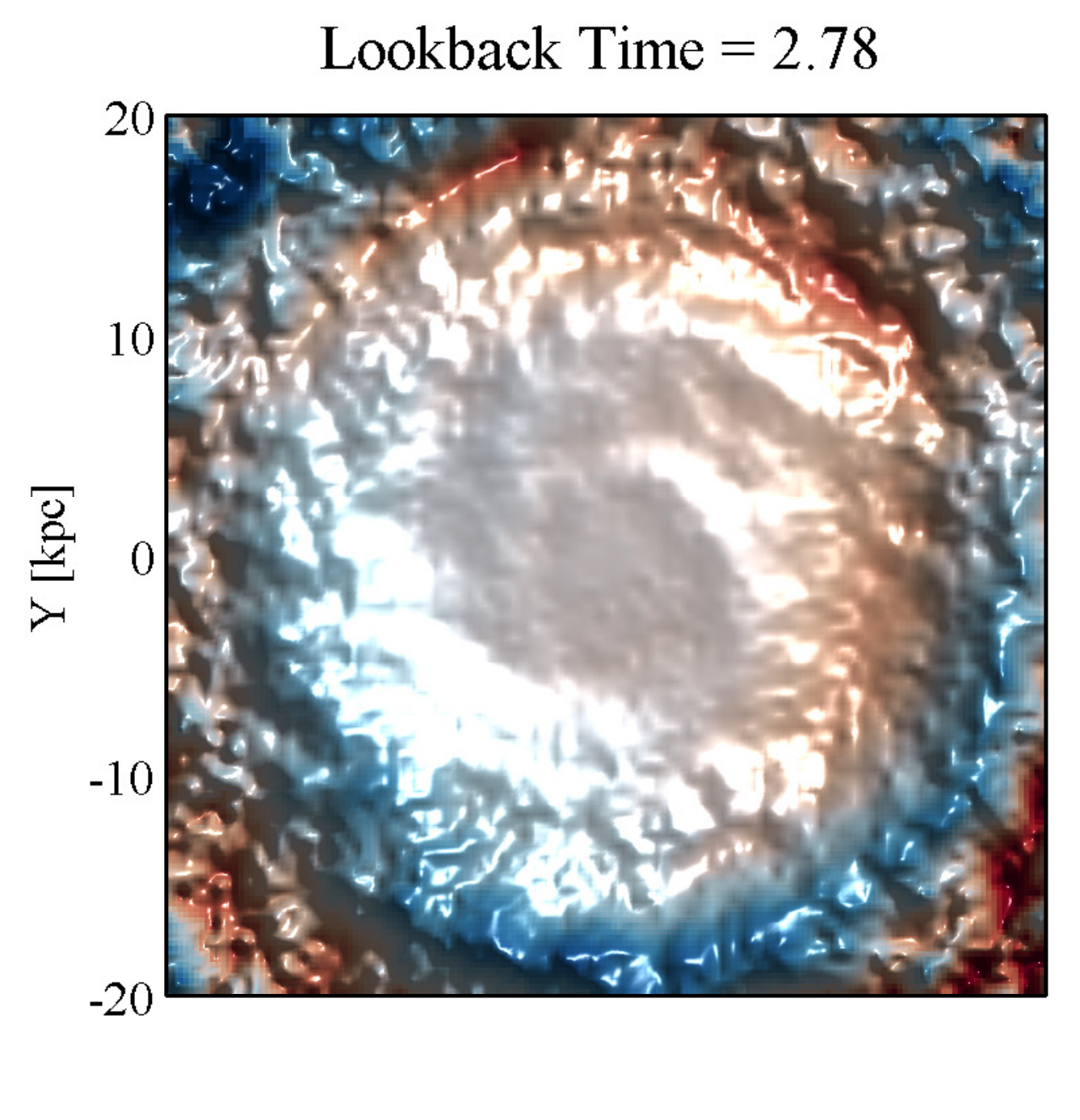}
\hspace{-0.18cm}
\includegraphics[width=33.5mm,clip]{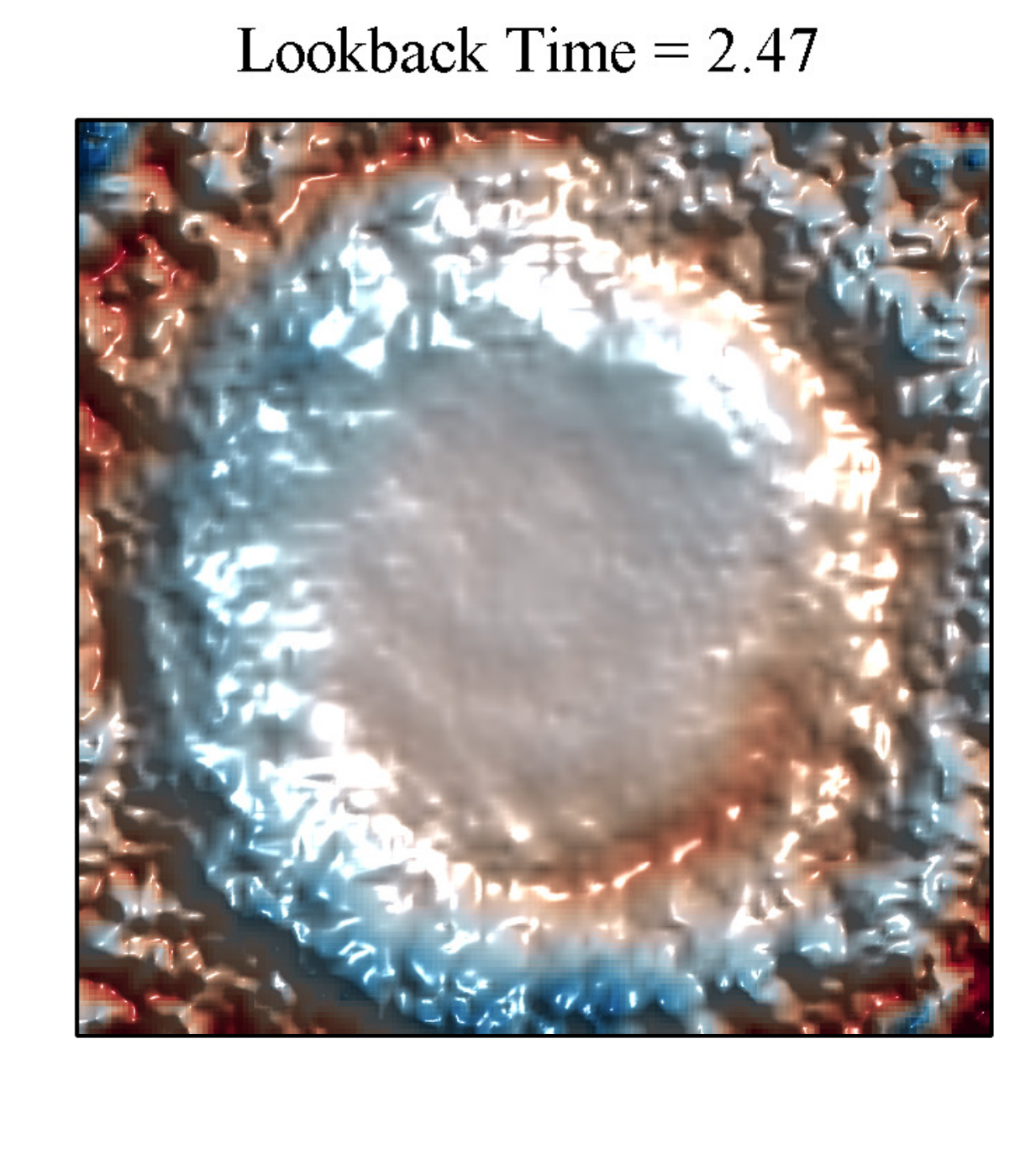}
\hspace{-0.21cm}
\includegraphics[width=33.5mm,clip]{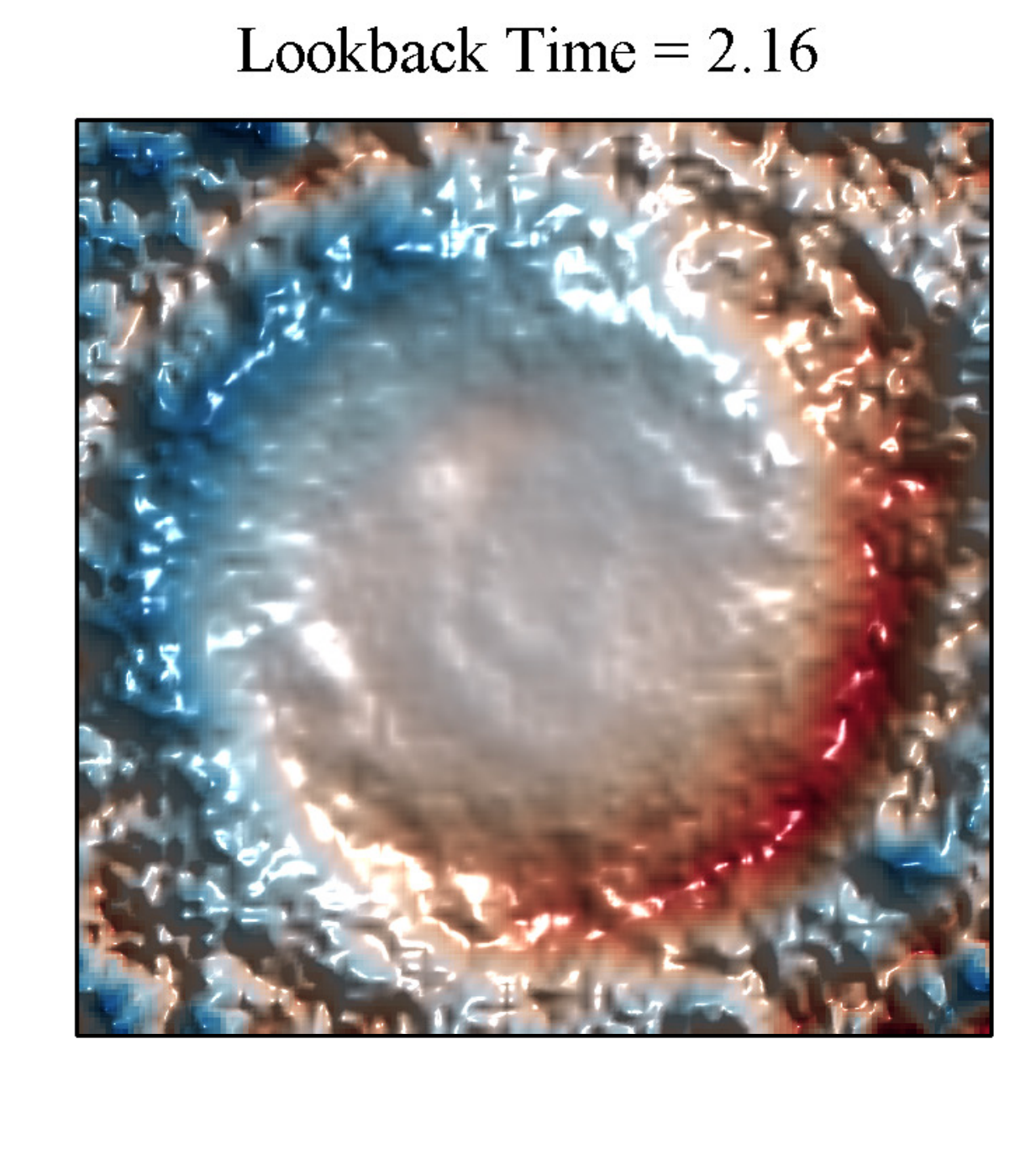}
\hspace{-0.21cm}
\includegraphics[width=33.5mm,clip]{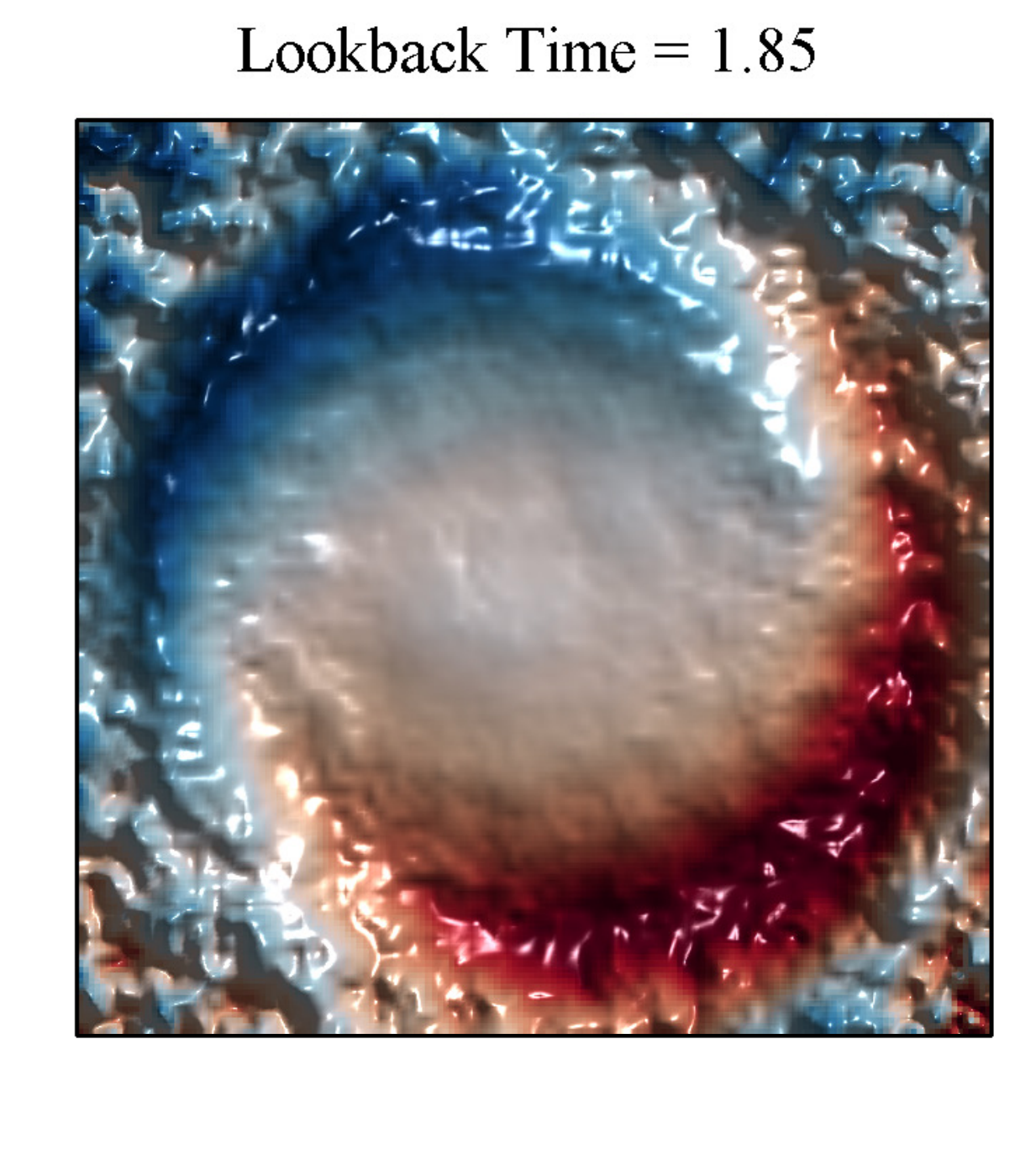}
\hspace{-0.21cm}
\includegraphics[width=38mm,clip]{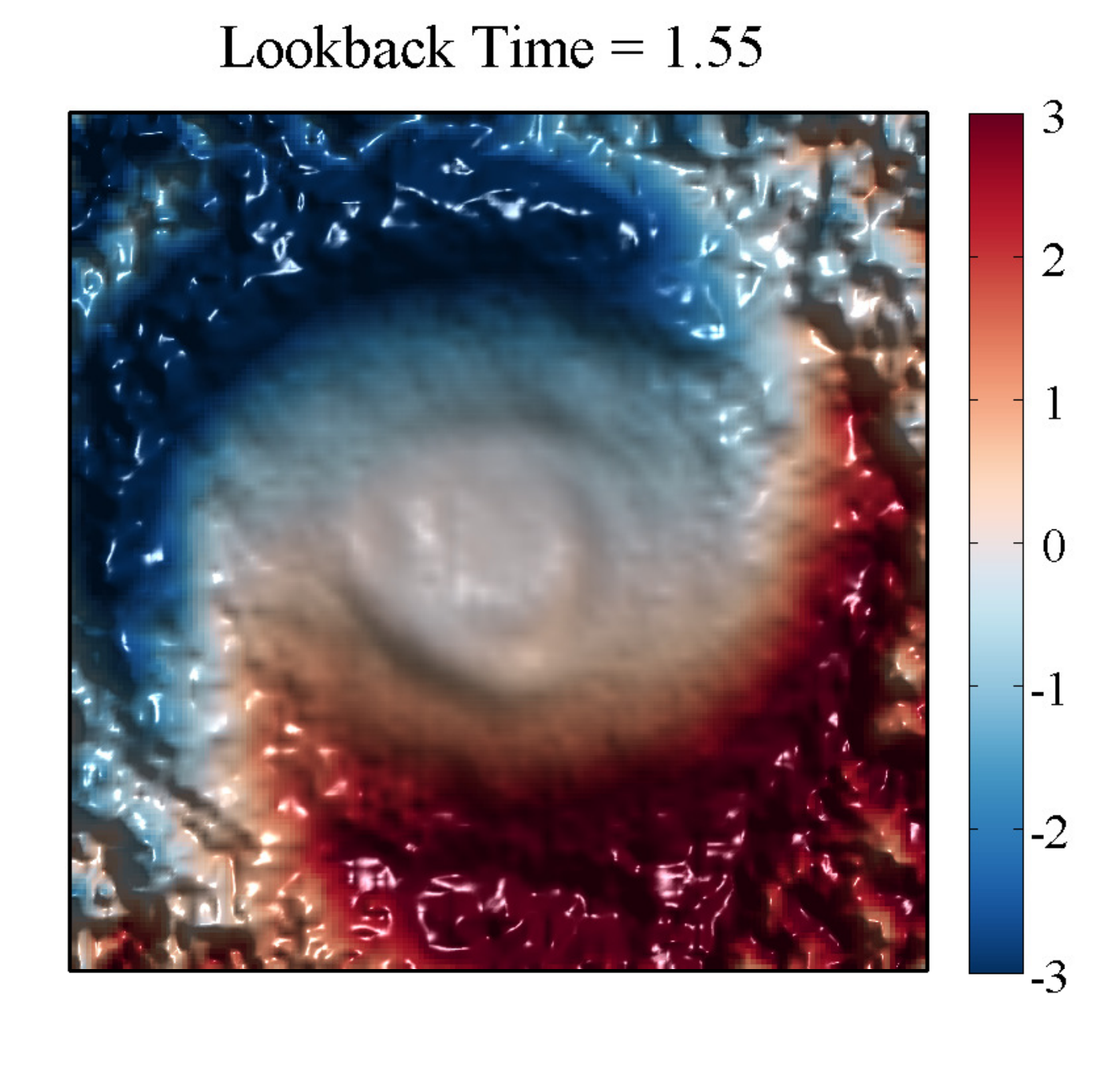}
\\
\includegraphics[width=36.5mm,clip]{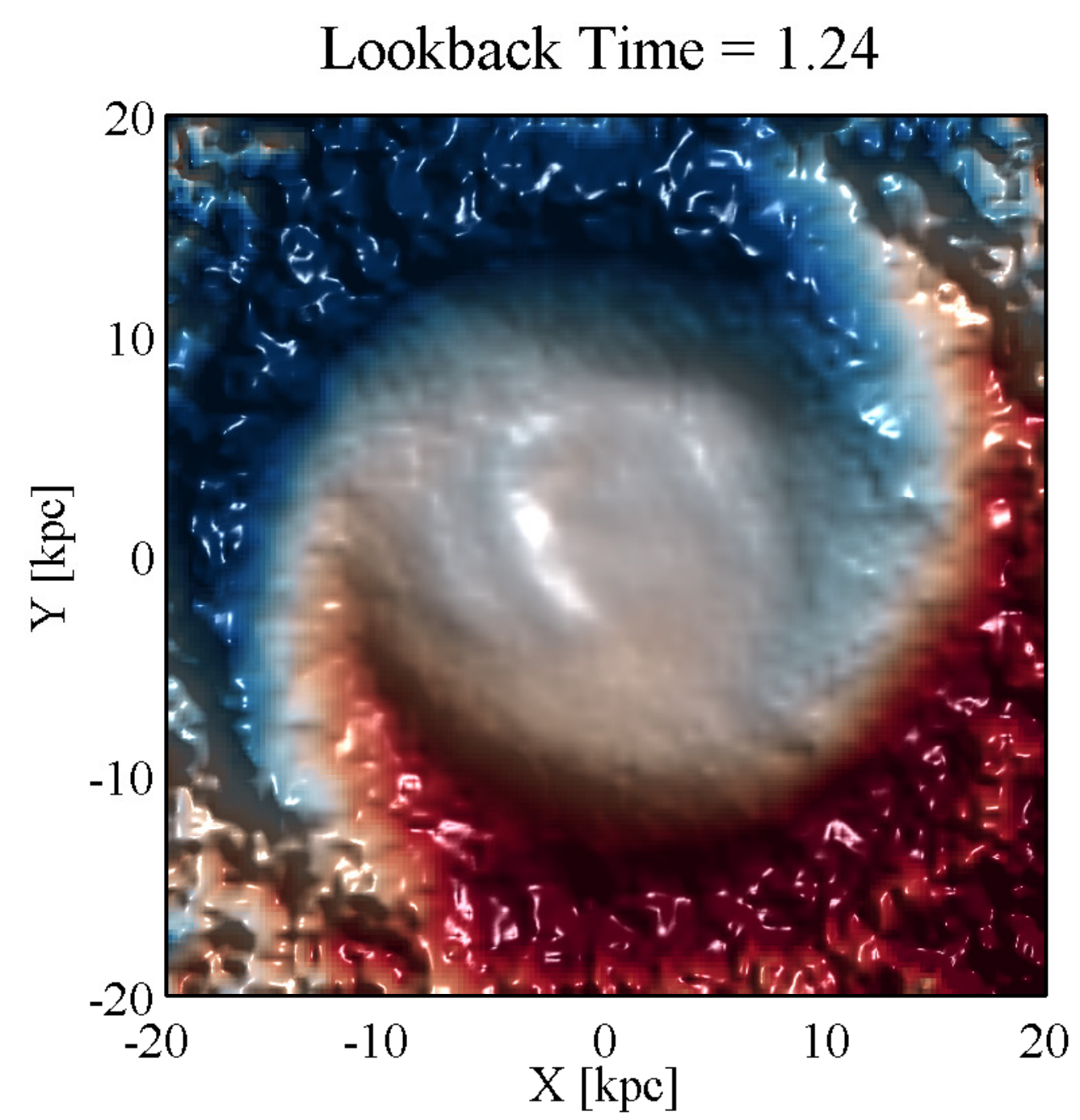}
\hspace{-0.18cm}
\includegraphics[width=33.5mm,clip]{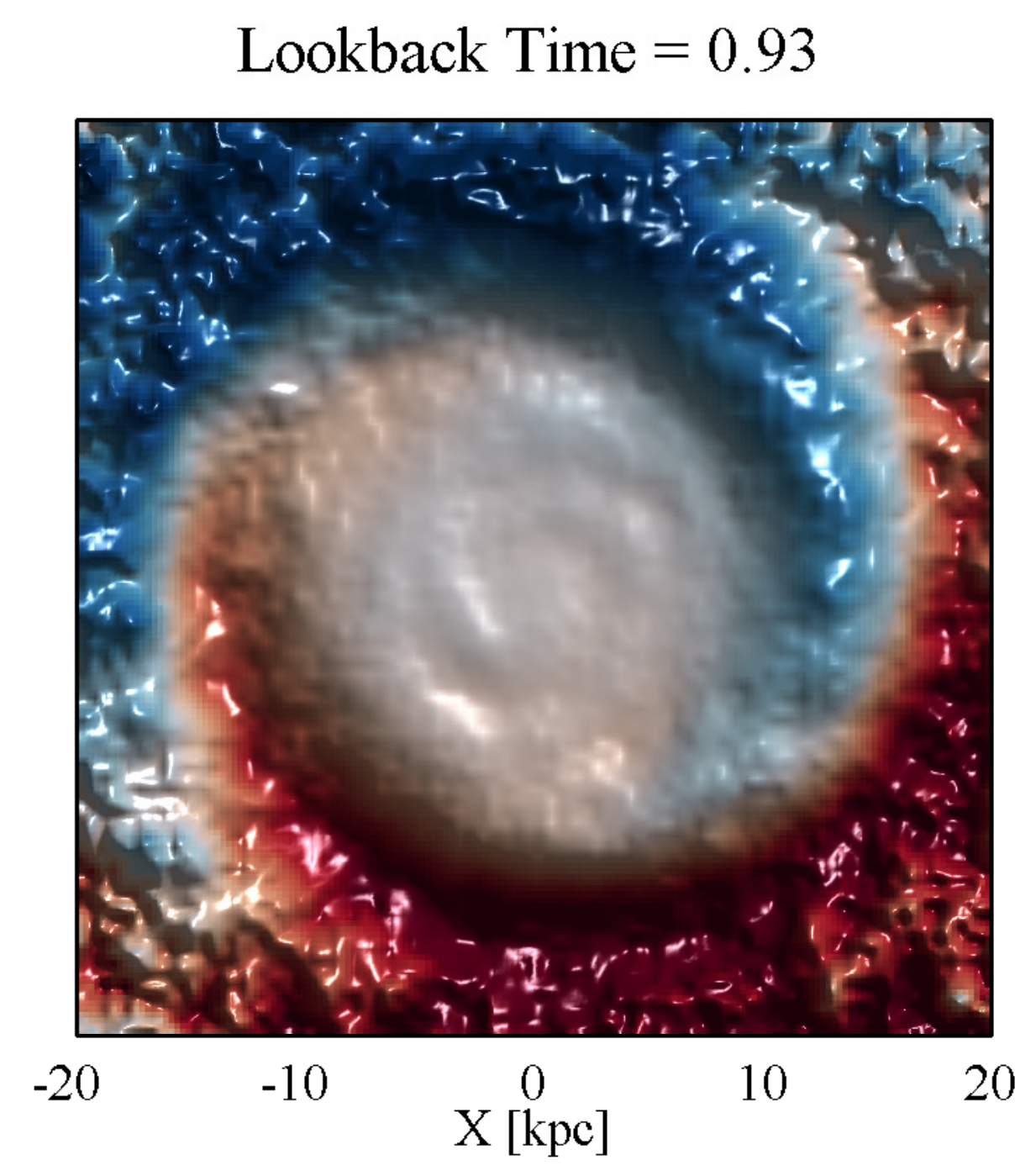}
\hspace{-0.21cm}
\includegraphics[width=33.5mm,clip]{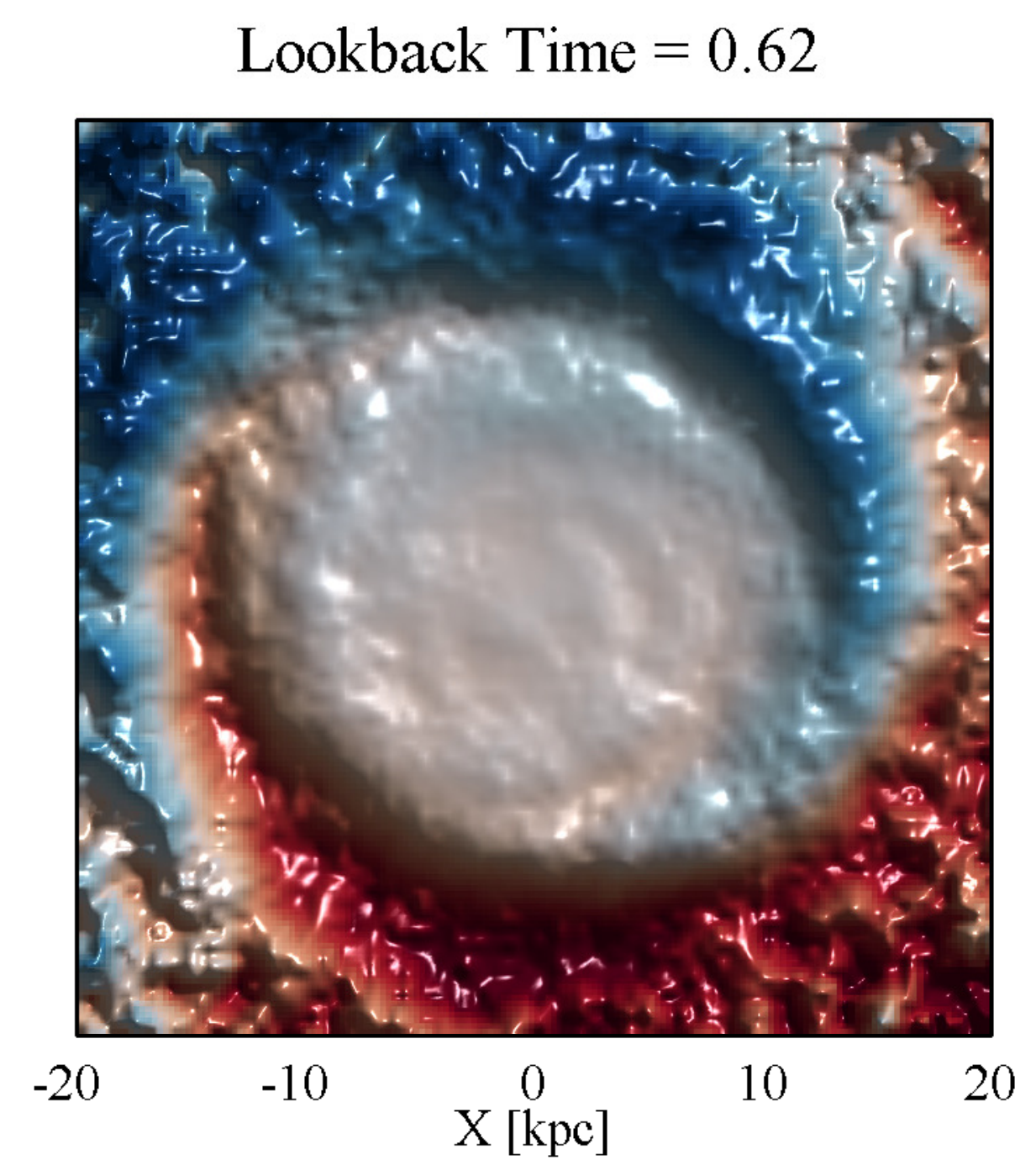}
\hspace{-0.21cm}
\includegraphics[width=33.5mm,clip]{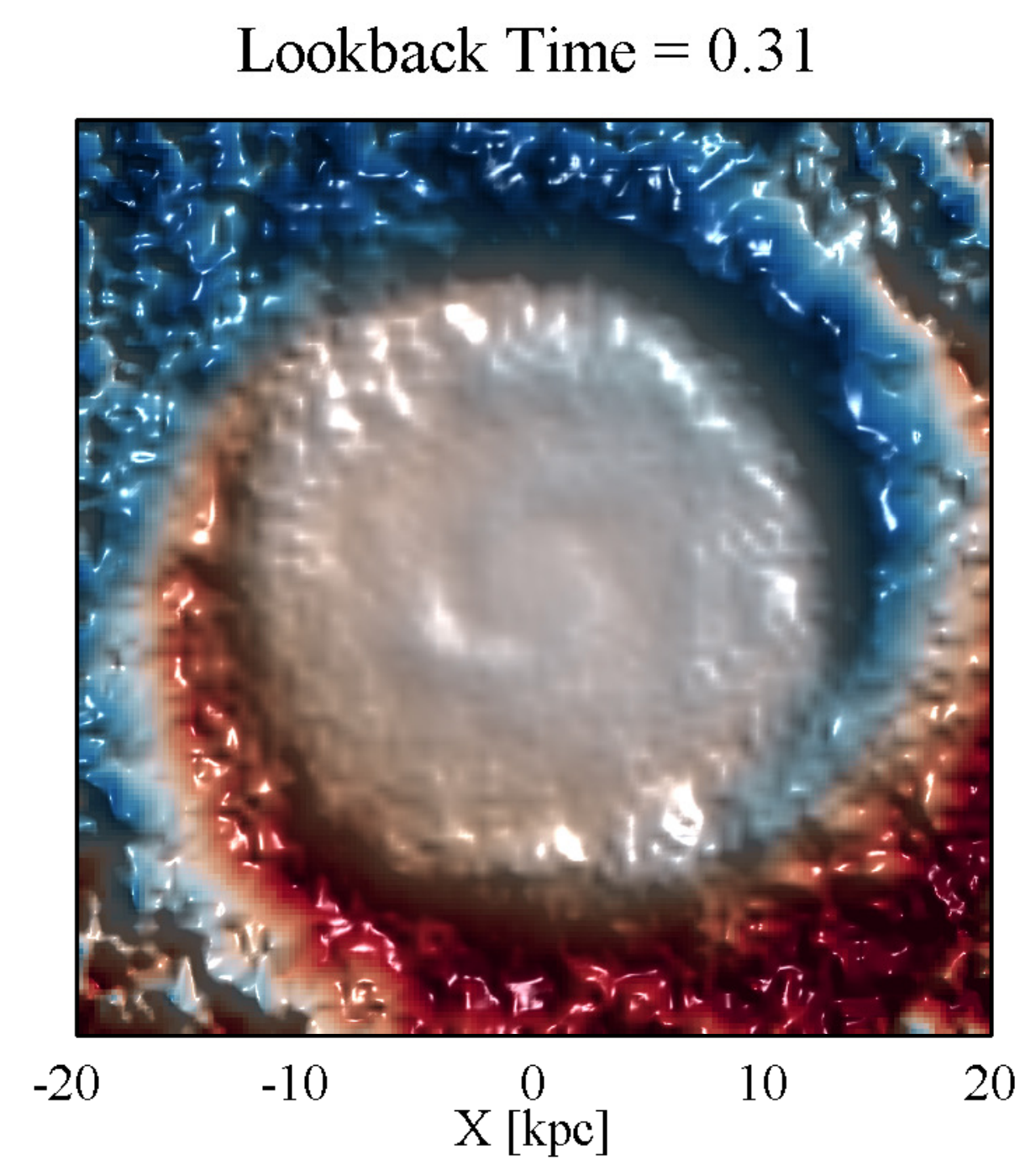}
\hspace{-0.21cm}
\includegraphics[width=38mm,clip]{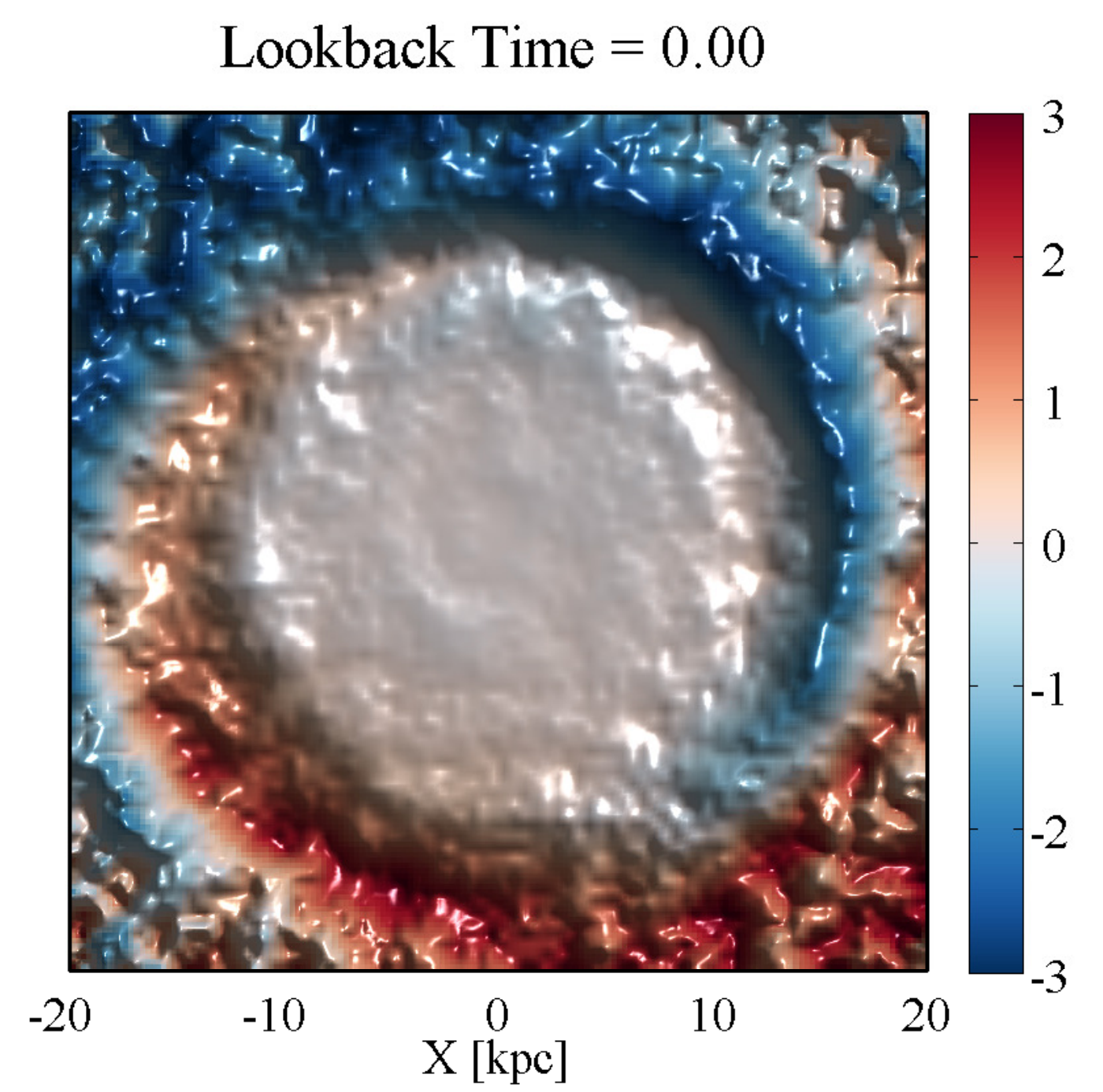}
\caption{Time evolution of the vertical structure of the disk over a period of $\sim 3$ Gyr. Each panel shows a map of 
mass-weighted $\langle {\rm Z} \rangle$  at different times. The colours and the relief indicate different values of $\langle {\rm Z} \rangle$  in kpc. The 
onset-time of the vertical perturbation  is $2.5 \lessapprox t_{\rm look}^{\rm onset} \lessapprox 2.8$ Gyr. The pattern is long-lived and 
coherent at later times. In this projection the galactic disk rotates counterclockwise.}
\label{fig:time_evol}
\end{figure*}

In Figure~\ref{fig:time_evol} we show the time evolution of the vertical disk structure over a period of $\sim 3$ Gyr. 
The onset of the perturbation  is at a lookback time 
$2.5 < t_{\rm look}^{\rm onset} \lesssim 3$ Gyr (see e.g. top second panel). The pattern is long-lived and coherent after this time. 
The vertical perturbation initially has the morphology expected for 
a typical ($m=1$) warp. As time goes by, the pattern slowly winds up to give rise to the spiral shape observed at the present day.

It is interesting to notice the small azimuthal displacement of the vertical pattern between consecutive snapshots. 
First of all, note that at $R = 12$ kpc, the radius at which the vertical pattern becomes noticeable, the azimuthal period of a star 
particle on a perfect circular orbit is $T_{\phi} = 2\pi / \Omega = 2 \pi R / V_{\rm circ}(R)  = 0.303$ Gyr. Here $\Omega$ is the circular 
frequency at the given $R$. This period coincides almost perfectly with the time between consecutive snapshots,
$t_{\rm snap} = 0.309$ Gyr. Thus, the small azimuthal evolution of the pattern indicates similar values for 
the star particle's angular and vertical frequencies in the outer regions of the disk. For example, assuming a spherical 
dark matter halo, in the disk's outer regions where its mass can be regarded as negligible, the circular, $\Omega$, and vertical, $\nu$, frequencies 
are expected to be equal \citep{2008gady.book.....B}. In a more general flattened potential, $\nu > \Omega$. To estimate a star particle's 
vertical period, let us assume locally a uniform density disk. In such a case, it can be shown that 
$T_{\rm z} = 2\pi / \nu = 2\pi / (4 \pi G \rho_{0})^{1/2}$, where $\rho_{0}$ is the total local density. 
At $R=12$ kpc we find $T_{\rm z} = 0.284$ Gyr, a very similar value to $T_{\phi}$, as expected. For $R \geq 16$ kpc we 
find $T_{\phi} \approx T_{\rm z}$. 

In spite of these similar angular and vertical frequencies, it is clear from Figure~\ref{fig:time_evol} 
that the initial $m=1$ pattern winds up significantly over a period of $\sim 2$ Gyr. In agreement with the empirically derived 
rules of \citet{1990ApJ...352...15B}, an initial warp gets distorted into a leading spiral pattern. The 
reason for this has been explained in great detail by \citet[][hereafter, SS06]{2006MNRAS.370....2S}. 
In this work, an initially relaxed disk is subjected to the torque of a 
massive outer torus, introduced to emulate the effect of a misaligned outer DM halo. The resulting torque causes the disk to 
 precess retrograde at a rate that is proportional to galactocentric distance, i.e. $w^{dm}_{p} \propto r$. Because of the disk's 
self-gravity the inner disk, which is strongly cohesive,
precesses slowly as a whole in a retrograde manner about the symmetry axis of the torus, while the outer disk precesses more 
rapidly generating a warp. The developing misalignment between the inner and outer disk causes the particles in the outer 
disk to feel an additional torque from the massive inner disk. 
SS06 showed that the inner disk's torque also causes the outer disk to precess retrograde. However, it does so at a rate that decreases with
galactocentric distance, i.e. $w^{id}_{p} \propto r^{-4}$. In isolation (i.e., neglecting the torque from the massive outer 
torus), the precession rate associated with the inner disk's torque 
would induce the formation of a leading spiral pattern. Thus, the orientation of the 
resulting spiral pattern (leading or trailing) depends on the relative magnitudes of the two torques\footnote{A trailing 
spiral pattern is obtained if the torque from the massive outer torus dominates over that from the inner disk.}. In our simulation 
the $m=1$ pattern  rapidly starts to wind up into a leading spiral. This indicates that 
the torque induced by the inner disk starts to dominate 
right after the pattern has been excited. We will show that this is indeed the case in Section ~\ref{sec:torque}.

\begin{figure*}
\centering
\includegraphics[width=180mm,clip]{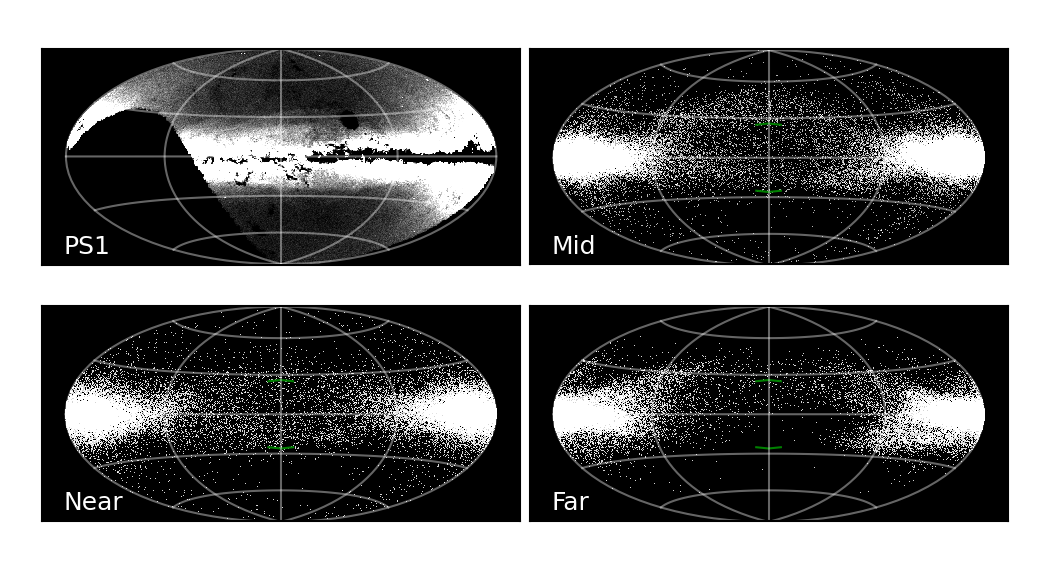}
\caption{Maps of star particle counts in Galactic coordinates for a reference system centred at $(X,Y,Z) = (0,-8,0)$ kpc. 
The centre of each map points towards the Galactic anticentre. The right and bottom left panels show maps for
for particles at different heliocentric distances. The ranges covered are $3.8 <  d_{\rm helio}  < 7.8$ kpc (bottom left), 
$5.5 <  d_{\rm helio}  < 13.8$ kpc (top right) and $11.5 < d_{\rm helio}  < 24$ kpc (bottom right). The observational data from PS1 are shown 
at the top left. Note that the different distance cuts overlap slightly to emulate the distance smearing due to the magnitude spread of 
MSTO stars.}
\label{fig:monoceros}
\end{figure*}

The $\langle {\rm Z} \rangle$ maps reveal that, at the present day, the vertical pattern spans a very large range in galactic longitude. 
(see Figure~\ref{fig:maps_t0}). For example, the arm above the mid-plane covers $\leq 180^{\circ}$ in galactic longitude, which is 
reminiscent of the Mon ring. We further explore this in Figure~\ref{fig:monoceros}, where we show maps of star particle 
counts in Galactic coordinates centred on the galactic anticenter.  For this analysis we have selected star particles with ages 
younger than 5 Gyr. This allows us to  avoid significant contamination from the stellar halo and the bulge. In addition, as we will show
later in Section~\ref{sec:bar_pert}, the vertical pattern is significantly better defined on this population of star particles.
We chose the Sun's location at $(X,Y,Z) = (0,-8,0)$ kpc. This particular choice is arbitrary and serves to enhance the similarities between our 
models and the observed morphology of the Mon ring. We divide the star particles into three three heliocentric distance bins in order to emulate the distance ranges covered by the three magnitude slices considered by S14:

\begin{itemize}
\item Near -- $3.8 <  d_{\rm helio}  < 7.8$ kpc

\item Mid -- $5.5 <  d_{\rm helio}  < 13.8$ kpc

\item Far -- $11.5 < d_{\rm helio}  < 24$ kpc
\end{itemize}

Note that, to emulate the distance smearing due to the magnitude spread of the MSTO stars, we allow the distance cuts to overlap slightly.
Many similarities between our mock data set and the Mon ring, as presented in Figure 4 of S14, are found. First and foremost, material
from the galactic disk can be found at galactic latitudes as large as $30^{\circ}$ in both the north and the south. This is true 
even for the far slice, revealing that the disk has been perturbed to very large heights in its outer parts. Clear arcs of stellar 
material are visible at various galactic longitudes, especially in the central range. These feature are reminiscent of the arcs seen in Figure 3 
of S14 (features A, B and C).
The arcs are also noticeable at the farthest distances, particularly on the north side of the disk. While in the near range the 
perturbed disk is dominated by 
material in the south, the north side of the disk becomes more prominent in the middle range, with the same general morphology as nearby. Most of 
these similarities are not reproduced by previous simulations that attempted to model the Mon ring either as tidal 
debris from a disrupting satellite or as a vertically perturbed pre-existing galactic disk (e.g., S14).

\section{The perturber and its perturbations}
\label{sec:pert}

In the previous section we have shown that our fully cosmological model of a Milky Way-like disk shows a well defined
and strong vertical pattern that can reproduce many of the features associated with the Mon ring. In this
Section we explore which agents are responsible for the excitation of this vertical pattern. 

As previously discussed in Section~\ref{sec:introduction}, vertical bulk motions on a disk can arise as a result of the global stellar 
response to spiral structure or a bar \citep{2014MNRAS.440.2564F,2014MNRAS.443L...1D,2015arXiv150507456M}. 
The vertical motions induced by such mechanisms are compressive (towards the mid-plane) and rarefactive (away from the mid-plane). 
As a consequence, these perturbations, known
as breathing modes, cannot explain the vertical oscillation of the disk's midplane either in the Milky 
Way or in our simulation. \citep[][]{2012ApJ...750L..41W,2013ApJ...777...91Y,2015ApJ...801..105X,2015arXiv150308780P}.  

This oscillatory behavior, usually described as a bending mode, is likely to be excited by some external perturbation, such as a 
misaligned outer DM halo or the accretion of cold gas or infalling satellites 
\citep[e.g.][]{1989MNRAS.237..785O, 1993ApJ...403...74Q, 1999ApJ...513L.107D, 1999MNRAS.303L...7J, 1999MNRAS.304..254V, 2006MNRAS.370....2S, 2012MNRAS.426..983D}. Note that satellites passing through the plane of the disk can excite both bending
and breathing modes \citep{2014MNRAS.440.1971W}. Which modes dominate in such a scenario depends strongly on the relative velocity of 
the satellite with respect to the galactic disk.

\begin{figure}
\centering
\includegraphics[width=80mm,clip]{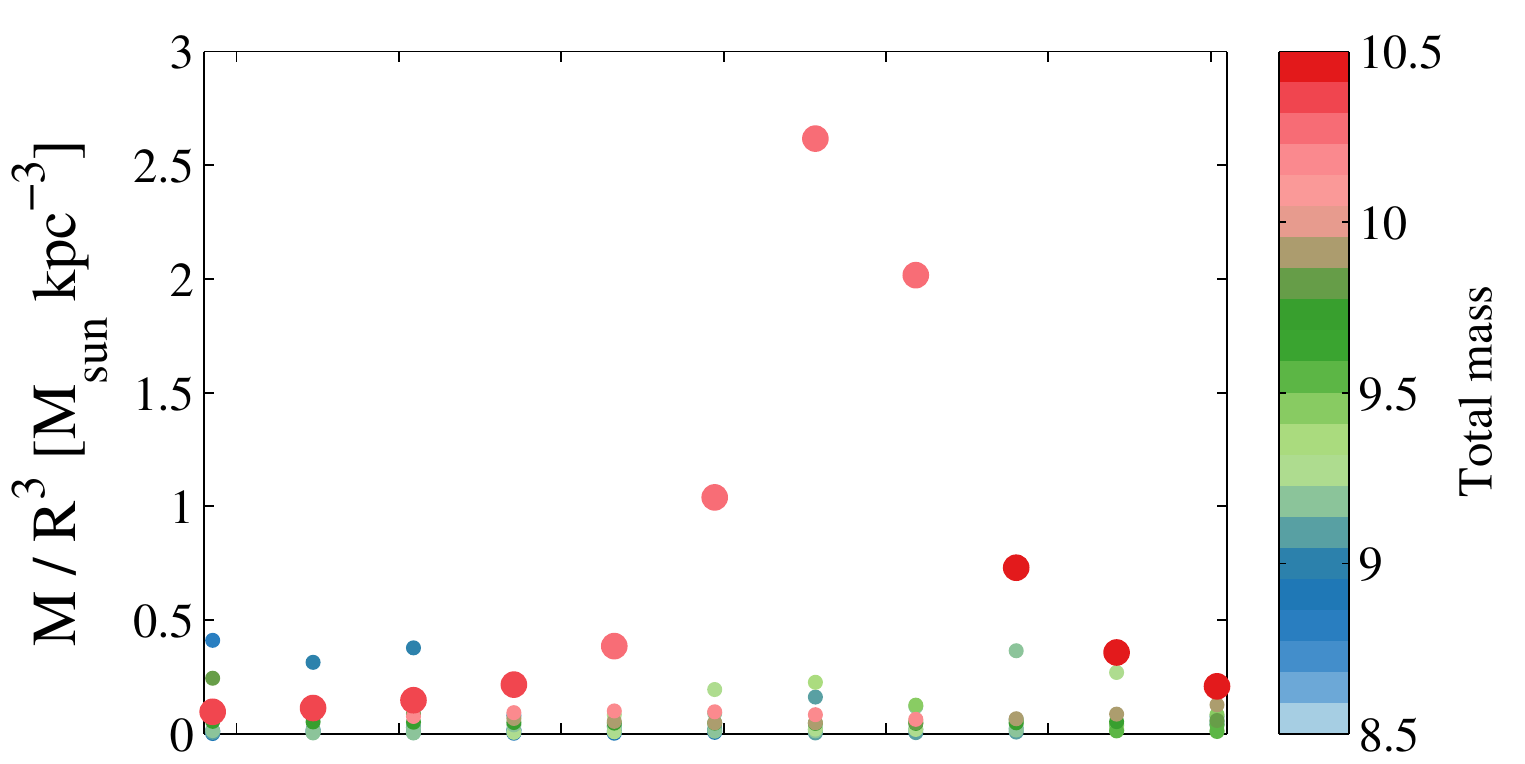}\\
\includegraphics[width=80mm,clip]{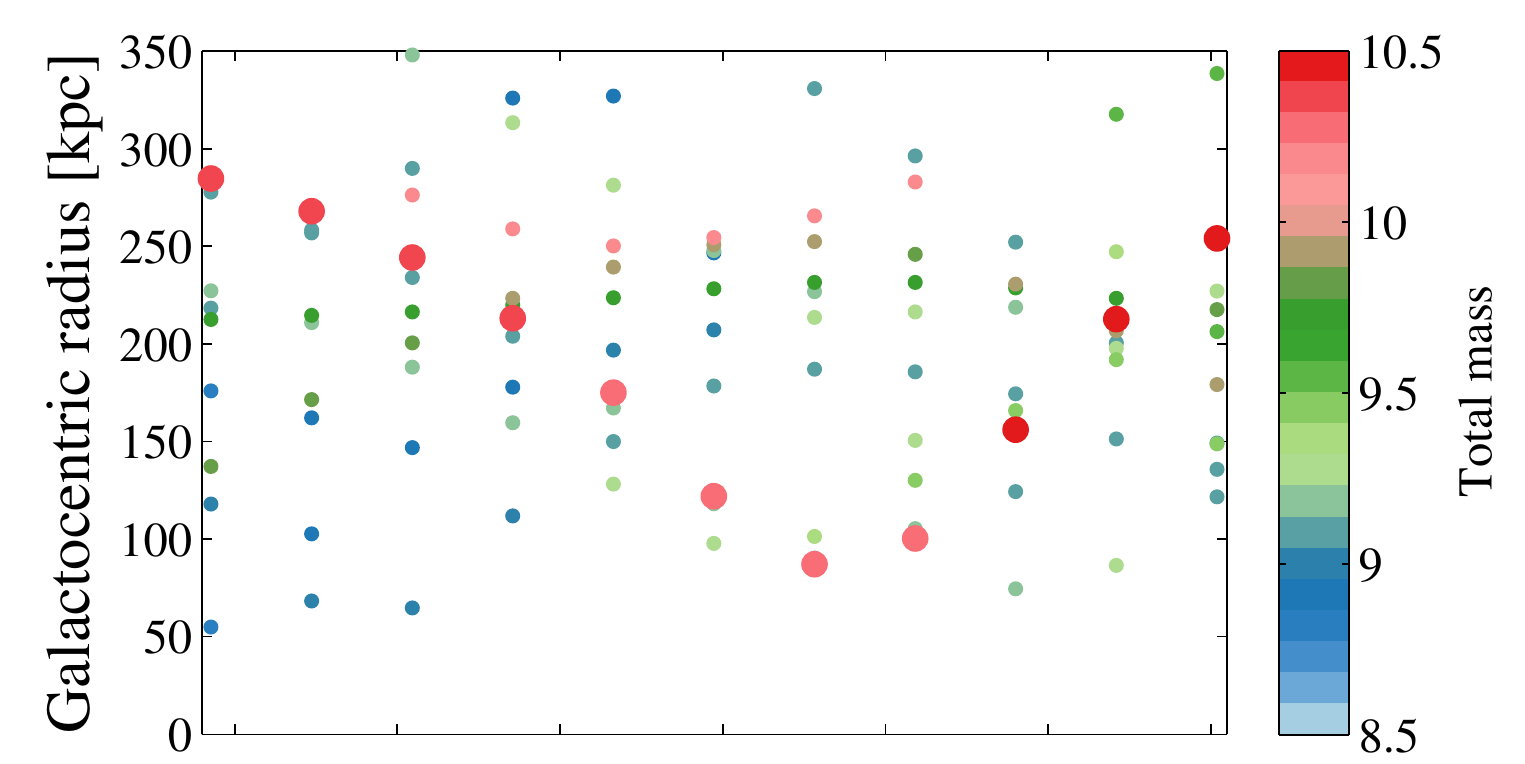}\\
\includegraphics[width=80mm,clip]{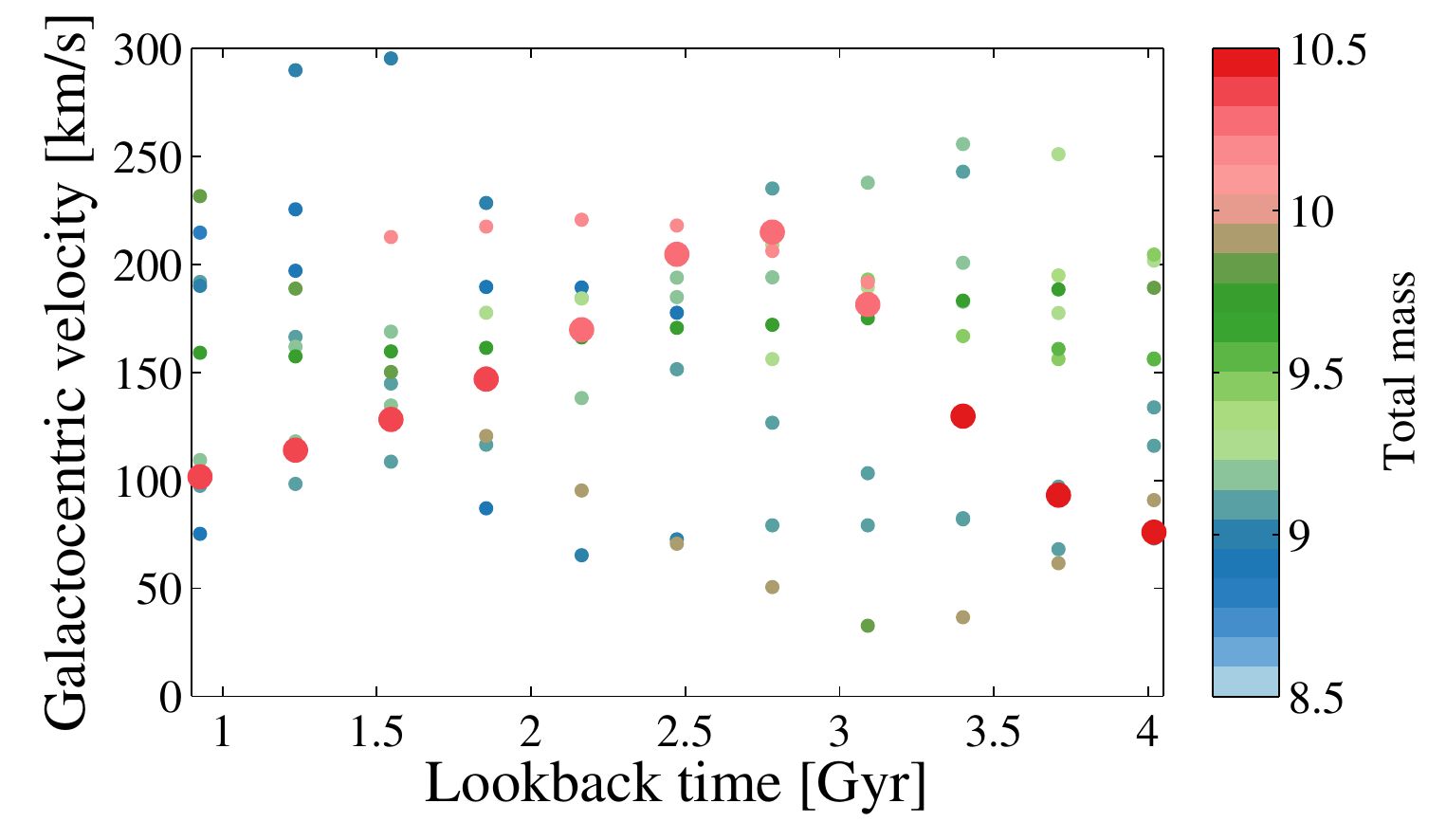}
\caption{{\it Top panel:} Tidal force exerted on the host by the 20 most massive satellites as a
function of time. Points are colour coded according to the mass of each satellite at the corresponding time.
{\it Middle panel}: Galactocentric distance of the 20 most massive satellites as a function of time. 
{\it Bottom panel}: As above, but for Galactocentric velocity. The most significant perturber during this 
period is indicated by thicker points.
Note that this satellite has a low-velocity encounter with the host at 
$t_{\rm look} \approx 2.7$ Gyr.}
\label{fig:orbits}
\end{figure}

\subsection{A putative perturbing satellite}
\label{sec:pert_sat}

We start by exploring which satellite might induce such a bending mode in our Milky Way-like disk. In the top panel of 
Figure~\ref{fig:orbits} we show the time evolution of the tidal force exerted on the host 
by the 20 most massive satellites as a function of time. Points are colour coded according to the mass of each 
satellite at the corresponding time. Motivated by Figure~\ref{fig:time_evol}, where we identify the onset of the perturbation  
as $2.5 < t_{\rm look}^{\rm onset} \lesssim 3$ Gyr, we focus our analysis on the time period $1 < t_{\rm look} < 4$ Gyr. 
It is evident from this top panel that there is only one plausible perturber interacting with the disk at 
$t \approx t_{\rm look}^{\rm onset}$. This satellite has a total mass at infall $M_{\rm sat} \approx 4 \times 10^{10}~\mo$
and pericentre radius of $r_{\rm peri} \approx 80$ kpc. This is shown in the middle panel of Figure~\ref{fig:orbits}, where 
we plot the simulated galactocentric distance of each satellite at different times. The pericentre passage takes place at 
$t_{\rm look} \approx 2.7$ Gyr. The other massive satellites during this 3 Gyr time span are either too low-mass  
($M_{\rm sat} < 10^{9} \mo$) or have pericentre passages at distances larger or similar to that of our candidate. The bottom panel of 
Figure~\ref{fig:orbits} shows the time evolution of the galactocentric velocity of the same satellite subset. The velocity at pericentre 
of the perturbing satellite is $\sim 215$ km s$^{-1}$. 

Note that the configuration of this disk-satellite interaction differs significantly from those considered in previous studies, 
where the satellites repeatedly plunge through the galactic plane. In fact, it is
very unlikely that this satellite's tidal field alone could account for the observed strong vertical pattern (see Section~\ref{sec:torque}). 
Instead, this interaction 
can be described as a low-velocity fly-by that crosses  the plane of the disk at a relatively large galactocentric distance.
 Interestingly, as discussed in detail by
\citet{2000ApJ...534..598V}, a low-mass, low-velocity fly-by that penetrates the outer regions of a galaxy can generate asymmetric 
features in the host DM halo density field. These perturbations can be efficiently transmitted to the inner parts of the primary system affecting 
the deeply embedded galactic disk. Furthermore, the excited modes are expected to be weakly damped, thus persisting well after the satellite's 
pericentric passage. We will explore whether this mechanism is acting in our host galaxy  in Section~\ref{sec:dm_wake}.

\subsection{Time evolution of the host's baryonic component} 
\label{sec:bar_pert}

\begin{figure*}
\includegraphics[width=160mm,clip]{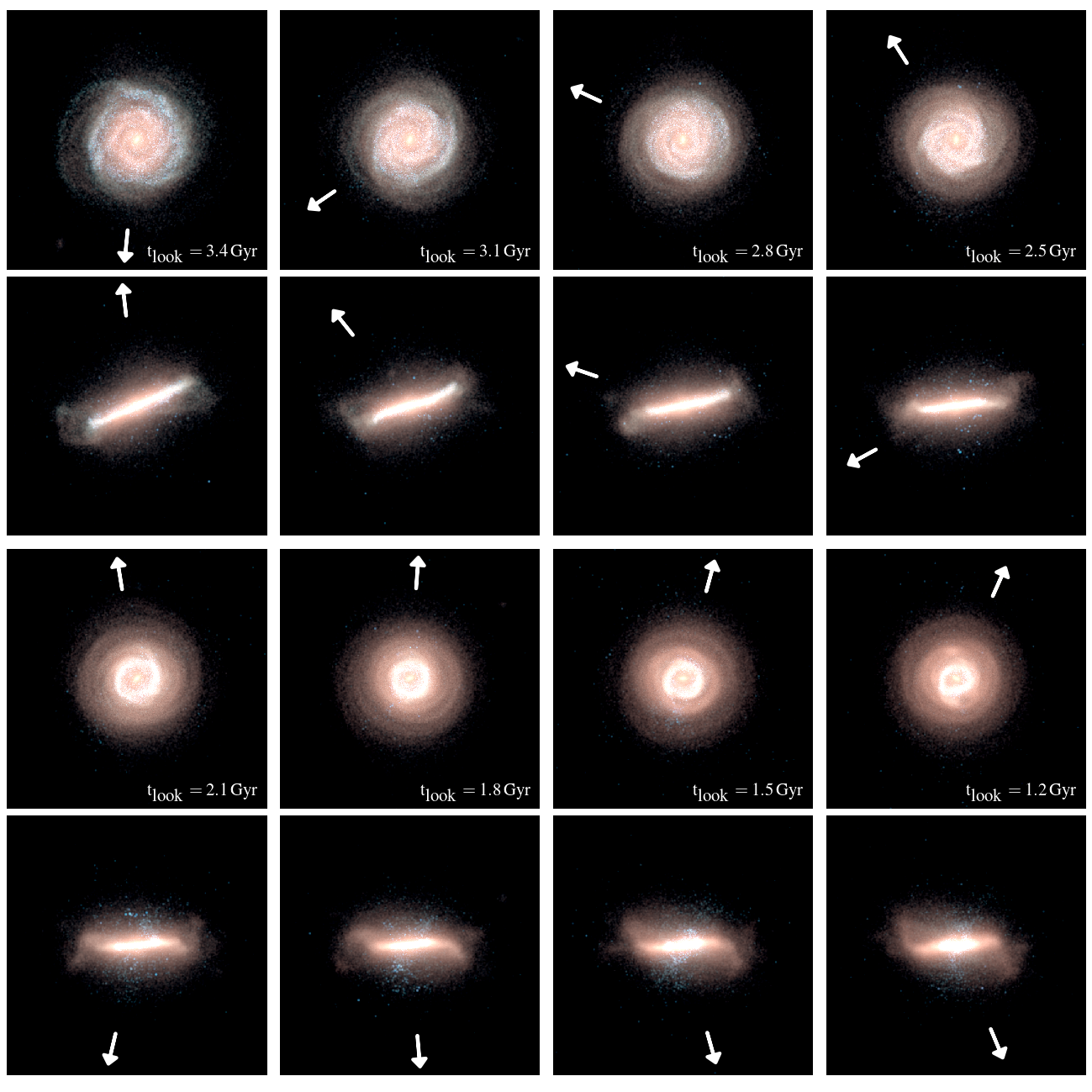}
\caption{Time evolution of the projected stellar density within a cube with side length of 80 kpc. 
The panels display the stellar mass distribution in two 
orthogonal projections that are kept fixed in time and were chosen such that at 
$t_{\rm look} = 1.5$ Gyr the disk is seen face-on and edge-on. The arrows point towards the location of the perturber 
identified in Section~\ref{sec:pert_sat}. Images were constructed by mapping the K-, B- and U-band luminosities 
to the red, green and blue channels of a full colour composite image. The same logarithmic mapping of stellar luminosity to image intensity 
is used in all panels. Note the tilt of the stellar disk during this period of time.}
\label{fig:star_lum}
\end{figure*}

In Figure~\ref{fig:star_lum} we follow the time evolution of the star component from $t_{\rm look} = 3.4$ Gyr to 
$t_{\rm look} = 1.2$ Gyr. These panels display the stellar mass distribution within a 80 kpc cube in two orthogonal projections that are 
kept fixed in time and were chosen such that at $t_{\rm look} = 1.5$ Gyr the disk 
is shown both face-on and edge-on.
As in M14, the images were constructed by mapping the K-, B- and U-band luminosities to the red, green and blue 
channels of a full colour composite image. Young 
stellar populations hence appear blue, old stellar components appear red. In all panels the same logarithmic mapping of stellar luminosity to image 
intensity is used. Note the evolution of the  stellar disk's morphology, as inferred  from its stellar
luminosity. As discussed by M14, during and after the satellite's closest approach, the 
star formation is affected  in disks  and is  restricted to a narrow inner ring. This changes the appearance of the disk.
As time goes by, the stellar disk  tilts significantly.

\begin{figure}
\centering
\includegraphics[width=85mm,clip]{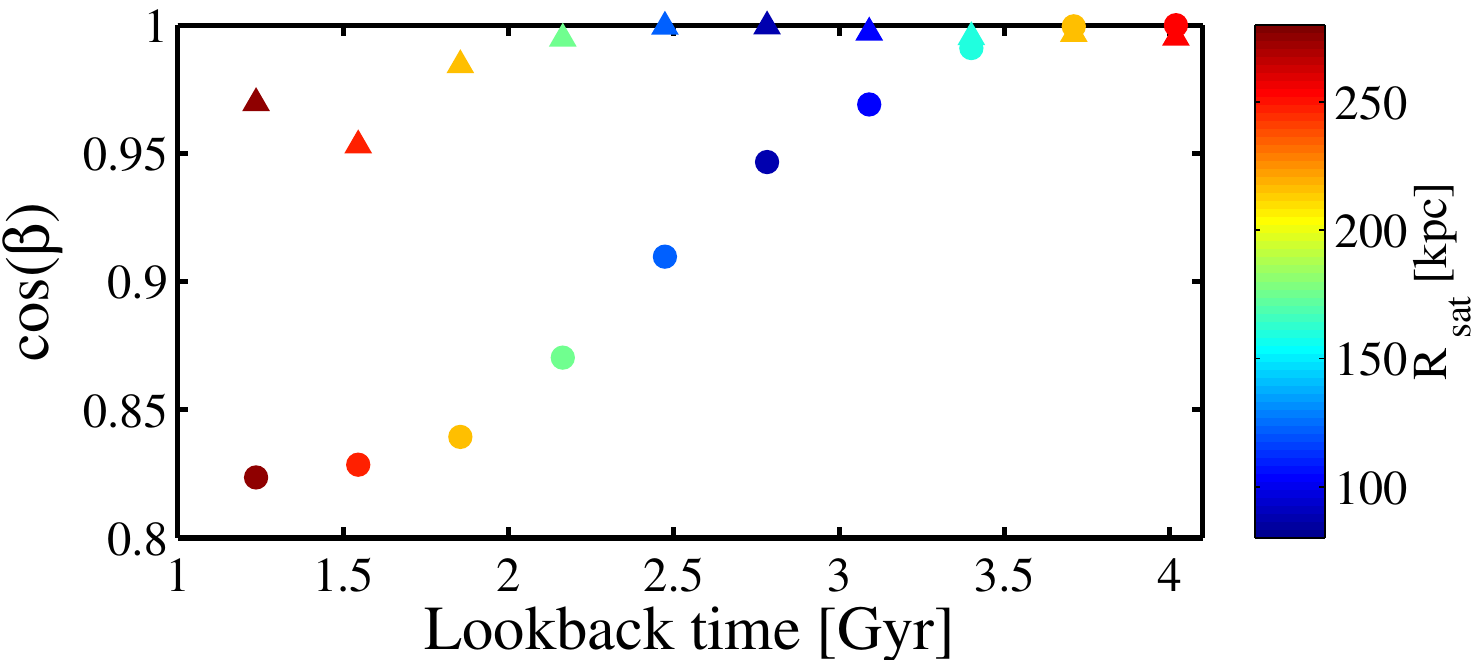}
\caption{The solid dots indicate the time evolution of the disk's angular momentum vector  with respect
to its initial orientation at $t_{\rm look} = 3.7$ Gyr. During this period the disk tilts a total of 
$\approx 35^{\circ}$. 
The solid triangles indicate the time evolution of the misalignment between the stellar and the cold gas disks. Note
that at all times these two galactic components are very well aligned with each other. In both cases, the colour coding indicates
the galactocentric distance of the satellite galaxy introduced in Section~\ref{sec:pert_sat}.}
\label{fig:ang_orients}
\end{figure}

The dots in Figure~\ref{fig:ang_orients} show the time 
evolution of the stellar disk's angular momentum orientation with 
respect to its initial orientation at $t_{\rm look} = 3.7$ Gyr. To compute the orientation we 
select, at $t_{\rm look} = 3.7$ Gyr, all the star particles that are younger than 5 Gyr. 
These same star particles are used to compute the angular momentum 
vector at later times. 
Different criteria defining the angular momentum vector at each snapshot do not affect our results. During the fly-by
 the host stellar disk tilts by an angle of $\approx 35^{\circ}$. 
This shift in the angular momentum takes  place over a 
$\sim 2$ Gyr time period. Note that similar changes of the orientation of 
disk angular momentum vector have been previously measured in hydrodynamical simulations of Milky Way-like galaxies 
\citep[e.g.][]{2012MNRAS.426..983D,2013MNRAS.428.1055A,2014arXiv1411.3729Y}, where they are often not due to satellite effects.

As shown by \citet{2009MNRAS.396..696S}, a misalignment between the cold gas disk and the stellar disk can destabilize 
the system, triggering significant perturbations that can even destroy a pre-existing stellar disk. 
The filled triangles in Figure~\ref{fig:ang_orients} show 
the time evolution of the angle between the two disks. They remain well aligned throughout the relevant period of time. 
This suggests that this mechanism is not playing a role in exciting of the disk's vertical perturbation. 
Our results are in good agreement with those presented 
by \citet{2009MNRAS.396..696S}, where the same system was studied using a different hydrodynamical technique.

\begin{figure}
\centering
\includegraphics[width=42.5mm,clip]{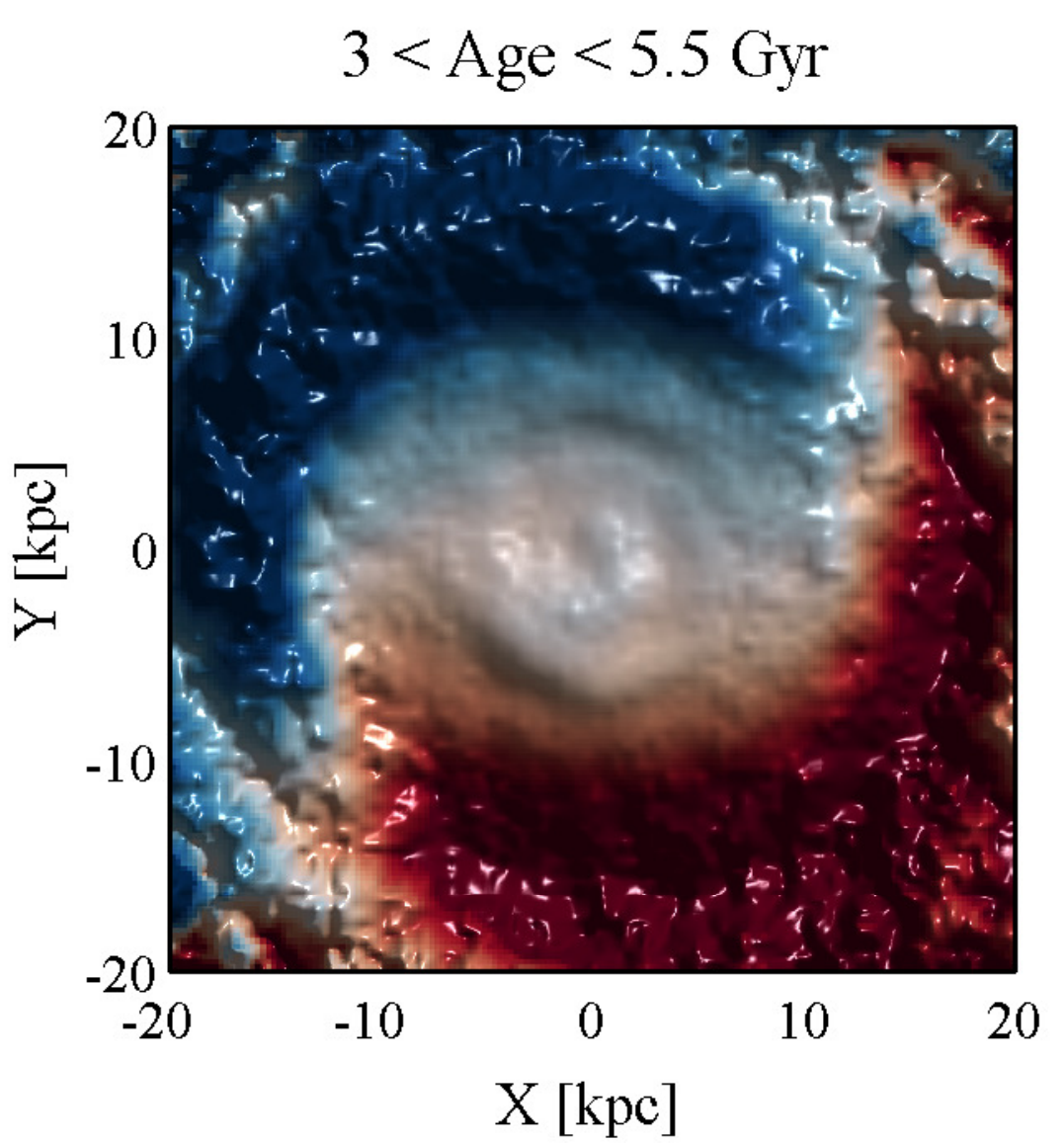}
\hspace{-0.3cm}
\includegraphics[width=42.5mm,clip]{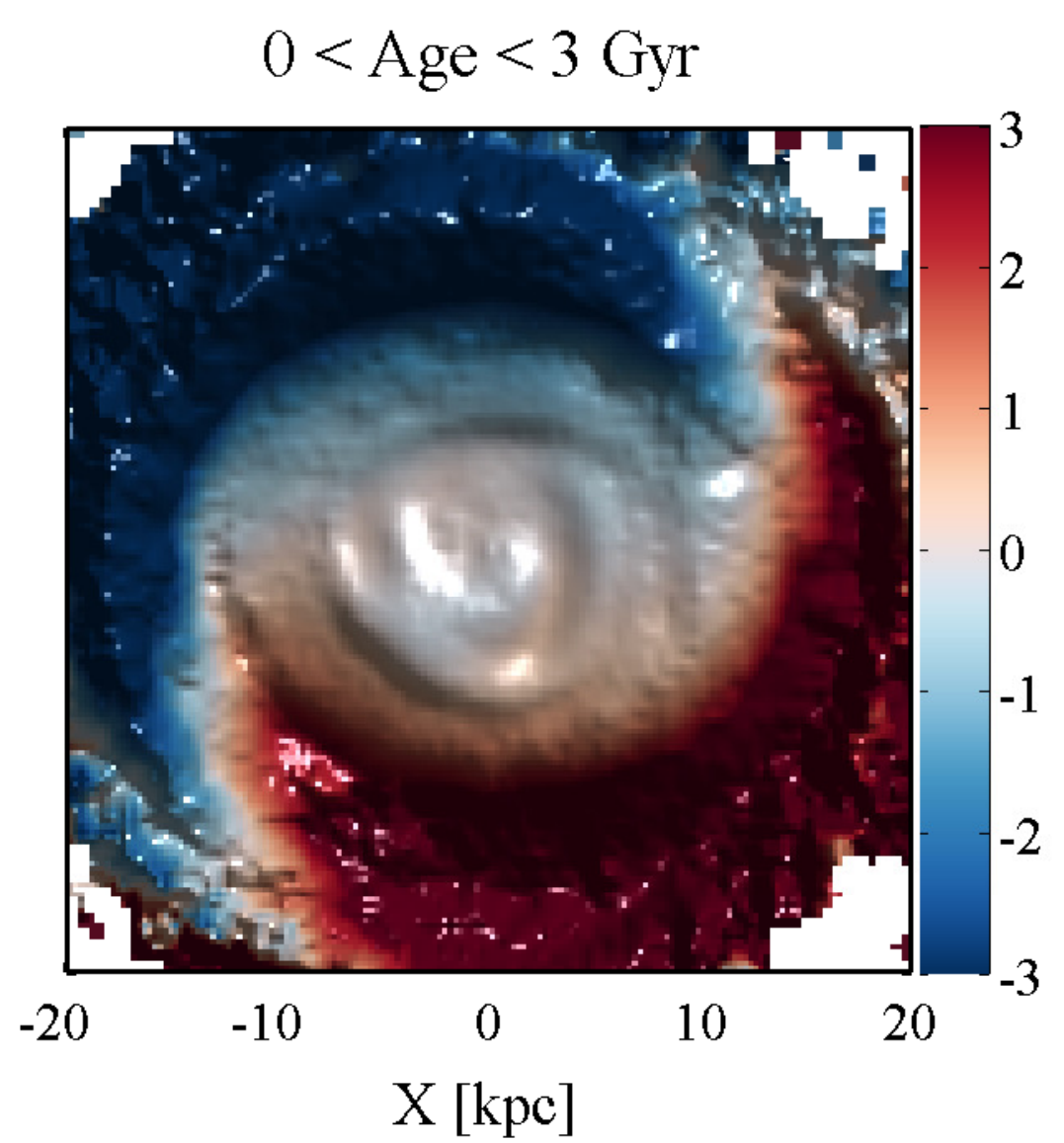}
\caption{Map of the mass-weighted $\langle {\rm Z} \rangle$ for the simulated galactic disk at $t_{\rm look}  = 1.55$ Gyr.
The right panel shows the $\langle {\rm Z} \rangle$ maps for the youngest star particles (Age $< 3$ Gyr 
at $t_{\rm look} = 1.55$ Gyr), whereas in the left panel only older stars are considered 
($3 <$ Age $< 5.5$ Gyr  at $t_{\rm look} = 1.55$ Gyr). Note that the two sets of star 
particles show almost identical vertical patterns, indicating a global response of the pre-existing disk.} 
\label{fig:old_vs_young}
\end{figure}

\citet[][hereafter R10]{2010MNRAS.408..783R} presented a possible mechanism for the formation of outer galactic warps based on the misalignment 
between inner stellar disk and the surrounding hot gas halo. As discussed by R10,  cold gas infalling onto a pre-existing disk can be strongly
torqued by the surrounding hot gas halo. By the time the cold gas reaches the star forming regions of the disk, its angular
momentum is aligned with that of the hot gas. If the spin of the hot gas halo is not aligned with that of the
inner disk, a misaligned outer disk forms composed of newly accreted material. The main characteristic of this mechanism is that it does not 
involve significant warping of the pre-existing disk. The outer stellar warp is 
heavily dominated by newly formed stars. In their simulations, old stars in these outer regions can also be seen at large distances from the 
midplane but their morphology is that of a thickened population.  
If this mechanism is playing a role in our system, the $\langle {\rm Z} \rangle$
of the old population should lie close to 0 kpc at all radii. The left panel of Figure~\ref{fig:old_vs_young} shows the $\langle {\rm Z} \rangle$
map of star particles that, at $t_{\rm look} = 1.55$ Gyr, have ages $3 < {\rm Age} < 5.5$ Gyr. For comparison, the right panel 
shows the $\langle {\rm Z} \rangle$ map of the star particles that, at this time, have an age $< 3$ Gyr. 
It is clear that both sets of star particles exhibit nearly the same vertical perturbation. 
This indicates that the mechanism introduced by R10 is not playing a significant role in our system.

\begin{figure*}
\includegraphics[width=37mm,clip]{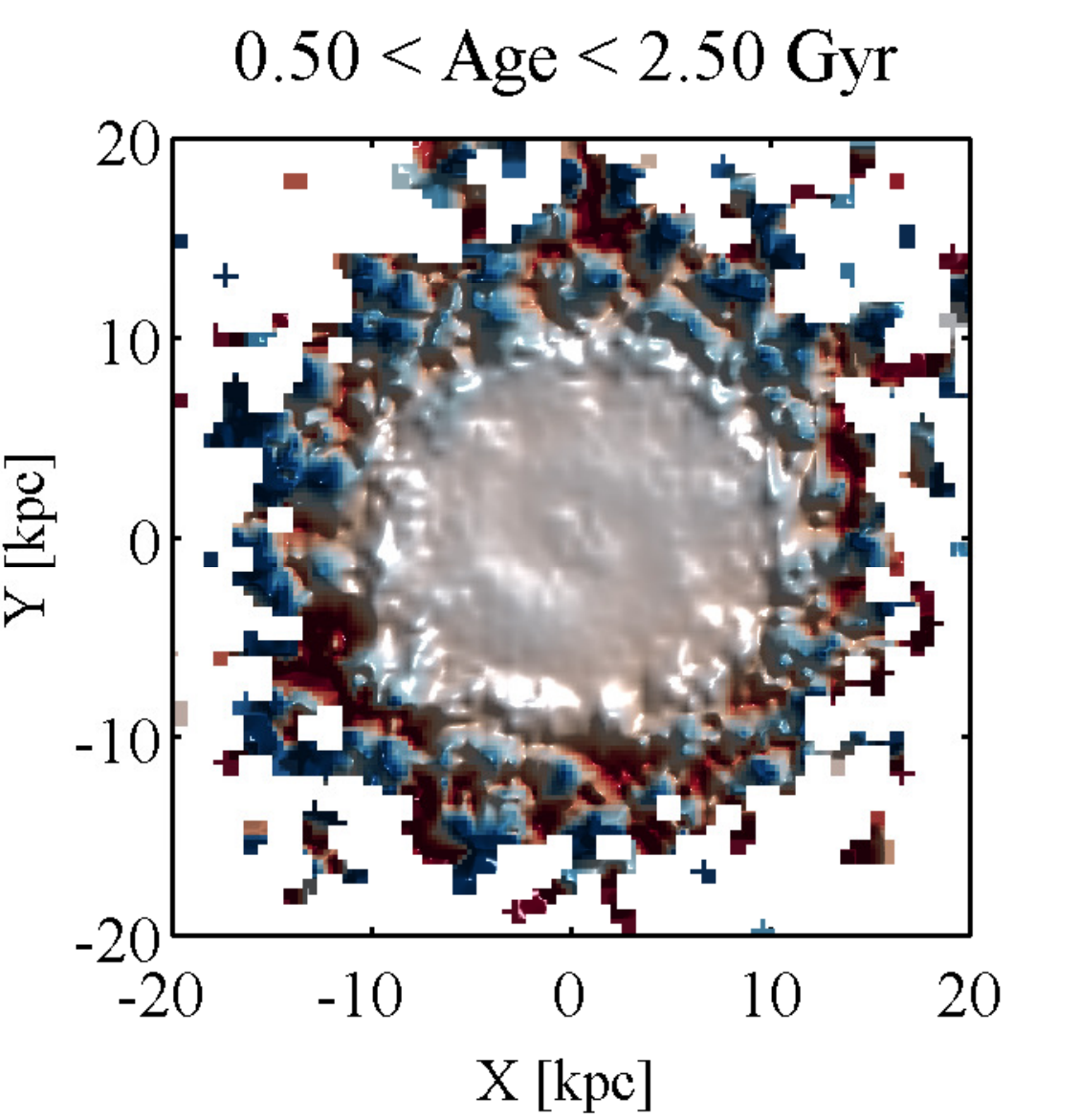}
\hspace{-0.45cm}
\includegraphics[width=37mm,clip]{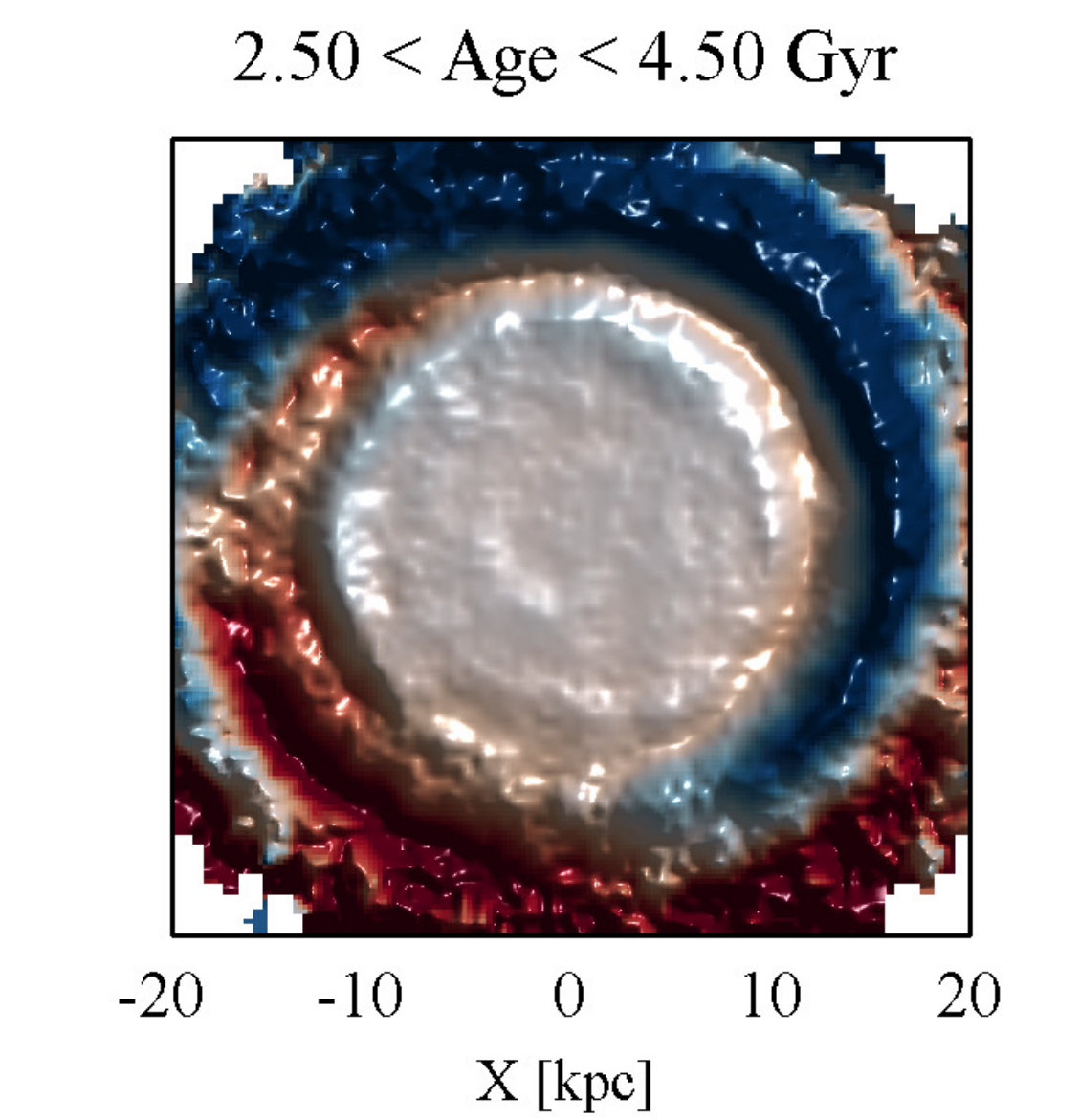}
\hspace{-0.45cm}
\includegraphics[width=37mm,clip]{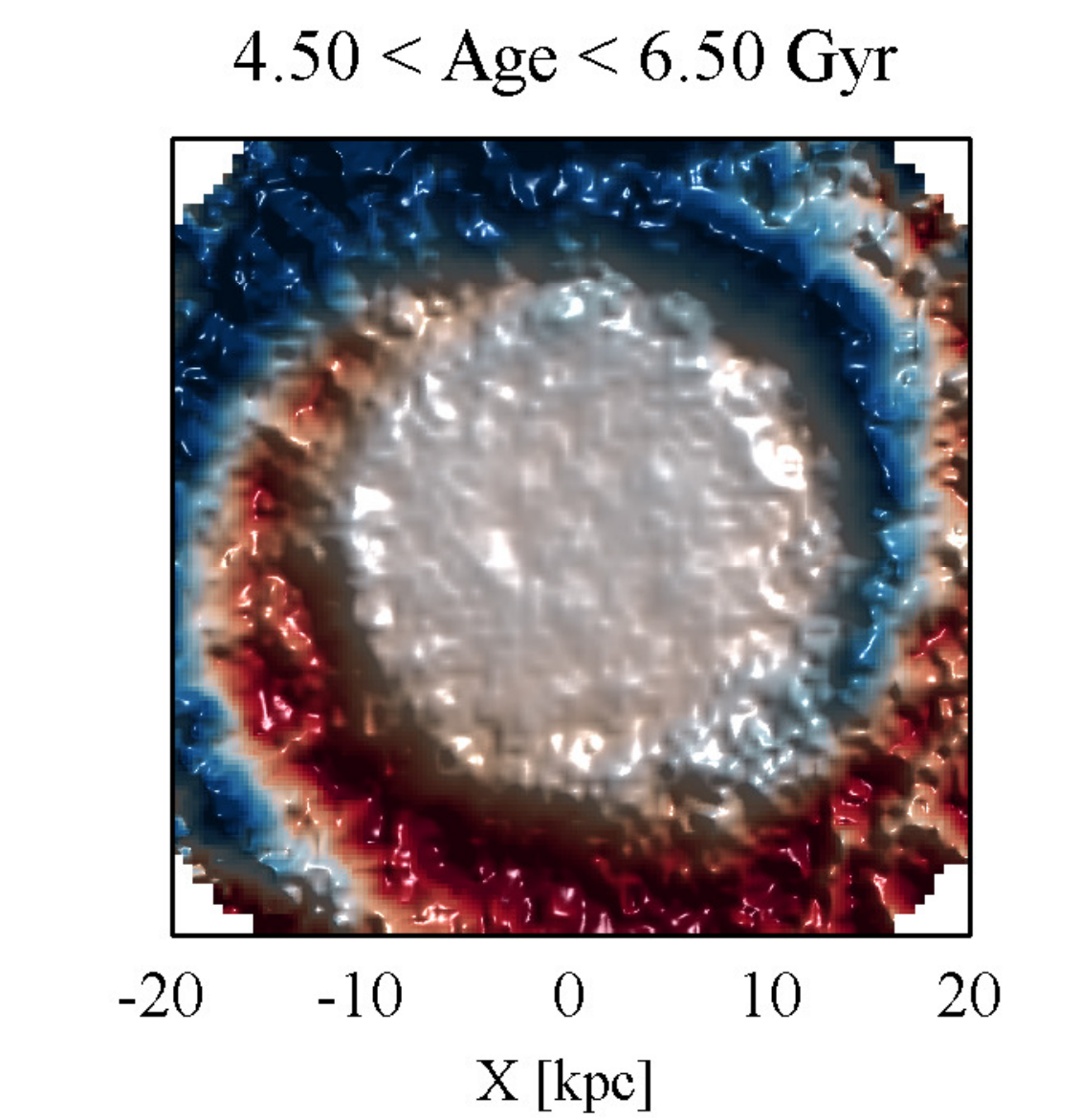}
\hspace{-0.45cm}
\includegraphics[width=37mm,clip]{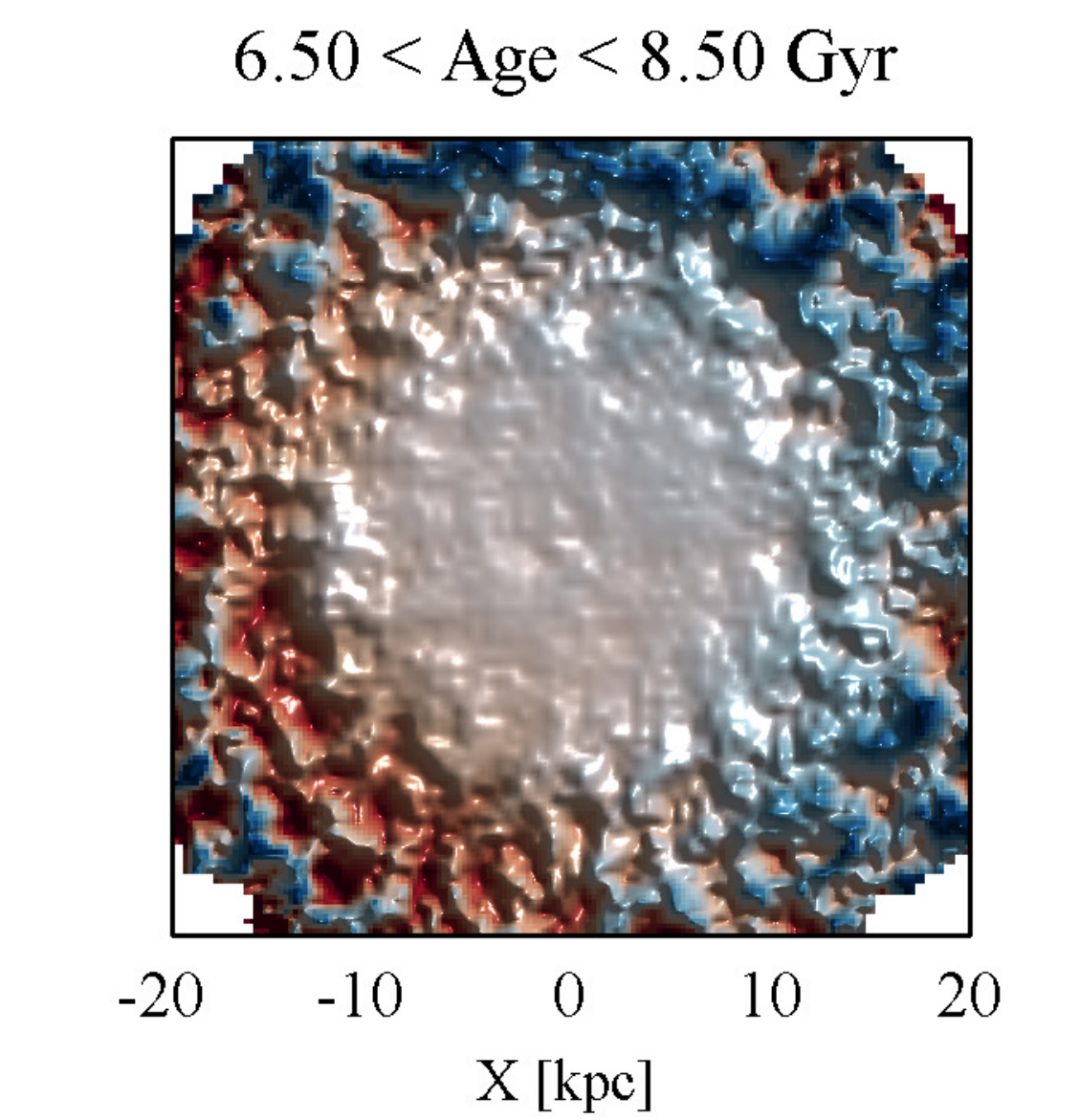}
\hspace{-0.2cm}
\includegraphics[width=37mm,clip]{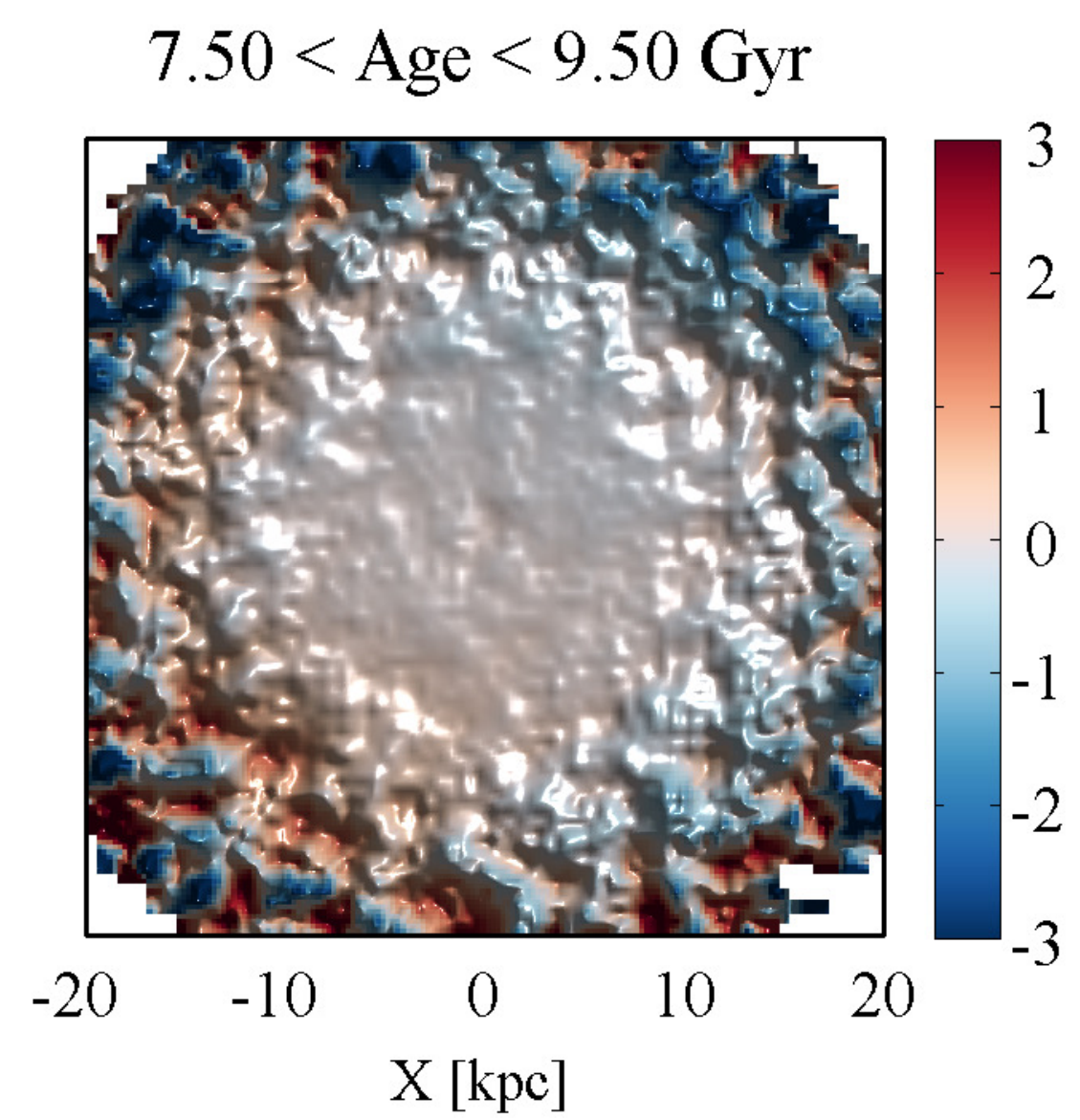}
\caption{present day mass-weighted $\langle {\rm Z} \rangle$ maps obtained from subsets of star particles born within various 2 Gyr time intervals, 
as indicated on top of each panel. The vertical pattern is best resolved when considering star particles in the age range 
$2.5 < {\rm Age} \leq 4.5$ Gyr.}
\label{fig:aging}
\end{figure*}

In Figure~\ref{fig:aging} we dissect our present day galactic disk by selecting  star particles that were born within different 2 
Gyr time intervals.
Each panel shows the $\langle {\rm Z} \rangle$ map obtained from a different subset of star particles. As expected from our previous 
discussion, star particles younger than $2.5$ Gyr show no sign of  vertical perturbations. These stars are born after the onset of the observed
vertical pattern ($2.5 < t_{\rm look}^{\rm onset} < 3$ Gyr) and  within the inner $\lesssim 12$ kpc; 
a region where the disk has not been vertically perturbed. Interestingly, the vertical pattern is best resolved when 
considering star particles within the age range $2.5 < {\rm Age} \leq 4.5$ Gyr. Recall that this
time interval approximately corresponds with $t_{\rm look}^{\rm onset}$ as well as with the most significant perturber pericentre passage. 
When considering older subsets of star particles we find that the vertical pattern gradually degrades until it almost disappears for 
star particles older than $7$ Gyr. 

\begin{figure}
\centering
\includegraphics[width=80mm,clip]{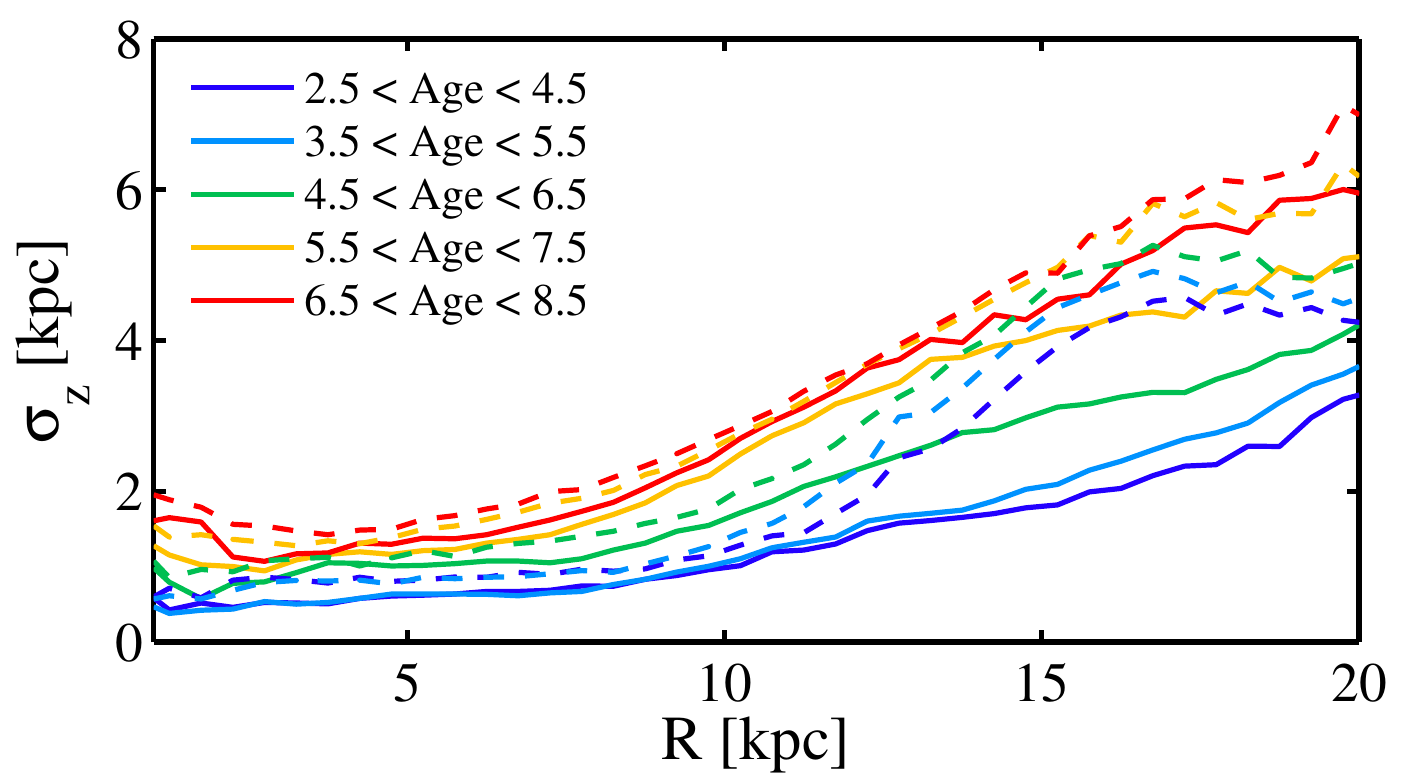}
\caption{Mass-weighted vertical dispersion, $\sigma_{\rm Z}$, for subsets of star particles selected 
within various 2 Gyr age bins, as indicated by the legend. 
The dashed lines show the present day $\sigma_{\rm Z}$ as a function of radius for each subset. 
The solid lines show the radial dependence $\sigma_{\rm Z}$ for
the same subsets of particles, but now 2.5 Gyr ago.}
\label{fig:heating}
\end{figure}

The relationship between the age of star particles and the vertical structure of the disk can be understood from 
Figure~\ref{fig:heating}. Here we show the mass-weighted vertical dispersion, $\sigma_{\rm Z}$, for subsets of star particles binned by age. 
This quantity provides a measure of disk scale-height as a function of galactocentric radius. 
The present day age of the star particles within each bin is indicated in the top left corner of the figure. 
The solid lines show $\sigma_{\rm Z}$ for disk stars selected at $t_{\rm look} = 2.5$ Gyr. In other words, the stellar age bin 
$2.5 < {\rm Age} \lesssim 4.5$ Gyr corresponds to star particles that, at $t_{\rm look} = 2.5$ Gyr, were born during the last 2 Gyr. 
This recently born population exhibits a very cold distribution, with values of $\sigma_{\rm Z} < 0.7$ kpc within the inner 8 kpc. Note that the 
contiguous stellar age bin, with ages $< 5.5$ Gyr, shows nearly the same value of $\sigma_{z}$ at all radii. This indicates that the 
disk has not suffered any violent heating episodes during this period. For older stellar
populations, the value of $\sigma_{\rm Z}$ increases systematically at all radii, indicating heating of the disk  either by secular 
evolution or by tidal interactions with other satellite galaxies. The dashed lines show, for 
the same subset of stellar 
particles, the present day value of $\sigma_{\rm Z}$. It is evident that, for $R > 12$ kpc, the younger stellar populations have 
been strongly vertically heated during the last $\sim 2.5$ Gyr. The older age bins are composed of stellar populations that, 
at $t_{\rm look} = 2.5$ Gyr,  were progressively hotter with increasing age. Thus, they are less susceptible to further heating than their 
younger counterparts. Note that the hotter the population was at $t_{\rm look}^{\rm onset}$, the faster the vertical pattern mixes.

\subsection{Where is the torque coming from?}
\label{sec:torque}

We now characterize the source of the external perturbation that, at $t = t_{\rm look}^{\rm onset}$, is exerted on the galactic disk 
and induces the observed vertical pattern. As previously discussed, the required torque could arise from 
a misaligned DM halo with respect to the disk or a close encounter with a companion galaxy.

To explore whether there is a significant misalignment between the disk and the DM halo we compute, at every snapshot time, the DM 
halo inertia tensor,
\begin{equation}
\label{eqn:mten}
\mathcal{M}_{\alpha,\beta} = \frac{1}{M} \sum_{i=1}^{N_{p}} m_{i} r_{i,\alpha}r_{i,\beta},
\end{equation}
considering particles located within different spherical shells. Here $N_{p}$ indicates the total number of particles within each shell. 
Four different shells, defined as $0 < r < 10$ kpc, $10 < r < 25$ kpc, $25 < r < 50$ kpc and $50 < r < 80$ kpc are selected in order to 
characterize separately the behaviour of the DM halo's inner and outer regions. We will refer to them as ${\rm dm}(a,b)$, 
where $a$ and $b$ indicate the lower and upper radial limits of each shell. Once these tensors are computed they can be diagonalized 
to obtain the directions of the principal axes and their relative strengths.

\begin{figure}
\centering
\includegraphics[width=85mm,clip]{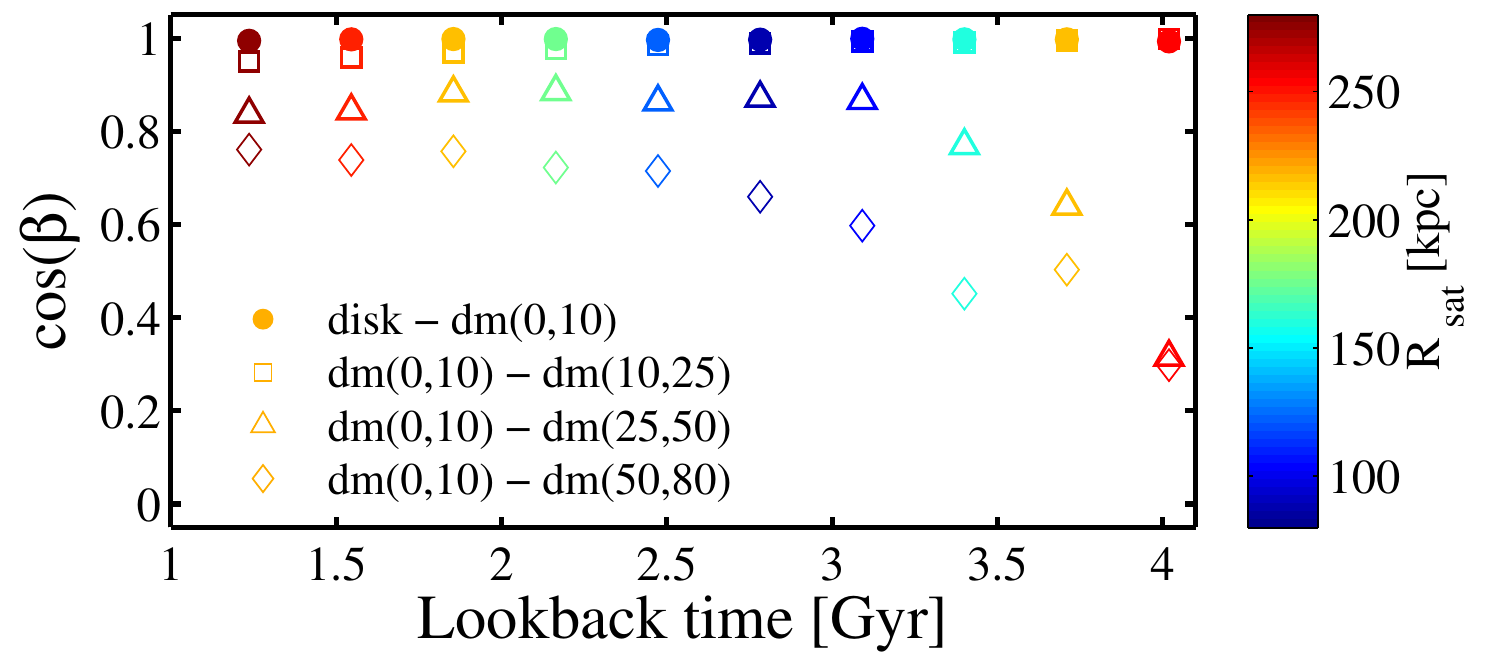}
\caption{The solid dots indicate the time evolution of the angle between the inner disk and the semi-minor axis of the inner DM halo, i.e. dm(0-10). 
Open symbols show the time evolution of the angle between the semi-minor axes of different DM shells, as indicated in the legend, In all cases, 
the colour coding indicates the galactocentric distance of the perturber.}
\label{fig:dm_orients}
\end{figure}

In Figure~\ref{fig:dm_orients} we show, with solid dots, the time evolution of the angle between the angular momentum of the galactic disk\footnote{Using the inertia tensor of the disk to determine its orientation yields the 
same results.} and the semi-minor axis of the innermost DM shell, i.e., ${\rm dm}(0,10)$. The colour coding indicates the galactocentric 
distance of the satellite galaxy identified in Section~\ref{sec:pert_sat}. We find that the inner DM halo
and the galactic disk are very well aligned at all times. The open squares show the time evolution of the angle between the semi-minor
axes of the two innermost DM shells, ${\rm dm}(0,10) - {\rm dm}(10,25)$. Note that these two shells are also very well aligned at all times. 
As we move into the outer regions of the DM halo, a misalignment becomes noticeable. The open triangles and diamonds show the angles 
between ${\rm dm}(0,10) - {\rm dm}(25,50)$ and ${\rm dm}(0,10) - {\rm dm}(50,80)$, respectively. Note that the two outer DM 
shells are also misaligned with respect to each other. The  outermost shell shows the largest misalignments. Our results are in 
very good agreement with previous studies \citep[e.g.][]{2005ApJ...627L..17B,2012MNRAS.426..983D,2013MNRAS.428.1055A}. Using 
cosmological hydrodynamical simulations, 
\citet{2005ApJ...627L..17B} find that, while the inner regions of DM halos, $r < 0.1 R_{\rm vir}$, are well aligned with 
the disk, the outer regions are unaffected by the disk's presence and are often differently oriented. 
In our simulations, $0.1 R_{\rm vir} \approx 24$ kpc. 

For the shell ${\rm dm}(25,50)$ the misalignment with the inner shell, ${\rm dm}(0,10)$,
decreases as the satellite approaches. It reaches a minimum value  before the satellite's pericentre passage and then it remains approximately 
constant until $t_{\rm look} = 1$ Gyr. A similar behaviour is observed for ${\rm dm}(50,80)$, although the misalignment stops decreasing 
just after pericentre. In contrast, we find that, at all times, these four DM shells are very well aligned along their semi-major axes.

From the previous analysis it remains unclear whether the misalignment along the semi-minor axis of the outer DM halo could be the main 
driver of the disk's vertical pattern. Recall that the vertical pattern has a relatively sudden onset 
at $t = t_{\rm look}^{\rm onset}$. As previously discussed, the DM shell ${\rm dm}(25,50)$ reaches a constant orientation 
with respect to the inner halo prior to this onset. In fact, the misalignment is largest before the onset. 
The outermost shell, ${\rm dm}(50,80)$, seems too distant to exert a torque that could vertically perturb the disk.

To clearly identify the main torquing source we compute the time 
evolution of the torque exerted by the previously defined DM shells on a given ring of disk particles. In addition, to use 
as a normalization, we compute the net torque exerted by all DM halo particles within $r < 80$ kpc. 

We focus our attention on the ring of disk particles located within
$14 < r < 15$ kpc since, at this galactocentric radius, the vertical pattern can 
be clearly observed at all times (after $t = t_{\rm look}^{\rm onset}$). 
In order to isolate the torque exerted by the DM particles contained in each individual shell, 
we proceed as follows. First, at every simulation snapshot, disk particles within $14 < r < 15$ kpc are selected. By 
computing and diagonalizing the mass tensor associated with this particle subset, we obtain the orientation of the ring 
with respect to an inertial frame. The whole system is rotated such that the ring's plane is aligned with the X-Y plane.
Subsequently, a set of 2000 test particles\footnote{The number of test particle is chosen 
such that convergence is found in our results.} are uniformly azimuthally distributed at $r = 14.5$ kpc. 
A mass $m_{i} = M_{\rm ring}/2000$
is assigned to each test particle. Here $M_{\rm ring}$ represents the disk stellar  mass enclosed within the ring. 
Finally, the resulting torque on the ring is computed as
\begin{equation}
\mathbf{\tau}_{\rm dm}^{\rm shell} = \sum_{i=1}^{2000}  \mathbf{r}_{i} \times \mathbf{F}_{i}^{\rm shell},
\end{equation}
where $\mathbf{r}_{i}$ represents the galactocentric distance to each test particle and
\begin{equation}
\mathbf{F}_{i}^{\rm shell} = \sum_{j=1}^{N_{\rm shell}} \mathbf{F}_{ij}.
\end{equation}
Here, $N_{\rm shell}$ represents the number of DM particles enclosed within a given spherical shell. Note that since we are only interested in
identifying the torque responsible for the ring's tilting, from now on we will  focus on the torque's component  perpendicular 
to the ring's angular momentum (i.e., in the X-Y plane; the Z-component of the torque only affects the magnitude of the 
angular momentum).   

\begin{figure}
\centering
\includegraphics[width=80mm,clip]{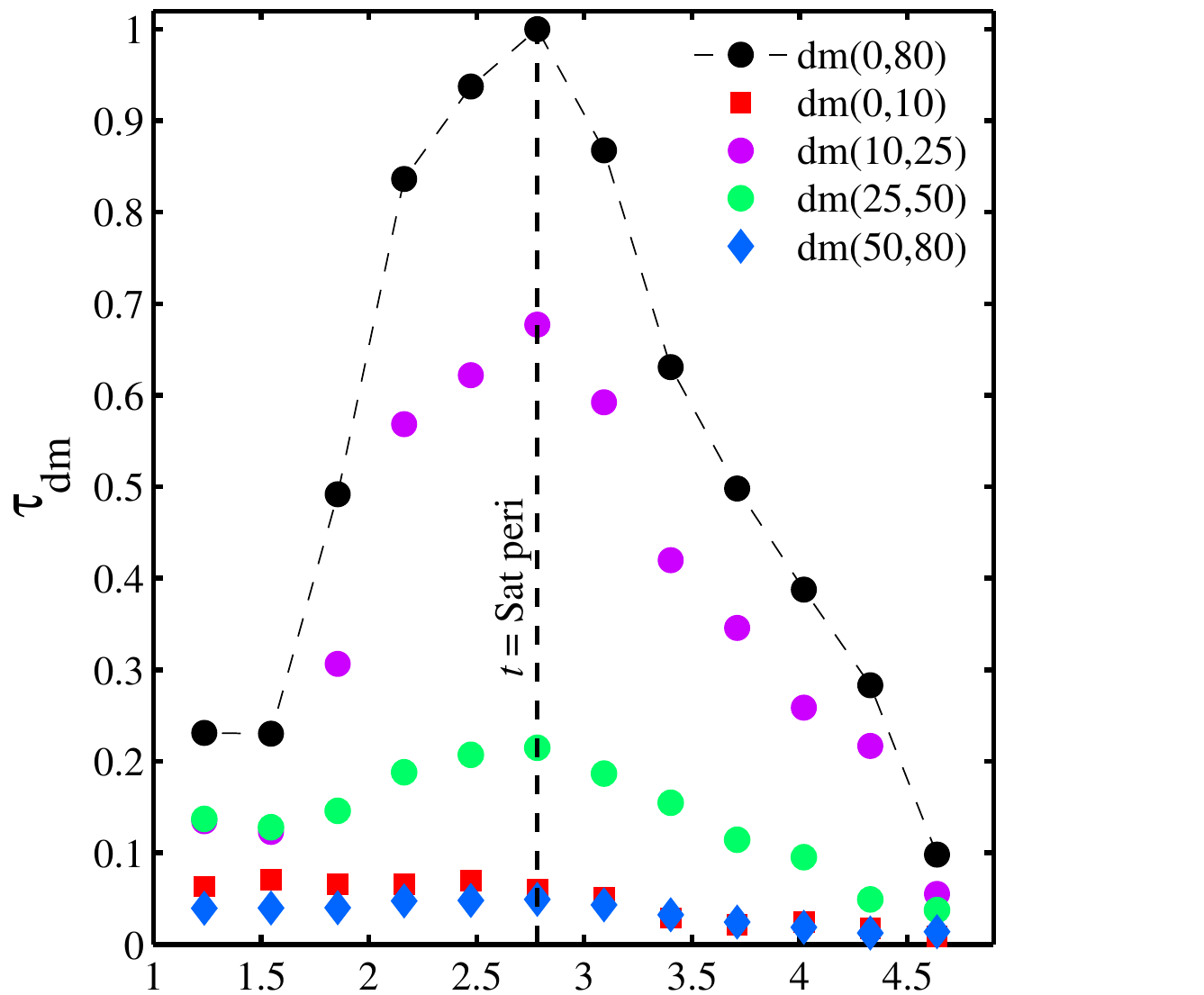}\\
\includegraphics[width=80mm,clip]{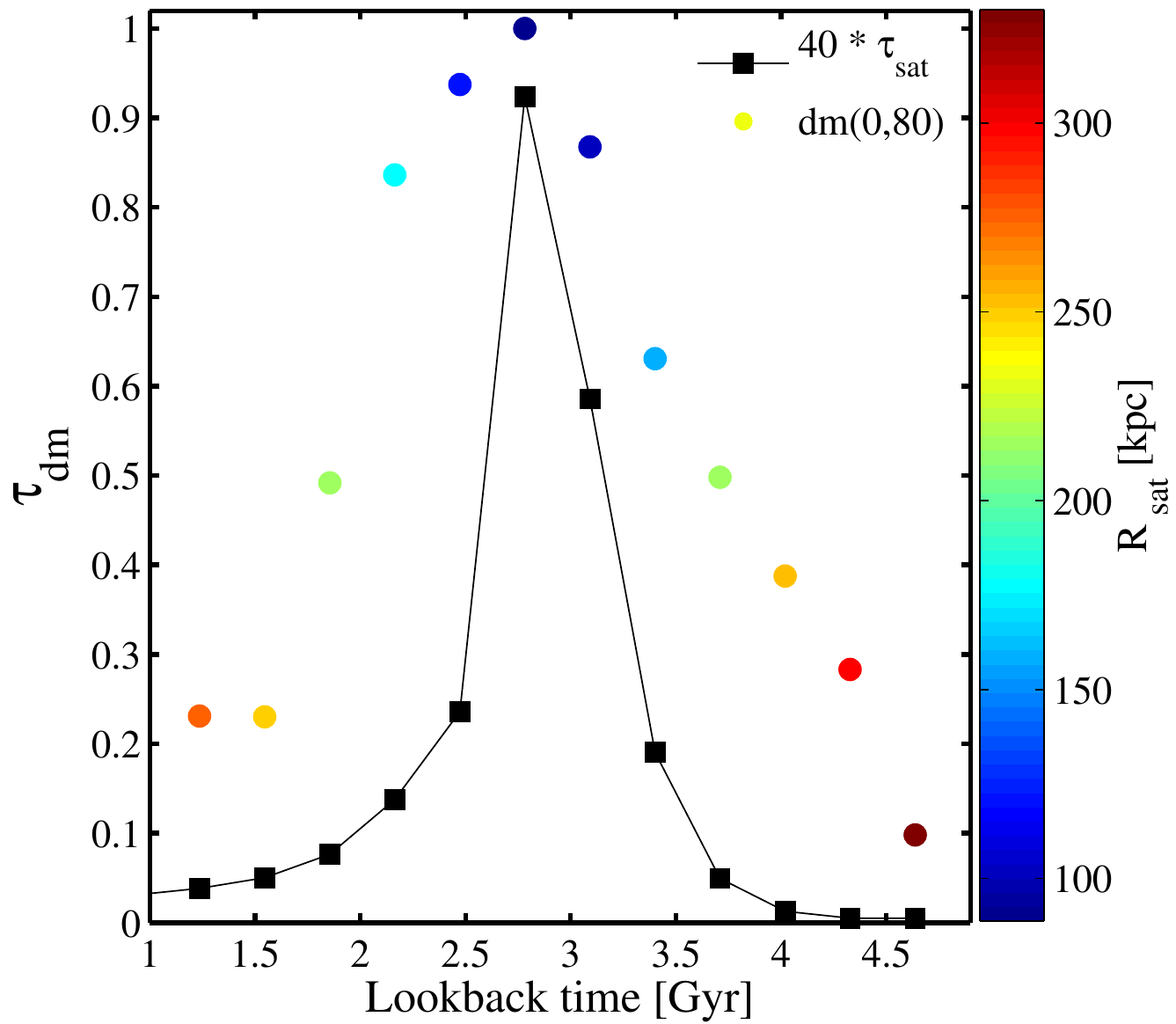}
\caption{{\it Top panel}: Time evolution of the torque exerted on the disk's fraction located within $14 \leq R \leq 15$ kpc. The different symbols 
indicate the torque associated with four different spherical shells of DM, as indicated in the legend. 
In addition, the overall torque exerted by all DM particles within 80 kpc is also shown. The vertical dashed line indicates the fly-by 
pericentre passage time. {\it Bottom panel}: Comparison between the time evolution of the torque exerted by the overall DM halo within 80 kpc 
and that exerted by the perturbing satellite. The colour coding indicates the satellite's galactocentric distance. To allow a direct comparison, the
magnitude of the torque exerted by the satellite has been multiplied by a factor of 40.}
\label{fig:dm_torqs}
\end{figure}

The black dots in the top panel of Figure~\ref{fig:dm_torqs} show the time evolution of the torque associated with the
overall DM halo within 80 kpc, dm(0-80). This figure clearly reveals that the torque increases rapidly 
from $t_{\rm look} \approx 4.5$ Gyr, where it is almost vanishing, to $t_{\rm look} \approx 2.7$ Gyr, where its magnitude peaks. 
Interestingly, this time not only coincides with the onset of the vertical pattern, $t = t_{\rm look}^{\rm onset}$, but
also with the satellite pericentre passage, as indicated by the vertical dashed line. After the torque reaches its 
maximum, it slowly decays again over a period of 2 Gyr to small values. Note that this behaviour does not correlate with the time evolution of 
the DM halo misalignment, at least within the inner 50 kpc. As previously discussed, 
the two innermost DM shells, dm(0-10) and dm(10-25), remain
very well aligned, at least, until $t_{\rm look} = 2$ Gyr. In fact, the shell dm(10-25) becomes slightly more misaligned 
at late times, where we observe the intensity of DM torque to continue its decay. 
The misalignment of the DM shell dm(25-50) ceases to evolve before the pericentre passage, remaining constant for the rest of 
this period. 

To further explore this we show, with differently colored symbols, the contribution to the total DM torque from each shell. 
The red squares show the time evolution of the torque associated with the inner DM shell, dm(0-10). Clearly, this part of 
the DM halo exerts a negligible torque on the ring. The torque exerted by the outer DM shell dm(50-80), indicated with blue diamonds, is also negligible. The green dots show the torque associated with dm(25-50). We find that this shell contributes $\sim 20\%$ of the total DM torque. 
Its evolution follows that of the overall DM halo torque, and is 
uncorrelated with the time evolution of this shell's misalignment. As shown by the magenta dots, $\approx 70\%$ of the DM
torque is coming from the inner shell dm(10-25), a shell that is well aligned with the disk and the inner halo during this period. 
The bottom panel of Figure~\ref{fig:dm_torqs} compares the time evolution of the DM halo torque with that due to the satellite
itself. As before, the dots show the torque exerted by the overall DM halo. The colour coding indicates the satellite's 
galactocentric distance. The black solid squares shows the torque exerted by the satellite\footnote{The torque associated
with the gaseous galactic component is negligible at all times.}. Note that the time evolution
of these two is well correlated, peaking at the satellite's pericentre passage in both cases. However, in this 
panel the satellite's torque has been scaled up by a factor of 40 to allow a direct comparison.

These results show that the torque exerted on the galactic disk is driven mainly by the response of the DM halo to the 
satellite's fly-by, rather than to a  halo - disk misalignment or to the satellite itself. In other words, the DM halo is acting as an
amplifier of the satellite's perturbation. This type of halo - satellite interaction has been studied in the past by
several authors. In particular, \citet{2000ApJ...534..598V} showed that a low--mass, low--velocity fly-by can cause
significant distortions in its host halo that can subsequently perturb an embedded stellar disk. 
In what follows we explore whether this perturbation, known as a DM halo wake, is indeed present in our system.

\subsection{Dark matter halo wake}
\label{sec:dm_wake}

In the previous Section we have shown that a ring of disk particles, at $R=14.5$ kpc, is subject to a strong and transient torque that 
plausibly induces the disk's vertical pattern. The driving force behind this torque is the DM halo, especially its 
inner regions. More than $\gtrsim 70\%$ of the torque comes from the DM distribution in $10 < R < 25$ kpc, whereas the DM halo particles 
outside $R > 50$ kpc have a negligible effect. We have also shown that this transient torque is not correlated with a misalignment between the disk
angular momentum vector and the corresponding spherical DM halo shell.  
Instead, we find its behaviour to be strongly correlated with the orbital phase of the fly-by encounter. 
The DM halo torque peaks exactly at the satellite's pericentre, simultaneously with the torque exerted by the satellite itself.

Our results suggest that the DM halo is acting as an amplifier of the perturbation exerted by the satellite, increasing its 
magnitude by a factor of $\sim 40$. It has been long known  that a satellite galaxy orbiting its host can induce strong distortions 
on the host's DM halo density field. The response of the DM halo to the satellite perturbation is commonly described as a wake and is governed 
by resonant dynamics \citep[see][for a detailed and clear review]{2007PhDT........22C}. 
Resonances between the orbital frequencies of a satellite galaxy and the host halo can induce an exchange of energy and angular momentum. 
For simplicity, let us assume a satellite orbiting on a circular orbit about
a spherically symmetric non-rotating host. A resonance in such system is defined as
\begin{equation}
m \Omega_{\rm sat} = l_{1}  \Omega_{1} + l_{2} \Omega_{2} + l_{3} \Omega_{3}.
\end{equation}
Here $\Omega_{1}$ and $\Omega_{2}$ are the radial and azimuthal orbital frequencies of a dark matter particle, $\Omega_{3}$ the frequency of 
the azimuth of the ascending node and $l_{1},l_{2}, l_{3}$ and $m$ are integer numbers. In a host as defined above 
$\Omega_{3} = 0$ and $|l_{2}| \leq l = m$. The quantity $l$ specifies the order of the resonance. In general, the strongest
resonances are those associated with corrotation, i.e., $l_{2} = m,~l_{1} = 0$. Furthermore, the power of
a resonance decays as $\propto 1/r^{l}$, where r is the distance from the satellite 
\citep[e.g.][]{1989MNRAS.239..549W}. Thus, the lower the order of the 
resonance, the stronger the response. The most relevant resonances are the dipole, $l=m=1$ and the quadrupole, $l=m=2$.
The resulting wake on the DM overdensity field can be thought of as the superposition of different modes excited by the 
resonant interaction between the DM halo and the orbiting satellite. Interestingly \citet{1994ApJ...421..481W} showed 
that the lower order modes, especially $l=m=1$, are very weakly damped and thus, such modes can persist for long periods of 
time \citep[see also][]{2002ApJ...568..190I}. The perturbation induced on the DM halo during the interaction can be 
efficiently transmitted to the inner regions of the halo, where it can affect the structure of an embedded stellar disk 
\citep{1995ApJ...455L..31W,1998MNRAS.299..499W,2000ApJ...534..598V}. 

\citet[][hereafter VW00]{2000ApJ...534..598V} studied the response of a DM halo to a fly-by encounter. This study showed 
that even an unbound satellite that penetrates the outer regions of a host's DM halo can lead to strong asymmetries in the 
inner regions of the host. The strength of the response depends on both the velocity of the satellite and its pericentre
distance. Low velocity and close encounters induce stronger perturbations. For example, a satellite with a mass 
$M_{\rm sat} =0.05~M_{\rm host}$, and a pericentric 
velocity and distance of 200 km/s and 54 kpc respectively, can induce a perturbation with an energy that is comparable to 
the total (background) energy of a MW-like host. The peak of the response occurs after the satellite pericentre passage and  
decays slowly thanks to the weakly damped modes. 

The configuration studied by VW00 is similar to the fly-by discussed in Section~\ref{sec:pert}. In what follows, 
we explore whether this fly-by induces an overdense wake in the host DM halo. Identifying such a wake in a fully cosmological
triaxial DM halo is not a trivial task. Because of its underlying triaxial structure, an overdensity pattern relative to 
a spheroid or axisymmetric ``mean'' DM halo is expected, even without an external perturbation. Furthermore, noise associated with  numerical 
resolution could significantly affect the structure of any induced wake \citep{2007PhDT........22C}. Fortunately, the particle 
resolution in our host DM halo ($N_{\rm dm} \approx 6 \times 10^{6}$)  is large enough to resolve the important resonances, 
at least within the relevant radial range \citep[i.e. $R \gtrsim 5$ kpc,][]{2007PhDT........22C, 2007MNRAS.375..425W}.

As previously discussed, the strongest  $l=m=1$ resonance corresponds to a dipolar response of the DM density field which 
should dominate the morphology of the wake. Note however that higher modes modify the structure so that it 
differs from a perfect dipole. The dipolar response of the halo density field 
can be thought of as a displacement of its center of mass with respect to the center of density or central cusp 
\citep[e.g.][]{1983ApJ...274...53W, 1989MNRAS.239..549W}. The distribution of such asymmetries in the inner regions of dark halos, 
and their correlation to the accretion of external material, have been identified and characterized in detail by 
\citet{2006MNRAS.373...65G} using the Millennium Simulation \citep{2005Natur.435..629S}.

To identify such dipolar structure in our host DM halo, we define a system of reference based on the phase-space coordinates of the 
host's most bound DM particle \citep[see, e.g.,][]{2006MNRAS.373...65G,2007PhDT........22C}. We then obtain a DM overdensity, 
field as follows:
\begin{itemize}

\item We generate a three dimensional grid in polar coordinates, centred on the bottom of the system's potential well. The 
spacing between the grid elements is defined as $[\Delta \theta, \Delta \phi, \Delta r] = [\pi/20,\pi/20,1]$, where 
$\Delta r$ is in kpc. Here $\theta$ and $\phi$ represent galactocentric latitude and longitude, respectively. 

\item For every grid point, we select all DM particles located within a sphere of radius $r_{\rm sp} = 
3 r \sin(\Delta \phi/2)$ and 
compute a local DM density, $\rho_{\rm loc}(\theta,\phi,r)$. The radius of each sphere is allowed to increase 
with galactocentric radius to account for the uniformly spaced  grid in polar 
coordinates. In addition, it serves to smooth out noise due to particle resolution, thus enhancing plausible global 
overdensity features.

\item For every grid point at a given distance $r$, we also select all DM particles within a galactocentric spherical shell 
defined as $r \pm r_{\rm sp}$, and compute a spherical averaged density, $\langle \rho \rangle_{|r}$.

\item Finally, an overdensity field, $\hat{\rho}(\theta,\phi,r)$, is obtained as  
\begin{equation}
\hat{\rho}(\theta,\phi,r) = \frac{\rho_{\rm loc}(\theta,\phi,r) - \langle \rho \rangle_{|r}}{\langle \rho \rangle_{|r}}.
\end{equation}
\end{itemize}

\begin{figure*}
\includegraphics[width=170mm,clip]{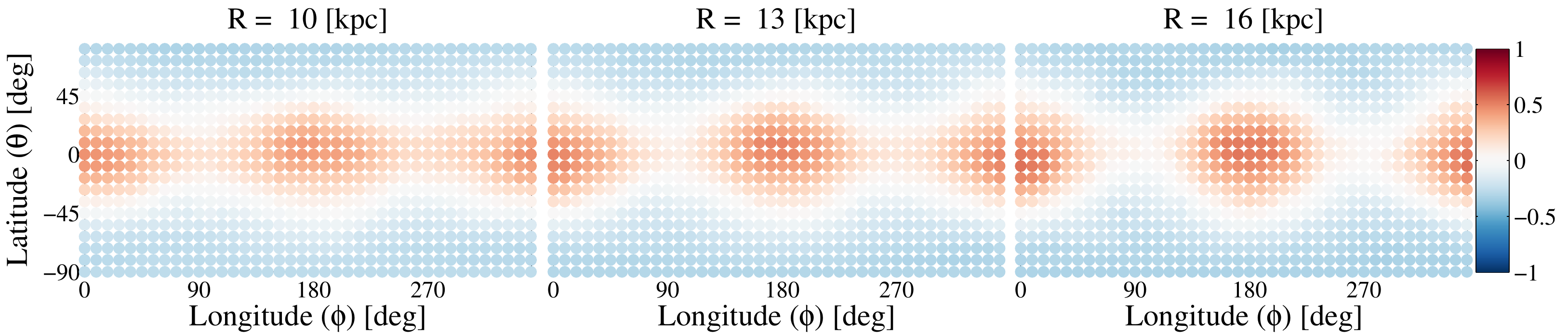}
\\
\includegraphics[width=170mm,clip]{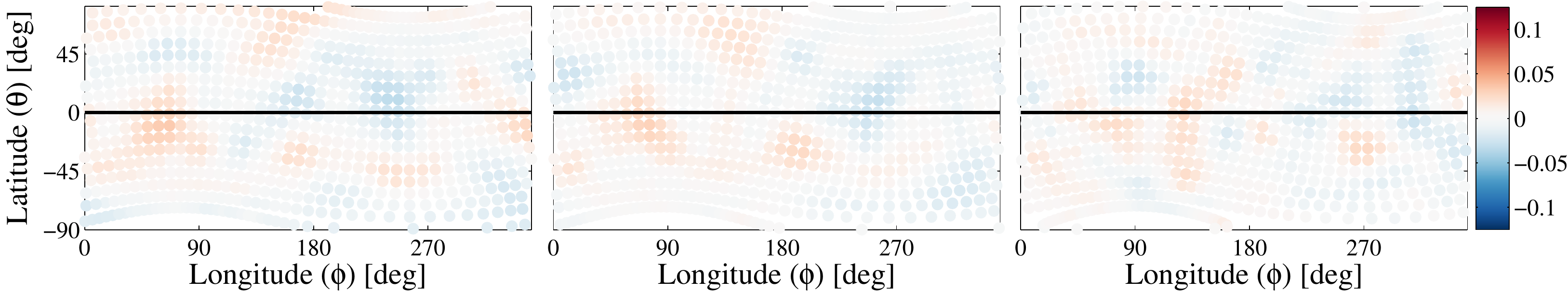}
\\
\includegraphics[width=170mm,clip]{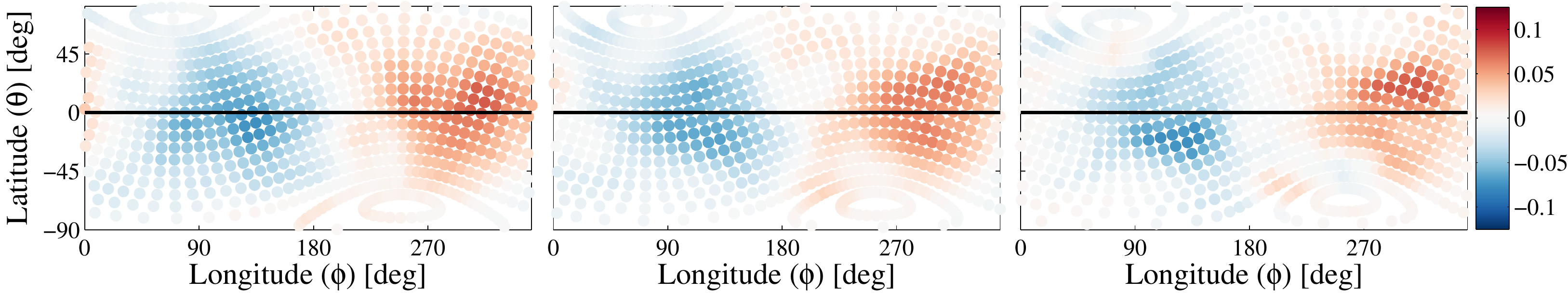}

\caption{{\it Top panels}: Overdensity maps, $\hat{\rho}$, obtained from three spherical shells centred at 
different galactocentric distances. The maps are computed $\approx 1$ Gyr prior the satellite's pericentre passage 
($t_{\rm look} \approx 3.7$ Gyr) and oriented with respect to the direction of the shell's 
principal axes. The triaxial shape of the DM halo can be clearly appreciated in these maps, with the
semi major, intermediate and minor axis pointing in the directions $(\theta,\phi) = (0,180^{\circ})$, $(0,90^{\circ})$ and $(90^{\circ},0)$,
respectively. {\it Middle panels}: Overdensity maps, $\hat{\rho}_{\rm dipole}$, obtained  after {\it i)}  rotating $\hat{\rho}$ 
by $180^{\circ}$ in $\phi$, {\it ii)} flipping the resulting maps about $\theta = 0^{\circ}$, and {\it iii)} 
subtracting the resulting rotated maps from  $\hat{\rho}$. As above,  maps are 
obtained at ($t_{\rm look} \approx 3.7$ Gyr). The goal of this procedure is to erase quadrupolar features while enhancing any 
plausible underlying dipole associated with the DM halo wake. {\it Bottom panels}: As in middle panels for the snapshot just after 
the satellite pericentre passage ($t_{\rm look} \approx 2.5$ Gyr). Note the strong dipolar signature at this time. 
Middle and bottom maps have been re-oriented such that the $\theta = 0^{\circ}$ plane is perpendicular to the inner stellar disk angular momentum 
vector.}
\label{fig:local_dipole}
\end{figure*}

\begin{figure*}

\includegraphics[width=37.3mm,clip]{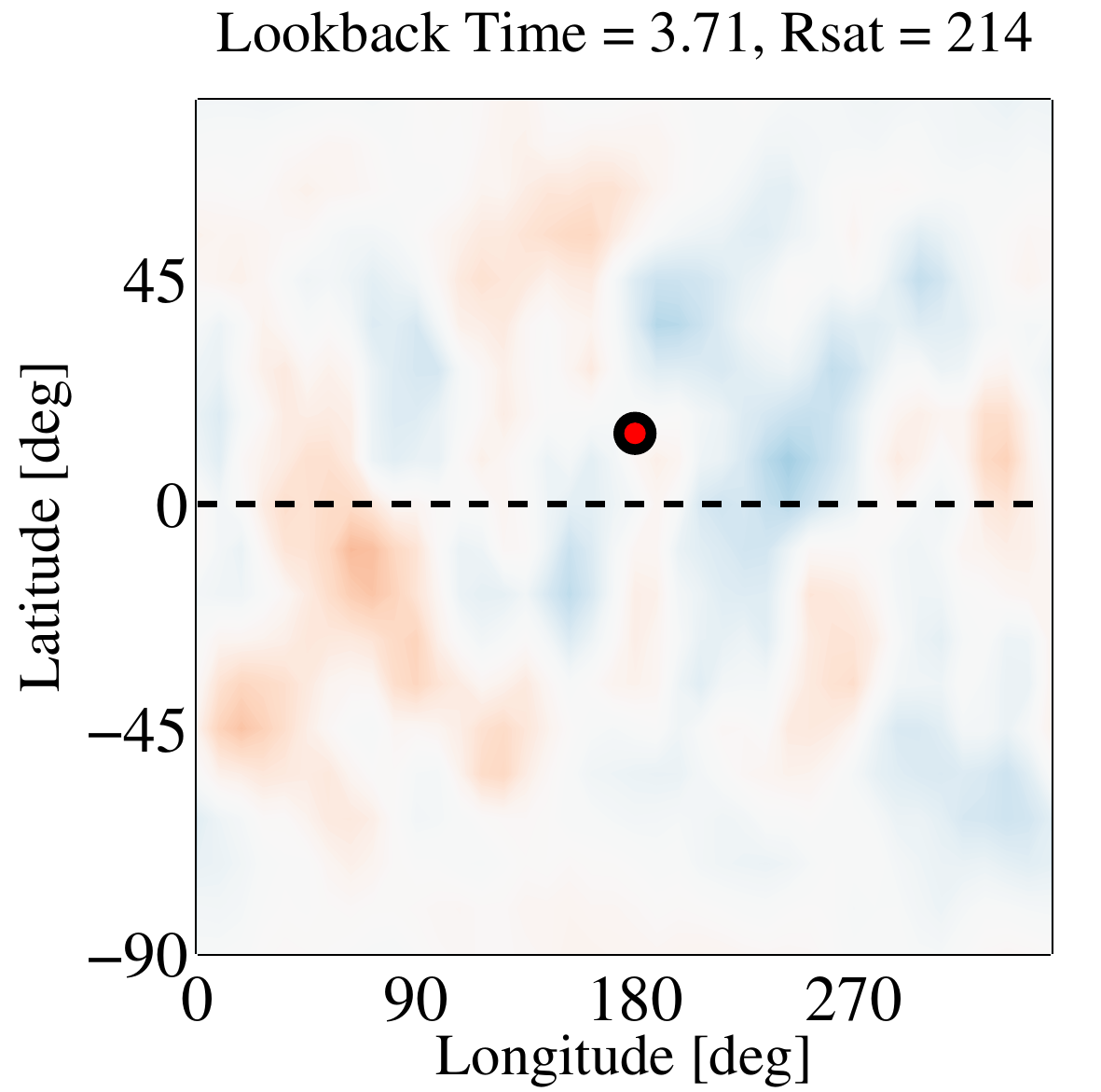}
\hspace{-0.1cm}
\includegraphics[width=32mm,clip]{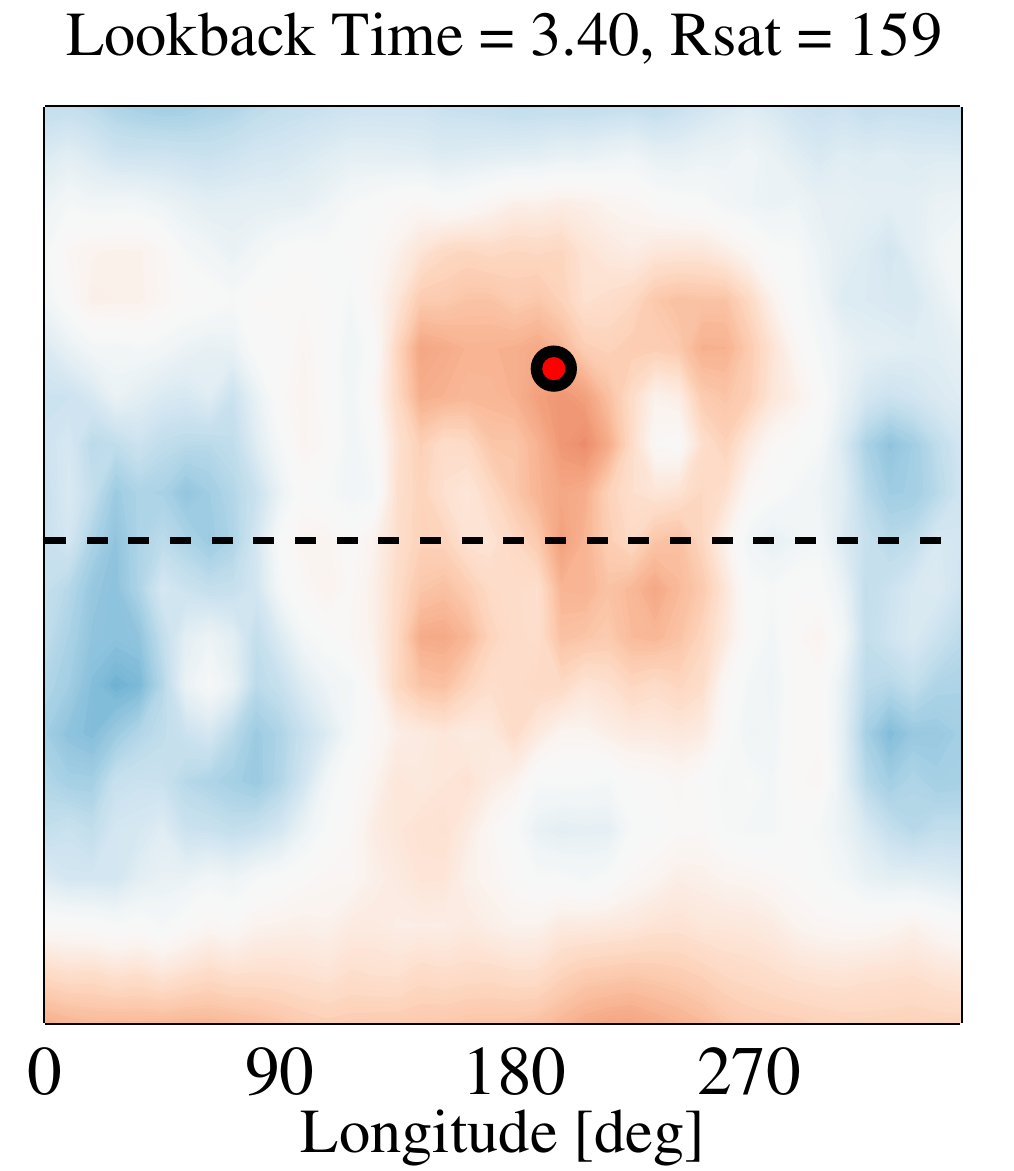}
\hspace{-0.1cm}
\includegraphics[width=32mm,clip]{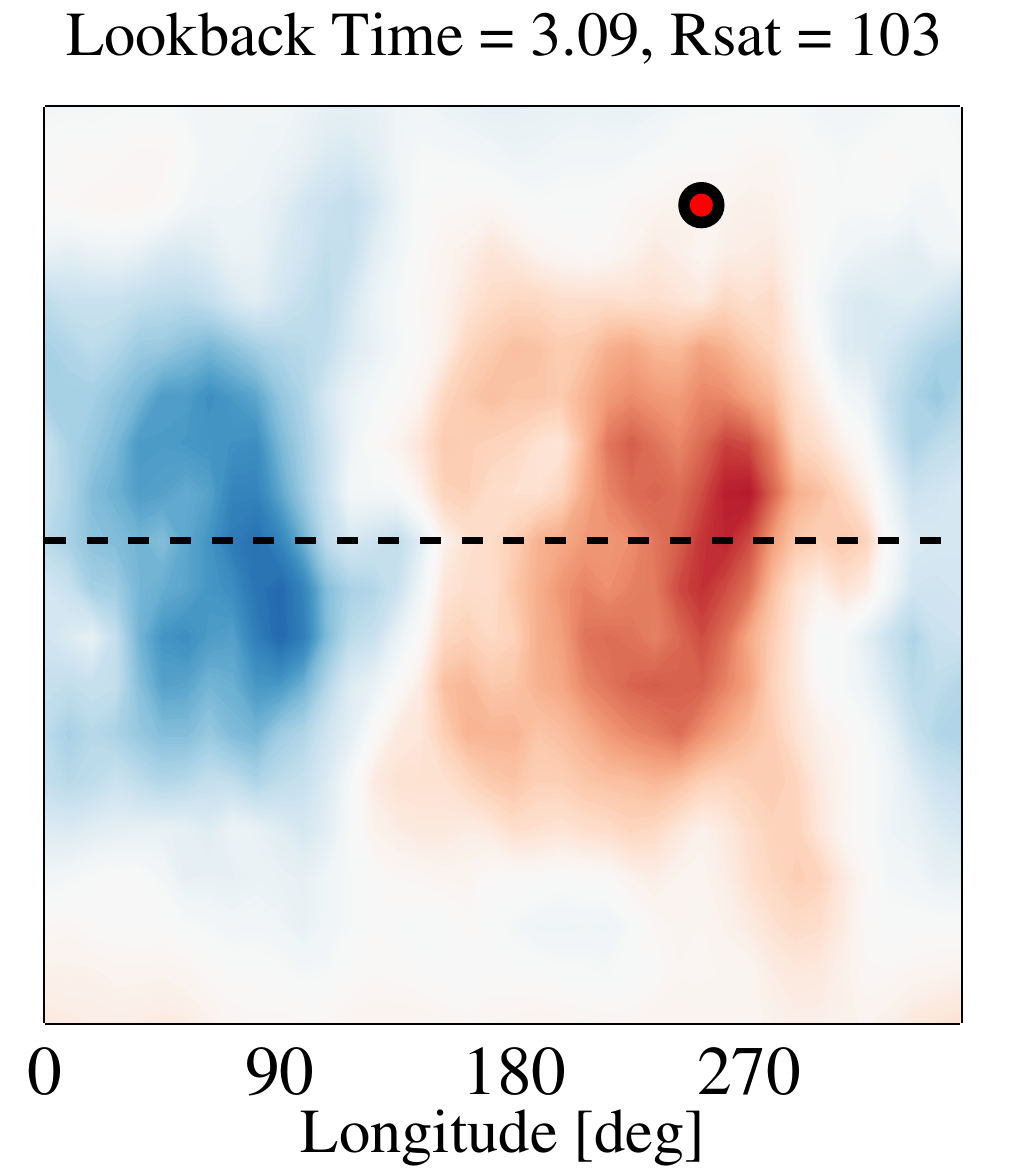}
\hspace{-0.1cm}
\includegraphics[width=32mm,clip]{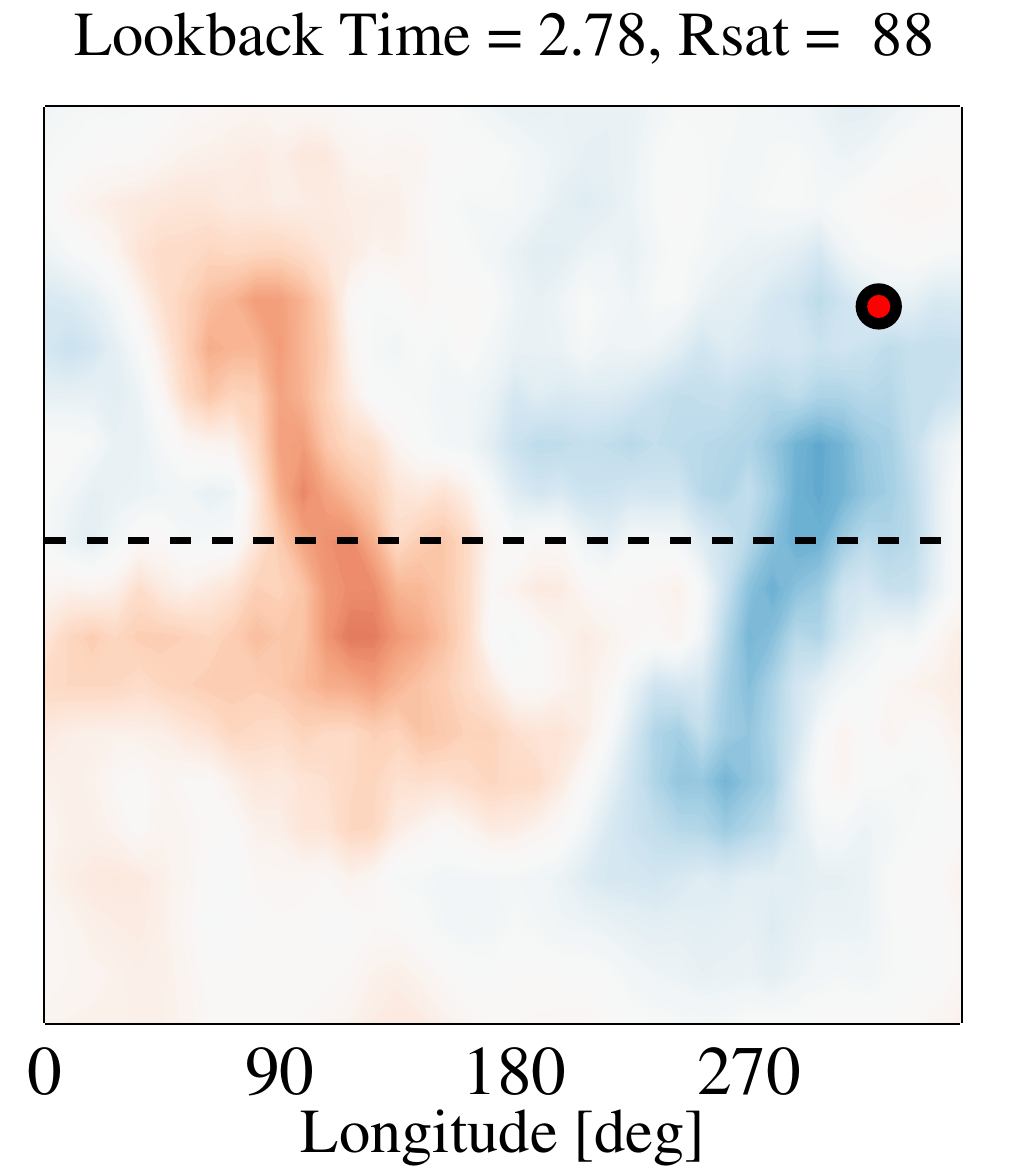}
\hspace{-0.1cm}
\includegraphics[width=38.9mm,clip]{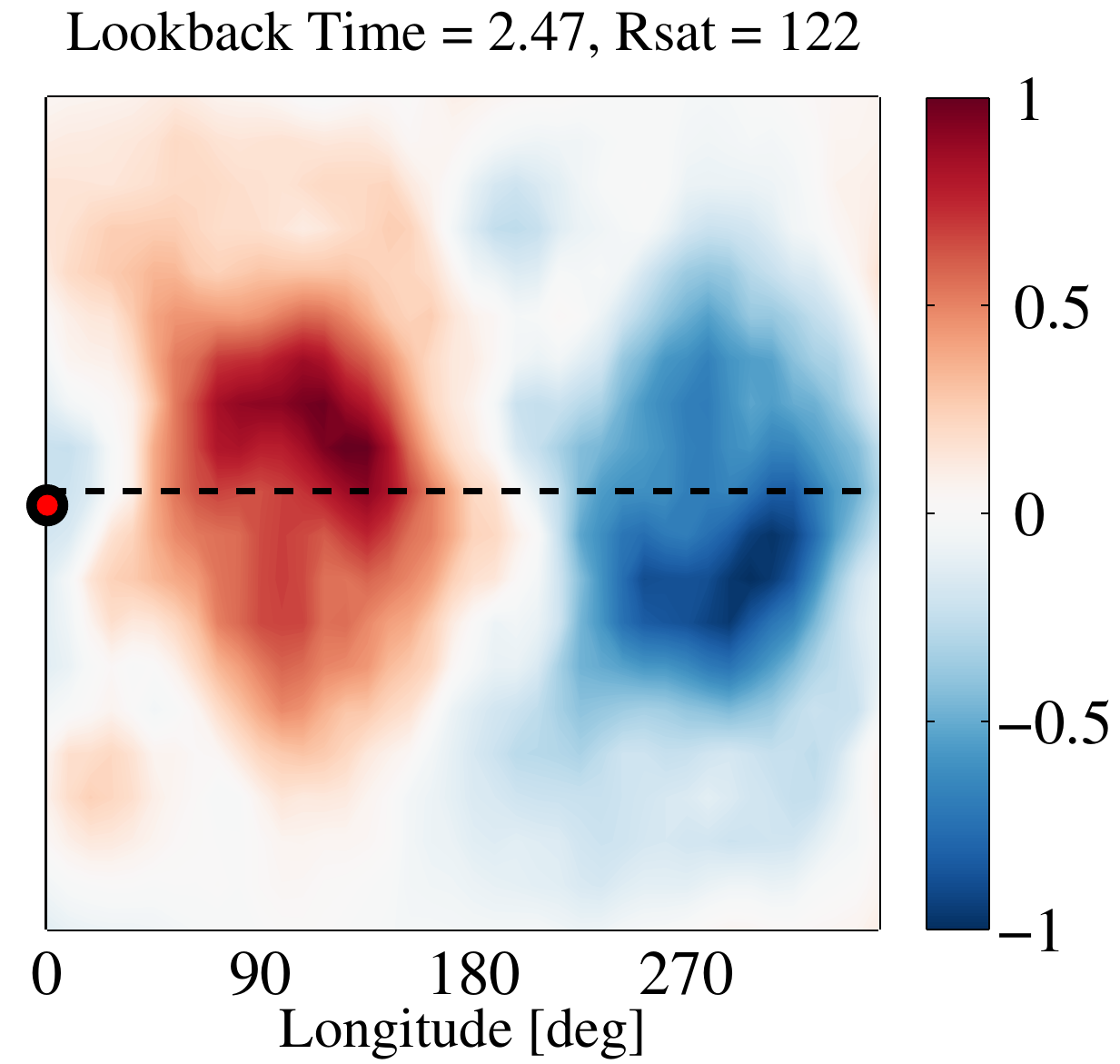}
\\
\includegraphics[width=37.3mm,clip]{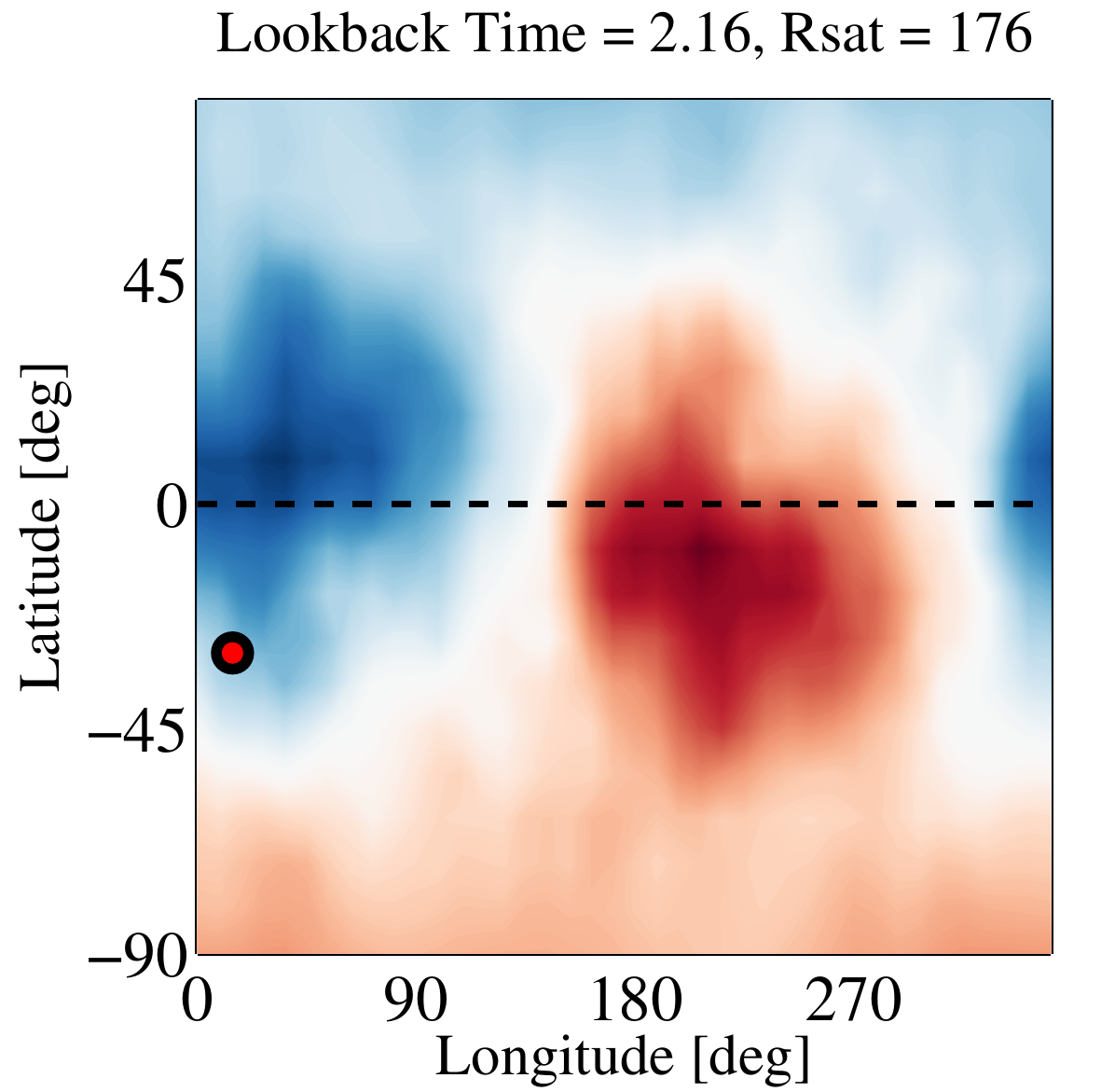}
\hspace{-0.1cm}
\includegraphics[width=32mm,clip]{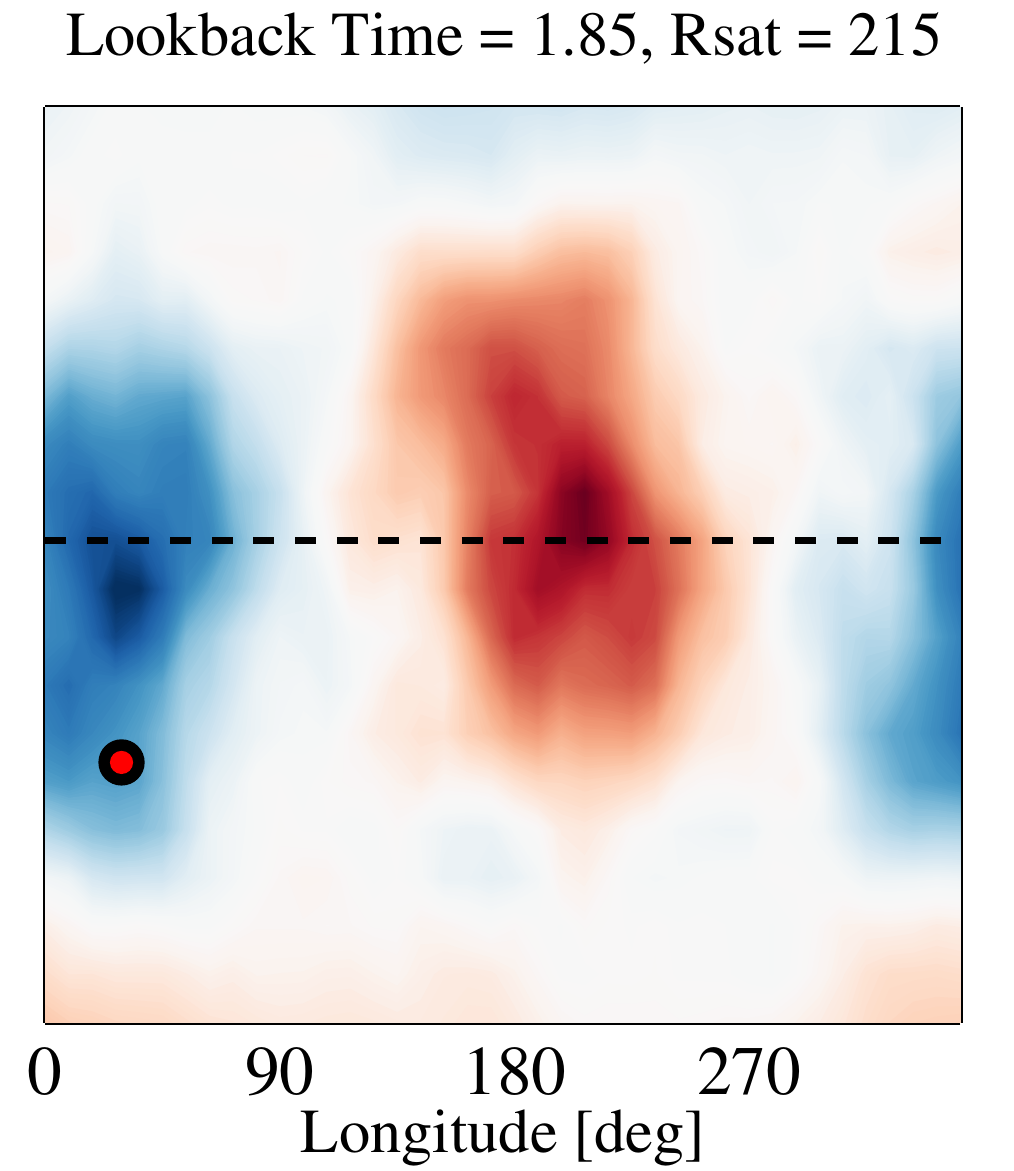}
\hspace{-0.1cm}
\includegraphics[width=32mm,clip]{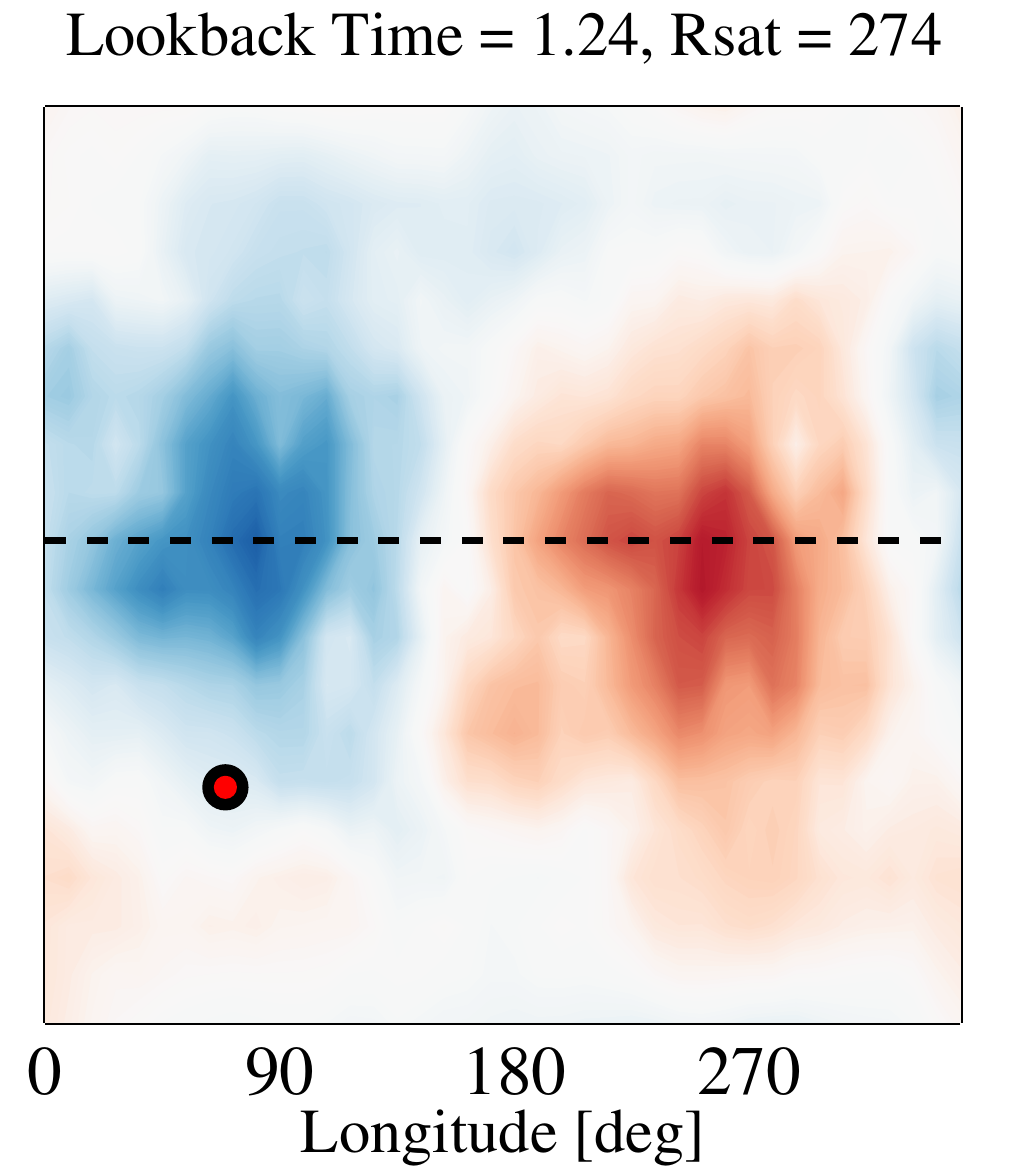}
\hspace{-0.1cm}
\includegraphics[width=32mm,clip]{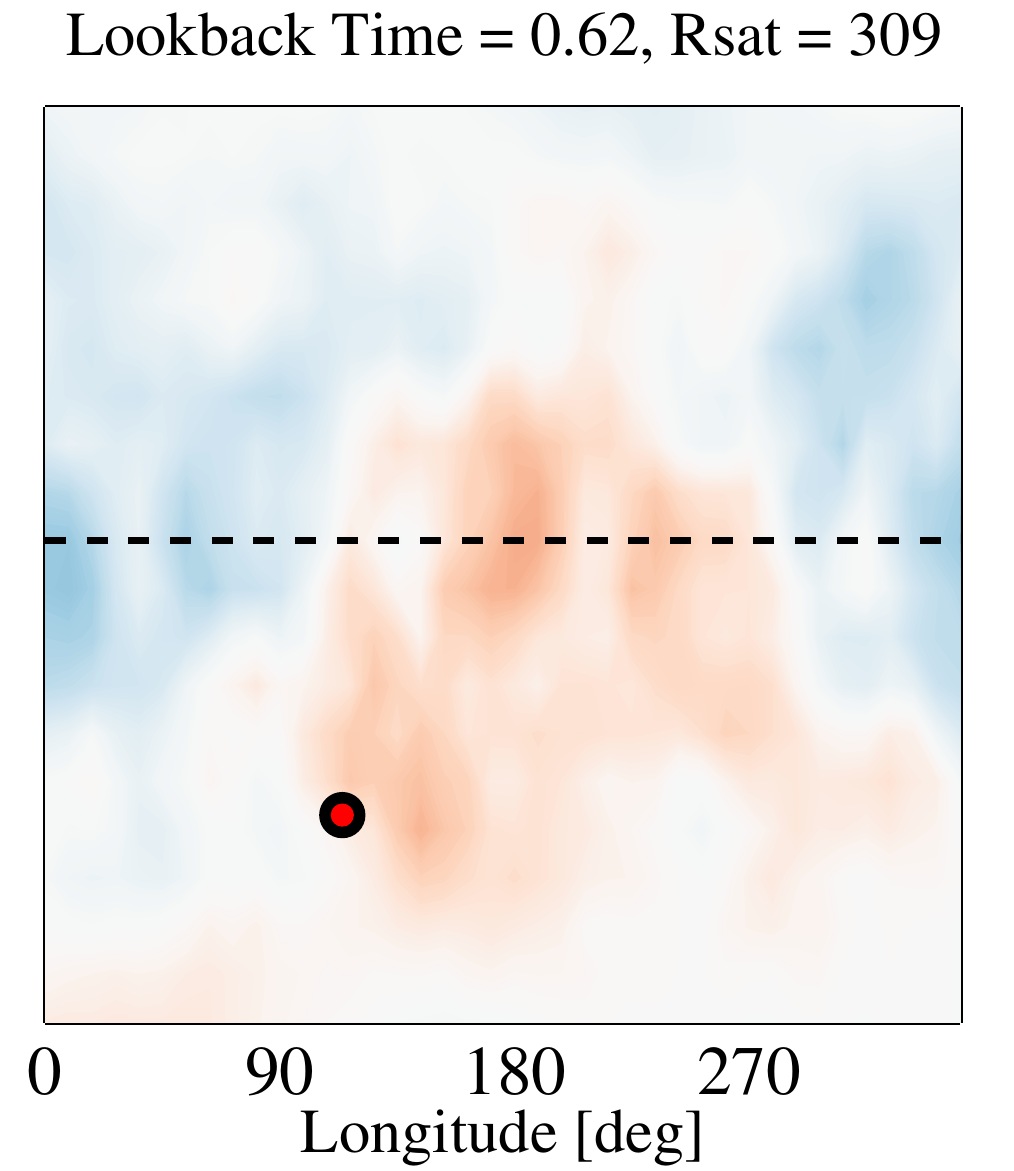}
\hspace{-0.1cm}
\includegraphics[width=38.9mm,clip]{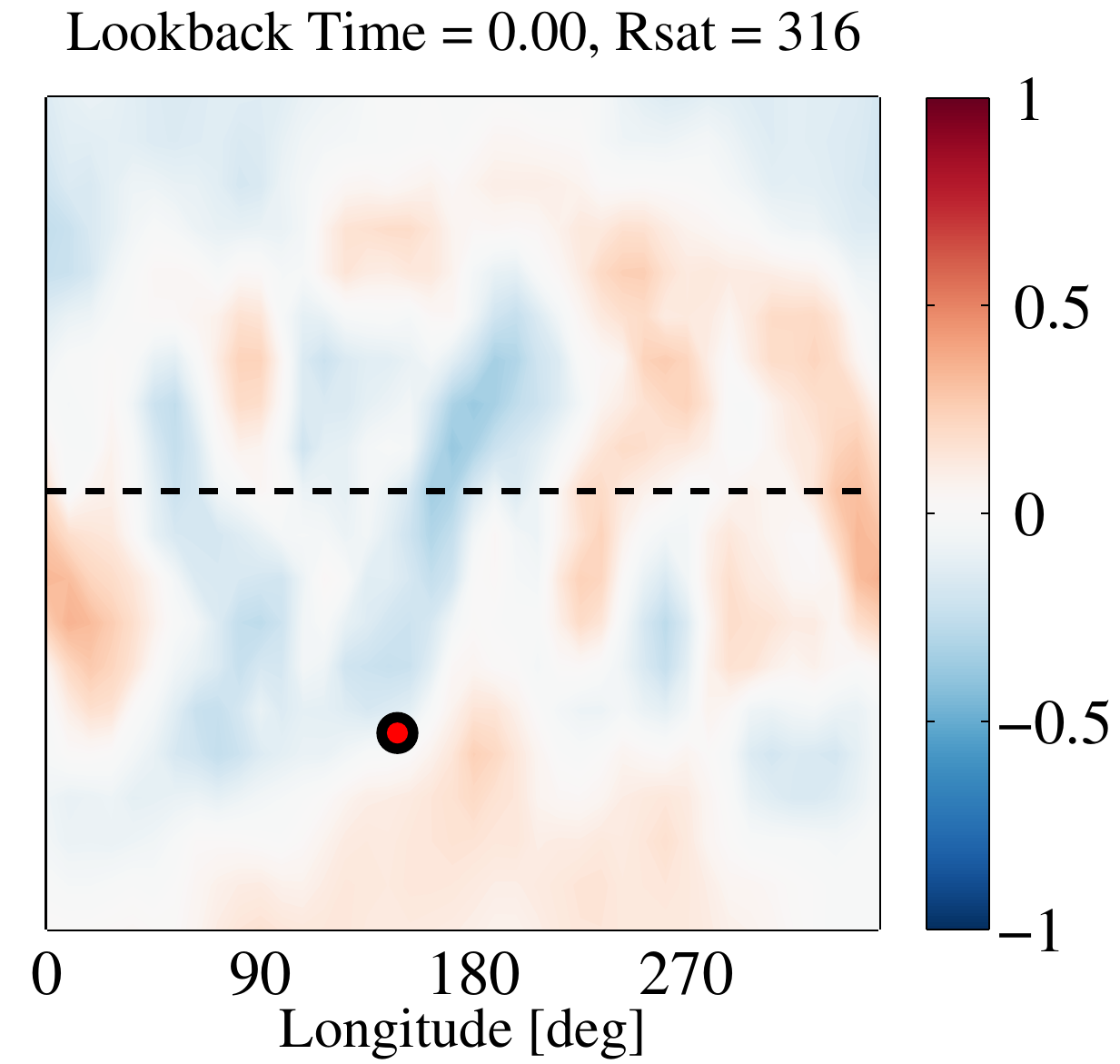}

\caption{Time evolution of the dipolar mode excited on the host DM halo due to the interaction with the fly-by encounter. 
These maps are the result of stacking, at every snapshot, the resulting $\hat{\rho}_{\rm dipole}$ obtained 
from $r = 2$ to 20 kpc (see text for details). All maps are normalized to the same value and oriented with respect to the inner 
galactic disk. The colour coded dot indicates the location of the fly-by. The galactocentric distance of the fly-by, $R_{\rm sat}$, is
indicated on the top of each panel.}
\label{fig:dipole_evol}
\end{figure*}

The top panels of Figure~\ref{fig:local_dipole} show overdensity maps obtained 1 Gyr prior to the satellite's 
pericentre passage ($t_{\rm look} \approx 3.7$ Gyr) at three different galactocentric distances. 
Since our DM halo is triaxial, we have oriented the maps with respect to the direction of  the shell's 
principal axes. This orientation is derived from equation~(\ref{eqn:mten}) considering only DM particles enclosed
within the corresponding shell. The triaxial shape of this halo can be clearly appreciated in these maps, with the
semi major, intermediate and minor axis pointing in the directions $(\theta,\phi) = (0,180^{\circ})$, $(0,90^{\circ})$ and $(90^{\circ},0)$,
respectively. To explore whether a dipolar mode is concealed by the triaxial shape of the halo we first rotate each map 
by $180^{\circ}$ in  $\phi$,  then flip the resulting maps about $\theta = 0^{\circ}$, and finally compute
\begin{equation}
\hat{\rho}_{\rm dipole} = (\hat{\rho} - \hat{\rho}^{180}_{\rm rot})/2.
\end{equation}
The result of this procedure is to erase quadrupolar features while enhancing any plausible underlying dipole associated
with the DM halo wake\footnote{Note that any quadrupolar mode excited by the DM halo -- satellite interaction would also be
erased.}. The second row of panels shows $\hat{\rho}_{\rm dipole}$ at the same time as above. These maps have been re-oriented 
such that the $\theta = 0$ plane is perpendicular to the angular momentum vector of the inner stellar disk ($R \leq 5$ kpc).  
Note that, at these radii, we obtain $\hat{\rho}_{\rm dipole}$  maps without structure after removing the quadrupolar feature, except 
for some residual noise. In contrast, the third row of  panels shows the same $\hat{\rho}_{\rm dipole}$, 0.3 Gyr after the satellite
pericentre passage ($t_{\rm look} \approx 2.5$ Gyr). A very clear dipolar feature, with an amplitude of 
$\approx 0.15 \langle \rho \rangle_{|r}$ can now be observed. This is a direct indication that a wake has been excited in the host by
the satellite.
In Figure~\ref{fig:dipole_evol} we follow the time evolution of this dipolar feature. To enhance its signature we stack all 
the resulting $\hat{\rho}_{\rm dipole}$ obtained from $r = 2$ to 20 kpc. At every snapshot we look for the peak value of the 
resulting map. We use the maximum of these peak values to normalize all maps. As before, maps are oriented with respect to the inner 
galactic disk. The colour coded dot indicates the direction of the satellite, as seen from the galactic centre. 
Before the satellite pericentre passage (top leftmost panel), at $t_{\rm look} = 3.7$ Gyr, the map does not show any dipole signatures. As 
the satellite approaches, a clear signature grows. As expected from VW00, this peaks just after
 pericentre passage and then slowly degrades. By the present day (bottom rightmost panel), the dipolar signature has vanished. 

\section{Discussion and conclusions}
\label{sec:discussion}

In this work we have analyzed a fully cosmological hydrodynamical simulation of the formation
of a Milky Way-like galaxy. As discussed by M14, the final system is strongly disk--dominated with a 
realistic rotation curve, a surface density profile close to exponential, and both stellar age distribution and size
that are consistent with expectations from large galaxy surveys in the local Universe. 

We have focused our interest on the evolution of the vertical structure of the stellar disk. For this
purpose we have created maps of the mean height of the disk within $R \leq 20$ kpc at various times. Our 
results show the present day disk configuration to have a strong and well defined vertical 
pattern, with amplitudes as large as $\langle {\rm Z} \rangle \approx 3$ kpc. The pattern becomes noticeable at 
$R \approx 13$ kpc, and increases  its amplitude as we move away from the galactic center. The inner disk,
$R \leq 12$ kpc has a smoother vertical structure, with $\langle {\rm Z} \rangle \approx 0$ kpc. This 
configuration is reminiscent of what has been observed in the  outer Galactic disk 
\citep{2014ApJ...791....9S,2015ApJ...801..105X,2015arXiv150308780P}. Indeed, the outer Milky Way (MW) disk exhibits 
 an oscillating asymmetry whose amplitude increases with Galactocentric radius, with material visible at large Galactic
 latitudes ($\sim 30^{\circ}$).
The mean vertical velocity of our simulated disk, $\langle v_{\rm z} \rangle$, has a similar structure
 to $\langle {\rm Z} \rangle$, with values as large as 65 km/s in the outer disk. 
 By contrasting the phases of the $\langle {\rm Z} \rangle$ and the $\langle v_{\rm z} \rangle$ maps, we have clearly exposed 
 the oscillatory behaviour of the vertical pattern.

Maybe the most interesting feature of the disk's present day vertical structure is the large span in 
galactic longitude ($\approx 180^{\circ}$) covered by this pattern. By creating maps of star particle 
counts, we have compared its structure to the Mon ring as seen by Pan-STARRS1 
(S14). As shown by S14, the Mon ring exhibits a complex 
morphology with both stream-like features and sharp edges both north and south of the disk. 
A comparison  between our mock data set and the Mon ring reveals considerable similarities. 
Material from the simulated galactic disk can be found at the correct galactic latitudes
at all radii and on both side of the disk. 
Previous numerical attempts to reproduce the Mon ring based on models in which 
the disk is strongly distorted by an accretion event have failed to deposit material at the correct heights \citep[e.g.][]{2009ApJ...700.1896K,2013MNRAS.429..159G,2015arXiv150308780P}. Furthermore, well defined 
arcs of disk material can be found at large and intermediate galactocentric distances, in agreement with 
the structure observed by S14. Unlike previous  studies, where these arc-like structures were associated with
tidal debris from disrupted satellites \citep[e.g.][]{2006ApJ...651L..29G,2011ApJ...738...98G}, S14 showed 
that they  have positions and morphologies that are suggestive of a connection to the Mon ring. 

To characterize the origin of this vertical pattern in our simulated galaxy, we have followed the time evolution of the 
system and explored the main sources of perturbations. The onset of the pattern is relatively sudden and takes place approximately 
at $2.5 < t_{\rm look} \lesssim 3$. Initially, it shows an $m=1$ configuration but quickly starts to 
wind up into a leading spiral. This indicates that the external source of the torque, responsible for the onset
of the perturbation, has decayed over time leaving behind a misaligned outer disk. As discussed by SS06, after
this point the main torque on this misaligned outer material comes from the inner disk. This causes 
the outer disk to precess retrograde at a rate that decreases with galactocentric distance, giving rise to the 
leading spiral shape. 

At $t_{\rm look}^{\rm onset}$ all star particles in the disk are similarly perturbed, regardless of their 
age. This rules out a scenario in which the initial warp reflects the formation of a misaligned outer disk 
from newly accreted material \citep[e.g.][]{2010MNRAS.408..783R}. However, we find that, at the present day,
the vertical pattern is significantly more coherent in the subpopulation of star particles that were born just 
before $t_{\rm look}^{\rm onset}$. Older populations were previously heated by secular evolution or
previous accretion events and, as a result, the vertical pattern in their distributions more rapidly mixes. 

To identify the main source driving this vertical pattern, we have characterized the distribution of satellite galaxies
that interact with our host at $t_{\rm look}^{\rm onset}$. The most significant perturber is a low-velocity fly-by with a
pericentre passage at $t_{\rm look} \approx 2.7$ Gyr. This satellite has a total mass of $\sim 4 \times 10^{10} M_{\odot}$, a
pericentre at $\sim 80$ kpc and a velocity at pericentre of $\sim 215$ km/s. 
Even though it is not massive enough to directly perturb the galactic disk, 
we find that it significant distorts the halo density field which subsequently perturbs the embedded stellar disk. 

The mechanism underlying this satellite -- halo -- disk interaction was presented by \citet{2000ApJ...534..598V}. 
The response of the halo is determined by a resonant interaction with the satellite. 
Its strength depends not only on the mass of the perturber, but also
on its pericentre distance and velocity. The lower the velocity, the stronger the response, which manifest as a 
density wake that can be represented as a superposition of several resonant modes. 
The strongest of them are the dipole and quadrupole which are only weakly damped. 
In particular, the dipolar response of the halo can be thought of as a displacement of the halo center of mass with respect to the central
density cusp. To search for such a dipolar response, we created halo overdensity maps at various times, centering our coordinates on the 
bottom of the potential well. This analysis showed a clear and strong dipolar response in the inner halo ($R \leq 20$ kpc) 
that is well correlated with the satellite pericentre passage. 
As expected from VW00, the dipolar perturbation peaks just after the satellite's pericentre and can be observed for, at least, 1 Gyr after 
this moment. Indeed, we  find that the torque exerted on the stellar disk comes, almost completely, from the inner  halo. 
The time evolution of the halo torque and that exerted by the satellite itself are well correlated, even though the satellite is not itself 
massive enough to significantly perturb the disk. The halo is acting to strongly amplify the satellite's perturbation. 
Approximately $70\%$ of the torque  comes from halo material lying at radii between 10 and 25 kpc. This inner halo is well 
aligned with the galactic disk during the generation of 
the vertical pattern,  confirming that the torque is not primarily driven by a misalignment between the principal axes of the disk and halo. 

The mechanism driving the vertical pattern in our simulated disk could, at least, be partially responsible for the perturbed 
vertical structure of our Galactic disk. It is thus interesting to consider what satellite could have induced a density wake 
in the Galactic halo to perturb the outer Galactic disk. The first plausible candidate is the Sagittarius dwarf galaxy (Sgr). As previously
discussed, a model of the interaction between this galaxy and the Galactic disk can account for many  of the global and local morphological features 
observed in the Galactic disk \citep{2011Natur.477..301P,2012MNRAS.423.3727G,2013MNRAS.429..159G,2015arXiv150308780P}.
These models, however, fail to deposit enough material at the required height to reproduce the observed distribution
of main sequence turn-off stars and RR Lyrae in the outer MW disk \citep[S14,][]{2015arXiv150308780P}. An important assumption in these simulations 
relates to initial set-up of the Sgr-like models. The satellites are launched 80 kpc from the galactic centre in the plane of the MW-like disk,
with their DM halo mass profiles truncated at the instantaneous Jacobi tidal radius. This reduces the initial bound mass of the Sgr models,
of the order of $10^{11}~M_{\odot}$, by a factor of $\sim 3$. Perturbations on the host halo density field that could have occurred during  
the virial-radius infall and this ‘initial’ location are thus neglected. To study whether such perturbations (and their associated density wake) 
are strong enough to bring the structure of the simulated disk into better agreement with observations, new simulations that initially place 
Sgr models outside the host's virial radius are required. We defer this analysis to a follow up study.  

Another plausible perturber is the Large Magellanic Cloud (LMC). Recent studies based on the SMC-LMC orbital history before infall into the 
MW and on the dynamical state of galaxies in the Local Volume suggest a total LMC mass at infall of $\approx 2 \times 10^{11}~M_{\odot}$ 
\citep[][]{2012MNRAS.421.2109B,2015arXiv150703594P}. Thus, this galaxy could be massive enough to significantly perturb the Galactic disk due 
both to its own tidal field and to the subsequent halo density wake. Note however that current measurement of the LMC proper motion 
suggest that this galaxy is currently undergoing its first pericentre passage \citep{2007ApJ...668..949B, 2013ApJ...764..161K}. 
Thus, it remains to be studied whether the mechanism
discussed in this work has had enough time to act. Clearly, new numerical models of the Milky Way-LMC interaction are required in order
to reassess the impact of our most massive companion on the morphology of our own Galaxy.

\section*{Acknowledgements}
FAG would like to thank Monica Valluri, Eric Bell, and Victor Debattista for very useful discussions and suggestions. 
RG and VS acknowledge support through the DFG Research Centre SFB-881 'The Milky Way System' through project A1.
VS and RP acknowledges support by the European Research Council under ERC-StG grant EXAGAL-308037.

\bibliographystyle{mn2e}
\bibliography{cosmomono}

\label{lastpage}

\end{document}